\documentclass[aps,prl,showkeys,superscriptaddress,footinbib,longbibliography,reprint]{revtex4-2}

\usepackage{graphicx}
\usepackage{subfigure}
\usepackage{amsmath}
\usepackage{amssymb}
\usepackage{dcolumn}
\usepackage{times}
\usepackage{bm}
\usepackage{color}
\usepackage{lipsum}
\usepackage{multirow}
\usepackage{url,hyperref}
\usepackage[normalem]{ulem}

\hypersetup{
colorlinks=true,
linkcolor=blue,
citecolor=blue,
urlcolor=blue,
}

\begin{document}

\title{Asymmetry-induced radiative heat transfer in Floquet systems}

\author{Hui Pan}
\email[]{panhui@nus.edu.sg}
\affiliation{Department of Physics, National University of Singapore, Singapore 117551,
Republic of Singapore
}

\author{Yuhua Ren}
\affiliation{Department of Physics, National University of Singapore, Singapore 117551,
Republic of Singapore
}

\author{Gaomin Tang}
\email[]{gmtang@gscaep.ac.cn}
\affiliation{Graduate School of China Academy of Engineering Physics, Beijing 100193,
China
}

\author{Jian-Sheng Wang}
\email[]{phywjs@nus.edu.sg}
\affiliation{Department of Physics, National University of Singapore, Singapore 117551,
Republic of Singapore
}

\date{\today}

\begin{abstract}
Time modulation opens new avenues for light, heat control, and energy harvesting, yet the impact of nonequilibrium dynamics of microscopic particles remains largely unexplored.
We develop a microscopic theory to describe radiative heat transfer in such Floquet systems.
Significant heat transfer occurs due to differences in electronic properties between parallel metal plates, despite identical driving protocols and temperatures.
This arises from a unique exponential-staircase distribution of radiative photons, induced by nonequilibrium electronic fluctuations, and can be tuned via both microscopic properties and driving parameters.
Our work highlights the importance of nonequilibrium microscopic details, unlocking new opportunities for active cooling, thermophotovoltaics, thermal imaging and manipulation, and carrier dynamics probing.
\end{abstract}

\maketitle

\textit{Introduction.--}
Heat transfer through radiation, spanning the near- to far-field regimes, relies on the interplay of electromagnetic fields and charge fluctuations within matter, closely related to thermophotonics~\cite{Harder2003,Fan17} and Casimir physics~\cite{Klimchitskaya2009}.
A strong focus lies on the aim to better understand and control radiative heat transfer at the nanoscale~\cite{Fan18,Baranov2019}, which is crucial for numerous applications, including radiative cooling~\cite{Fan14}, solar energy conversion~\cite{Lenert2014}, heat management~\cite{Pan23,Xu2024}, nanoscopy~\cite{Vincent2021}, and spectroscopy~\cite{Habibi2025}.
In local thermal equilibrium, it is well described by fluctuational electrodynamics based on the fluctuation-dissipation theorem~\cite{Rytov, PvH, Callen1951_FDT}.

Floquet driving, involving periodic time modulation, offers vast potential for engineering novel properties in both quantum materials and optical systems~\cite{Shaltout19, Yin22, Galiffi22, Horsley2024}, including topological phases~\cite{Tsuji08, Kitagawa11, Refael15, Rudner20, Floquet-grahene-24-1, Floquet-graphene-24-2}, photonic time crystals~\cite{Sharabi21, Saha23, Lustig23, He23}, and optical nonreciprocity~\cite{Yu09, Hadad16, Sounas17, Huidobro19, Galiffi19, WangX20}.
Recently, periodic driven systems have provided emerging possibilities for controlling radiative heat transfer~\cite{coppens17, Kou18, Li19, Li21_ACS, Picardi23,Lozano23}, such as heat shuttling~\cite{Shuttling18}, active cooling~\cite{Fan20, Fan23,Fan24}, and nonreciprocal transfer~\cite{Li21, Biehs22,Biehs23_2}, where optical properties, such as permittivity, are dynamically modulated.
Since such systems may fail to obey the fluctuation-dissipation theorem, extending fluctuational electrodynamics has also become an active research topic~\cite{DMFT-RMP14, Lozano23, GT24}.

A key factor in controlling thermal radiation is the photon thermal distribution, which strongly depends on the statistical distribution of microscopic particles, such as electrons, within materials.
For example, a static bias voltage can induce an effective photon chemical potential, resulting in radiative cooling~\cite{Chen2015, Zhu2019}.
Recent works have shown that Floquet driving leads microscopic particles into unique nonequilibrium states~\cite{Refael15, Matsyshyn23, Floquet-Fermi, kumari_josephson-current_2023, Laura2021}, such as electron staircase occupation in metals~\cite{Floquet-Fermi} and phonon lasing in silicon nanocavities~\cite{Laura2021}.
However, such nonequilibrium effects on radiative photon distribution and heat transfer remain largely unknown.

\begin{figure}
  \centering
  \includegraphics[width=0.48\textwidth]{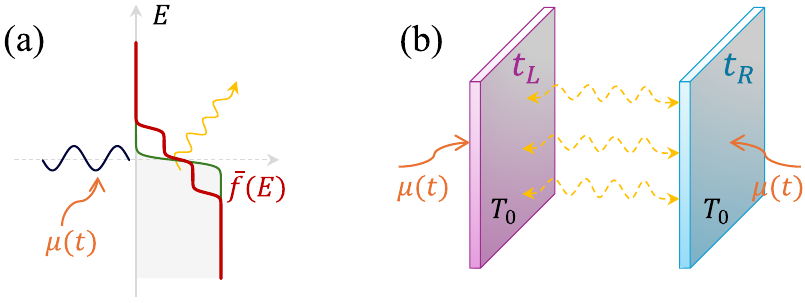}
  \caption{
  (a) Schematic of thermal emission controlled by electron staircase occupation (red line) driven by a time-varying potential $\mu \left( t \right) = 2{\mu _d}\cos \left( {\Omega t} \right)$ with frequency $\Omega$ and amplitude $2\mu_d$.
  (b) Illustration of near-field heat radiation between two driven metal plates with different hopping parameters $t_L$ and $t_R$ at the same temperature $T_0$.
  The interplay of differences in material properties and external drivings leads to an imbalance in photon occupation between the plates and, thereby, a finite heat flux.}
  \label{Fig:fig1}
\end{figure}

In this Letter, we study radiative heat transfer between dissipative metal plates with periodically time-modulated electron potentials (see Fig.~\ref{Fig:fig1}).
By combining the nonequilibrium Green's function formalism~\cite{JSW0, JSW1, JSW2, Wise22, Kamenev23, JSW23, JSW24} with Floquet theory, the electron dynamics is rigorously incorporated into thermal radiation.
We show that, even with identical driving protocols and temperatures, differences in material properties lead to significant heat transfer between the plates.
In contrast to the modulation of optical properties in previous studies~\cite{Fan20,Fan23,Fan24}, this arises from a unique photon nonequilibrium state driven by the electron staircase occupation, characterized by an exponential-staircase distribution, and thus does not require large permittivity modulations.
These findings open a promising avenue for nanoscale control of thermal emission and heat transfer via microscopic nonequilibrium dynamics.

\textit{Formalism.--}
We briefly describe the theory with details provided in the Supplemental
Material~\cite{SM}. 
We consider radiative heat transfer between two two-dimensional metal plates in a parallel configuration [see Fig.~\ref{Fig:fig1}(b)], despite that the theory applies to arbitrary geometries.
In such metal systems, it is the electronic fluctuations that primarily give rise to the exchange of radiative photons between subsystems.
The retarded Green's function of the electrons under a time-varying Hamiltonian satisfies~\cite{Haug_Jauho}
\begin{equation}
\bigl[ {i\hbar {\partial _t} - H(t) - {\Sigma ^r}} \bigr]{G^r}(t,t') = I\delta (t - t').
\end{equation}
In the tight-binding representation, the retarded Green's function is defined as $G^r_{jk}(t,t') = \theta(t-t') \bigl\langle c_j(t) c_k^\dagger(t') + c_k^\dagger(t') c_j(t) \bigr\rangle/(i\hbar)$, where $\theta(t)$ is the Heaviside step function,
$c_j$ ($c_k^\dagger$) is the annihilation (creation) operator at electron site $j$ ($k$),
and $\sum_{jk} c_j^\dagger H_{jk}(t)c_k$ is the many-body electron Hamiltonian.
The matrices $H$ and $G$ are indexed by the electron sites.
The identity matrix is denoted by $I$, and $\delta(t)$ is the Dirac delta function.
The last term on the left-hand side of the equation is a convolution in the time domain with $\int \Sigma^r(t,t'') G^r(t'', t') dt''$, where the retarded self-energy $\Sigma^r$ arises from the coupling to the bath.
This term is crucial for the thermal distribution in a nonequilibrium setting, as it describes how the system is coupled to the environment and exchanges energy and electrons.

The lesser component of the electron Green's function $G^<_{jk}(t,t') = (i/\hbar)\bigl\langle c_k^\dagger(t') c_j(t) \bigr\rangle$ describes the electron distribution.
According to the Keldysh theory, it is given by the equation $G^< = G^r \Sigma^< G^a$, where $\Sigma^<$ is the lesser self-energy from the bath, and $G^a = (G^r)^\dagger$.
The triple product denotes matrix multiplication in the site space and convolution in the time domain.
We assume each bath is in thermal equilibrium and has the usual fluctuation-dissipation theorem; in the energy domain, $\Sigma^<(E) = -f(E) \left[\Sigma^r(E) - \Sigma^a(E)\right]$ where $f(E) = 1/\left\{ {\exp \left[ {(E - {\mu _s})/({k_B}{T_0})} \right] + 1} \right\}$ is the Fermi function at temperature $T_0$ and chemical potential $\mu_s$.
We choose $\Sigma^r = - i \eta I$ with $\eta/\hbar$ the inverse of electron relaxation time, which implies that each site or mode is coupled identically to the bath.
It is certainly feasible to model the bath more explicitly, for example, by putting the system on a weakly coupled substrate.

Before discussing the photon Green's function, we first focus on general properties of the periodically driven system~\cite{Tsuji08, DMFT-RMP14}.
As noted above, the Green's functions depend on two time variables.
For steady-state systems with time-translation invariance, the Green's functions depend solely on the time difference, making it convenient to work in the frequency domain.
However, this convenience is lost for periodically driven systems.
Nonetheless, a residue symmetry persists for an arbitrary function $\mathcal{F}$ with two time variables when we are in a Floquet steady state, as $\mathcal{F}(t+2\pi/\Omega, t'+2\pi/\Omega) = \mathcal{F}(t,t')$ where $\Omega$ is the driving frequency.
This discrete time periodicity can be represented by a double Fourier series as
\begin{equation}\label{Eq:double-FFT}
\mathcal{F}(t,t') = \int_{-\Omega/2}^{+\Omega/2}\!\! \frac{d\omega}{2\pi} \sum_{mn}\mathcal{F}_{mn}(\omega) 
\,e^{ - i\omega_m t + i\omega_n t'},
\end{equation}
with $\omega_m = \omega + m\Omega$. We note that the Floquet matrix $\bm{\mathcal{F}}$ satisfies ${\mathcal{F}}_{mn}(\omega + k \Omega) =
{\mathcal{F}}_{m+k,n+k}(\omega)$.

To describe energy transport mediated by the electromagnetic field, we introduce the photon Green's function defined by vector potential ${\bm A}$~\cite{Lifshitz_book, JSW23},
\begin{equation}
  D_{\mu\nu}^r({\bm r},t;{\bm r}', t') = \frac{1}{i\hbar} \theta(t-t') \Big< \big[
  A_\mu({\bm r}, t), A_\nu({\bm r}', t') \big] \Big> ,
\end{equation}
with $\mu,\nu  = x,y,z$ the directions in a Cartesian coordinate system.
The operators are in the Heisenberg picture, and the square bracket denotes the commutator.
In this work, the particular temporal gauge~\cite{Lifshitz_book, kay_quantum_2021} has been chosen.
The retarded Green's function satisfies a Dyson equation, expressed as $D^r = v + v \Pi^r D^r$.
Here, $v$ is the free photon Green's function, relating electric current density to the vector potential via $\bm{A} = - v \bm{j}$, and the retarded photon self-energy $\Pi^r$ can be interpreted as the linear response of the induced current in matter, $\bm{j}^{\rm ind} = - \Pi^r \bm{A}$.

The lesser photon Green's function 
$D^<_{\mu\nu}(\bm{r},t; \bm{r}',t') = \bigl\langle
A_\nu(\bm{r}',t')A_\mu(\bm{r},t)\bigr\rangle/(i\hbar)$ 
describes the photon distribution, given by the Keldysh equation~\cite{JSW23}, 
\begin{equation}
D^< = D^r \left( \Pi^<  + \Pi^<_\infty \right) D^a. 
\end{equation}
There is a subtlety here as we have two self-energies: The first arises from the actual objects, while the second represents energy dissipation into the vacuum and is given by $\Pi^r_\infty = \bigl[- v^{-1} + (v^{-1})^\dag\bigr]/2$.
The photon self-energy from actual objects is derived through a diagrammatic expansion with the interaction from the Peierls substitution Hamiltonian, $\sum_{jk} c_j^\dag H_{jk}c_k \exp\bigl[ e_0/(i\hbar) \int_{\bm{r}_k}^{\bm{r}_j} \!\!\bm{A} \cdot d\bm{r} \bigr]$ with $e_0$ the elementary charge, taking the contributions linear in $\bm{A}$ with the random phase approximation and quadratic in $\bm{A}$ for the diamagnetic term.
The final expressions for $\Pi$ can be expressed as convolutions of two electron Green's functions in the Floquet space, such as $\Pi^< \propto G^< G^>$. Details are provided in the Supplemental Material~\cite{SM}.

The net energy emitted out of object $\alpha$ can be calculated either by integrating the Poynting vector over a surface enclosing the object or by performing a volume integral through the Joule heating formula, $-\bm{E}\cdot \bm{j}$.
We find the latter approach convenient for our purpose.
The current density of the object can be expressed in terms of the vector potential as $\bm{j} = - \bm{A} v^{-1}$ where $v^{-1} = - \epsilon_0
\left(\partial^2/\partial t^2 + c^2 \nabla \times \nabla \times \right)$ is a differential operator.
Here, $c$ is the speed of light and $\epsilon_0$ the vacuum permittivity.
The energy current can be expressed in terms of the photon
Green's function as~\cite{JSW23, SM}
\begin{align} \label{It}
  I_\alpha(t) 
  &= - \int_\alpha d\bm{r} \left< \bm{E} \cdot \bm{j} \right> \notag \\
  &= - \int_{\alpha} d\bm{r}\, \frac{\partial\ }{\partial t} \!\sum_{\mu,\nu} \langle A_\mu(\bm{r}t) A_\nu(\bm{r}'t') \rangle v^{-1}_{\nu\mu}(\bm{r}'t') \Big|_{ \bm{r}'t'=\bm{r}t } \notag \\
  &= \int_{\alpha} d\bm{r} \frac{\hbar}{2i} \frac{\partial\ }{\partial t}  \left[ D^K(\bm{r} t, \bm{r}' t') v^{-1}(\bm{r}'t') \right] \Big|_{ \bm{r}'t'=\bm{r}t }.
\end{align}
The Keldysh Green's function is $D^K = D^< + D^>$ with $D^>$ the greater photon Green's function.
Such a symmetric function ensures a real energy current.
Using the Dyson equation in the form $D^a v^{-1} = I + D^a \Pi^a$ and the Keldysh equation
$D^K = D^r \Pi^K D^a$, we find $D^K v^{-1} = D^r \Pi^K + D^K \Pi^a$.  
Inserting this into Eq.~(\ref{It}), we obtain the Meir-Wingreen formula
for the energy current out of object $\alpha$.
Since the space integral is restricted to object $\alpha$, we replace the total self-energy $\Pi$ by the self-energy of object $\alpha$ which is $\Pi_\alpha$.

For a periodically driven system, using Eq.~\eqref{Eq:double-FFT}, we can express the average energy current by the Floquet matrices with~\cite{SM}
\begin{equation}\label{Eq:energy-current}
  \bar{I}_\alpha = - \int_{-\Omega/2}^{+\Omega/2} \!\! \frac{d\omega}{4\pi} 
  {\rm Tr}\bigl[\bm{E} \left( \bm{D}^r \bm{\Pi}^K_\alpha + \bm{D}^K
  \bm{\Pi}^a_\alpha\right) \bigr].
\end{equation}
Here, $\bm{D}$ and $\bm{\Pi}$ are Floquet matrices of the photon Green's function and self-energy, respectively.
The entries of diagonal matrix $\bm E$  are $E_{mn} =\hbar \omega_m \delta_{mn}$.
The trace is over direction $\mu$, electron site, and Floquet index $m$.
Using the relation $\mathcal{F}_{mn}(\omega + k \Omega) = \mathcal{F}_{m+k,n+k}(\omega)$, Eq.~\eqref{Eq:energy-current} can be rewritten as
\begin{equation}\label{Eq:heat-flux-meir-wing-formula}
  {\bar I_\alpha } =  - \int_{ - \infty }^{ + \infty } \!\! {\frac{{d\omega }}{{4\pi }}{\hbar \omega}{\operatorname{Tr}}\bigl[ {{{\left( {{\bm{D}^r}\bm{\Pi} _\alpha ^K + {\bm{D}^K}\bm{\Pi} _\alpha ^a} \right)}_{00}}} \bigr]},
\end{equation}
where the integral of $\omega$ is expanded from $\left( { - \Omega /2, + \Omega /2} \right]$ to the entire frequency range, with the trace taken over the ``00'' block of the Floquet matrix.
This transform facilitates us in accessing the spectrum of heat current.
For simplicity, we use the notations $\bm{D}_{00} \to D$ and $\bm{\Pi}_{00} \to \Pi$ below.

\begin{figure}
  \centering
  \includegraphics[width=0.48\textwidth]{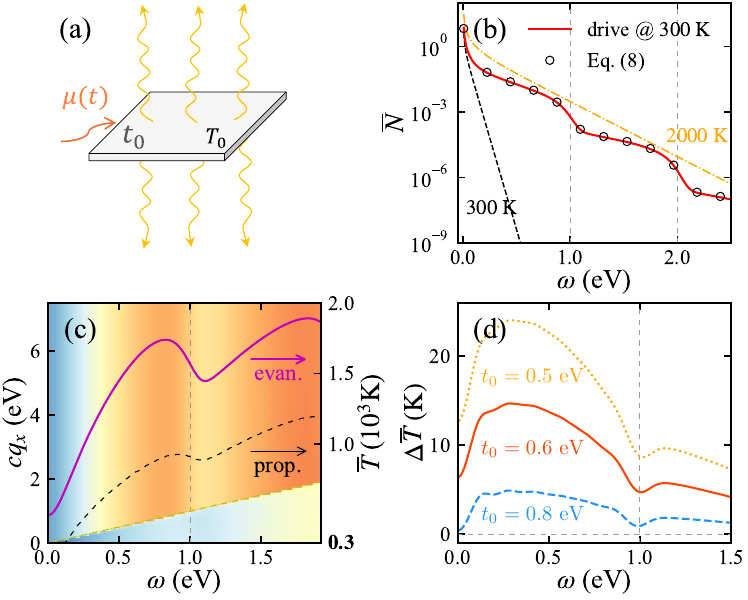}
  \caption{
  (a) Illustration of heat radiation from a metal plate with hopping parameter $t_0$, bath temperature $T_0$, and the electron chemical potential periodically modulated as $\mu \left( t \right)$.
  (b) Effective photon occupation distribution $\bar{N}$ for the metal plate in driven (solid line and circles) and undriven (dashed line for 300 K and dashed-dotted line for 2000 K) cases.
  (c) Spectral effective temperatures $\bar{T}$ for propagating (dashed line labeled ``prop.'') and evanescent (solid line labeled ``evan.'') modes in the driven case. Background map shows $\bar{T}$ in $(q_x,\omega)$ space.
  (d) Spectral temperature differences between the plates with different $t_0$, $\Delta \bar T\left( \omega  \right) = \bar T\left( {\omega ; 1~{\text{eV}}} \right) - \bar T\left( {\omega ; {t_0}} \right)$.
  }\label{Fig:fig2}
\end{figure}

\textit{Results.--}
We first explore the nonequilibrium statistical features of radiative photons from driven metals.
As illustrated in Fig.~\ref{Fig:fig2}(a), we consider a metal plate with hopping parameter $t_0$ and electron chemical potential periodically modulated as $\mu \left( t \right) = 2{\mu _d}\cos \left( {\Omega t} \right)$.
As an example, we use $t_0 = 1.0$ eV, $\eta = 0.1$ eV, $\mu_s = 0$, $\hbar\Omega = 1.0$ eV, and $\mu_d = 0.1$ eV in numerical calculations unless specified otherwise. 
Here, we temporarily exclude energy dissipation into the vacuum to simplify the analysis.
The effective photon occupation is given by $\bar N\left( {\bm{q},\omega } \right) \!=\! \operatorname{Tr} \left\{ {\operatorname{Im} \left[{{\Pi }}^ < \left( {\bm{q},\omega } \right)\right]} \right\}/2\operatorname{Tr} \left\{ {\operatorname{Im} \left[{{\Pi }}^r\left( {\bm{q},\omega } \right)\right]} \right\}$, with ${\bm{q}}$ the inplane photon wavevector.
The frequency-dependent distribution $\bar{N}(\omega)$ is obtained by summing over $\bm{q}$ in both the numerator and denominator.
As shown in Fig.~\ref{Fig:fig2}(b), in the undriven case  (i.e., $\mu_d = 0$), the effective photon occupation simply follows the Bose-Einstein distribution (black line), ${N_B}\left( {\omega ,T} \right) = 1/\left[ \exp{ \left(\hbar \omega /{k_B}T\right)} - 1 \right]$.
However, in the driven case, it exhibits an exponential-staircase distribution (red line), with step edges at integer multiples of $\Omega$.
Also, the photons reach an effective temperature of about 2000 K, indicating a pumping effect of the Floquet driving.
As a key characteristic  of radiative photons in Floquet states, this unique distribution also reflects the nonequilibrium electron dynamics in driven metals.
Indeed, we find that this distribution results from the electron staircase occupation $\bar{f}$ illustrated in Fig.~\ref{Fig:fig1}(a), as expressed by~\cite{SM}
\begin{equation}\label{Eq:Nbar-ana}
{\bar{N}}\left( {{\bm{q}},\omega } \right) = \frac{{\sum\nolimits_{\bm{k}} \!{{W_{{\bm{k}},{\bm{q}}}}\left( \omega  \right)\bar f\left( {{\varepsilon _{\bm{k}}}} \right)\left[ {1 - \bar f\left( {{\varepsilon _{\bm{k}}} - \hbar \omega } \right)} \right]} }}{{\sum\nolimits_{\bm{k}} \!{{W_{{\bm{k}},{\bm{q}}}}\left( \omega  \right)\left[ {\bar f\left( {{\varepsilon _{\bm{k}}} - \hbar \omega } \right) - \bar f\left( {{\varepsilon _{\bm{k}}}} \right)} \right]} }}.
\end{equation}
Here, ${W_{{\bm{k}},{\bm{q}}}}\left( \omega  \right) = {\left| {{V_{{\bm{k}},{\bm{q}}}}} \right|^2}\delta \left( {{\varepsilon _{\bm{k}}} - {\varepsilon _{\bm{k} - \bm{q}}} - \hbar\omega} \right)$ is the weight function with ${V_{{\bm{k}},{\bm{q}}}} = \left( {{{\bm{\upsilon }}_{\bm{k}}} + {{\bm{\upsilon }}_{{\bm{k}} - {\bm{q}}}}} \right)/2$, where $\varepsilon_{\bm{k}}$ and $\bm{\upsilon}_{\bm k}$ are the electron energy and group velocity, respectively.
In Fig.~\ref{Fig:fig2}(b), Eq.~\eqref{Eq:Nbar-ana} gives the effective distribution (black circles), closely matching that calculated from self-energies ${\Pi}$.

To show the key impacts of this unique photon distribution on radiative heat transfer, we define the spectral effective temperature $\bar{T}$ by ${N_B}( {\omega ,\bar T} ) = \bar N\left({\bm q}, \omega  \right)$.
Here, we calculate $\bar{N}$ from ${D}$ rather than ${\Pi}$ to reinstate the energy dissipation effects of propagating modes.
As shown in Fig.~\ref{Fig:fig2}(c), the light line separates the $\bar T\left( {{\bm{q}},\omega } \right)$ into the propagating ($c|\bm{q}| < \omega$) and evanescent ($c|\bm{q}| > \omega$) regions, with propagating modes exhibiting lower effective temperatures than evanescent modes due to heat dissipation into the vacuum.
Since the heat transfer is dominated by photons with energy below 1.0 eV under our chosen parameters, we restrict our analysis to this range.
A notable feature is the increasing effective temperature $\bar{T}$ with photon frequency $\omega$ (e.g., solid line), suggesting that a metal plate with a higher plasmon frequency possesses a greater $\bar{T}$.
To demonstrate this, Fig.~\ref{Fig:fig2}(d) shows the effective-temperature differences $\Delta \bar T$ between metal plates with varying $t_0$, relative to $t_0 = 1.0$ eV.
As expected, the periodic modulation results in a finite $\Delta \bar T$ (up to 10~K for $t_0 = 0.6$ eV) between different metal plates, even though the electron distribution $\bar{f}$ remains constant with $t_0$.
This phenomenon results in a remarkable consequence: Global periodic driving could induce finite radiative heat transfer between two asymmetric metal plates, despite identical bath temperatures.

We now examine the above inference by studying radiative heat transfer between two parallel metal plates separated with distance $d = 53$ nm, as illustrated in Fig.~\ref{Fig:fig1}(b), with the same bath temperature $T_0 = 300$ K but different hopping parameters ($t_L$ and $t_R$).
We use $t_L = 1.0$ eV and $t_R = 0.6$ eV unless stated otherwise.
\begin{figure}
  \centering
  \includegraphics[width=0.48\textwidth]{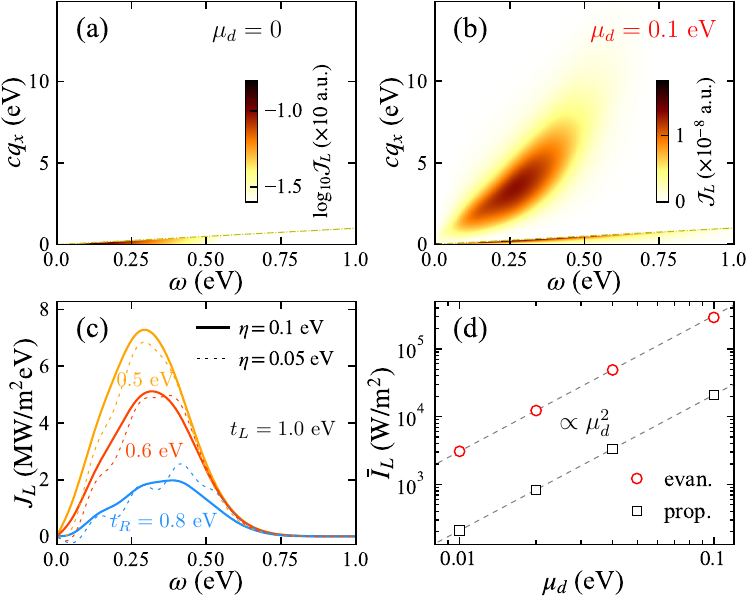}
  \caption{
  Radiative heat flux maps of the left plate $\mathcal{J}_L$ in the two-plate system for (a) undriven and (b) driven cases with separation distance $d = 53$ nm.
  (c) Spectral heat flux $J_L$ contributed by evanescent modes at various $t_R$ and $\eta$ in the driven case.
  (d) Heat fluxes $\bar{I}_L$ contributed by evanescent (red circles) and propagating (black squares) modes, both proportional to $\mu_d ^2$ (dashed lines) for $\mu_d \ll \hbar\Omega$.
  }\label{Fig:fig3}
\end{figure}
Figure~\ref{Fig:fig3}(a) shows the heat flux map from the left plate in the undriven case, calculated as $\mathcal{J}_L (q_x, \omega) = q_x \hbar\omega \operatorname{Tr}  {{{\left( {{{{D}}^r}{{\Pi }}_L^K + {{{D}}^K}{{\Pi }}_L^a} \right)}}} $.
Because the plates are in thermal equilibrium, only far-field radiation into the vacuum occurs, mediated by the propagating modes (below dashed line).
In contrast, time modulation induces a dominant heat flux contributed by the evanescent modes (above dashed line) as shown in Fig.~\ref{Fig:fig3}(b), indicating finite heat transfer between the plates, consistent with our analysis of the effective photon distribution, which suggests active refrigeration with low energy consumption.
To show the tunability of this effect, we further explore its dependence on microscopic details and driving parameters.
Figure~\ref{Fig:fig3}(c) demonstrates that increasing the difference in hopping parameters leads to a higher spectral heat flux, resulting from a greater $\Delta\bar{T}$ [see Fig.~\ref{Fig:fig2}(d)].
However, a further increase in this difference reduces the heat flux due to plasmon frequency mismatch~\cite{SM}.
Notably, although largely insensitive to electron damping $\eta$, the heat flux exhibits distinct spectral features at small hopping differences.
Figure~\ref{Fig:fig3}(d) indicates that both evanescent and propagating mode contributions to the heat flux scale as $\mu_d ^2$ for $\mu_d \ll \hbar\Omega$.
By applying Eq.~\eqref{Eq:Nbar-ana} for a two-level model, we demonstrate that this dependency results from $\bar{N} \propto {\mu_d ^2}$~\cite{SM}.

\begin{figure}
  \centering
  \includegraphics[width=0.375\textwidth]{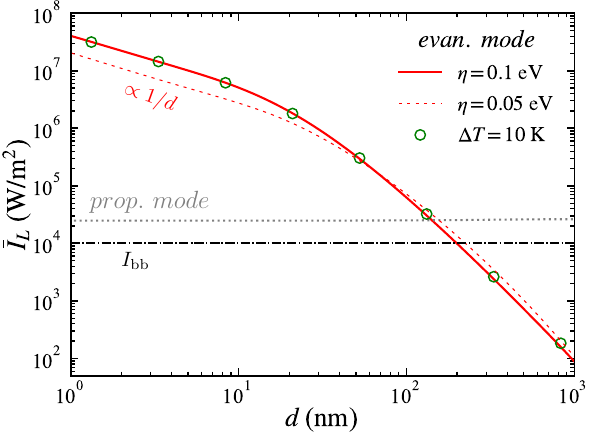}
  \caption{
  Distance dependence of heat flux between two driven metal plates (red and gray lines).
  For comparison, heat flux calculated using the Landauer formula with $T_{L(R)} = 1650\pm5~{\rm K}$ is shown as circles.
  The corresponding blackbody radiation heat flux $I_{\rm bb}$ is also shown according to the Stefan-Boltzmann law.
  }\label{Fig:fig4}
\end{figure}

In addition, we plot the distance dependence of this heat transfer in Fig.~\ref{Fig:fig4}.
To compare with the radiative heat transfer driven by a temperature difference in the same two-plate system, we use the Landauer formula~\cite{JSW23} to match the heat flux at $\eta = 0.1$ eV.
The results show that the heat flux driven by time modulation (solid line) is comparable to that of two undriven plates with an average temperature of 1650 K and a 10 K temperature difference (circles).

\textit{Discussion.--}
We remark that our findings are accessible in two-dimensional materials such as graphene, while applicable to arbitrary layered metallic systems. To observe the exponential-staircase distribution of radiative photons, both $\hbar\Omega \gg k_BT_0$~\cite{Floquet-Fermi} and $\mu_d > k_BT_0$ should be met, ensuring the first step in $\bar{N}(\omega)$ is visible. These have been realized in various platforms for Floquet dynamics~\cite{Floquet-grahene-24-1, Floquet-graphene-24-2}.
The photon occupation difference can be extracted as $\Delta {{\bar N}_{{\text{exp}}.}}(\omega ) = J (\omega )/{\hbar\omega \Xi (\omega )}$, where ${J}$ and ${\Xi}$ are, respectively, the heat flux and transmission spectra from spectral measurements of thermal emitters~\cite{Babuty2013, Saman2019}.
For near-field radiative heat transfer between two metal plates, the driving frequency should be less than the cutoff frequency of the effective transmission, i.e., $\Omega < \omega_c$.
The cutoff frequency is approximated as ${\omega _c} = {\omega _p}\sqrt {a/d}$ with $\omega_p$ the plasma frequency~\cite{Hwang2007} at separations of $d \gg$ $a$.
As for the asymmetry-induced heat transfer, advances in measuring near-field radiative heat transfer between planar materials~\cite{Hu2008, Song16, Zhang2024, Li2024} have created favorable experimental conditions for verification, and recent experiments on radiative thermal transistors~\cite{Thompson2020, Lim2024} may enable observations in temporally modulated systems.
Global periodic modulation of the electron chemical potential can be realized by applying an ac gate voltage. Since the intriguing phenomena reported in this Letter are closely tied to the evanescent modes, the separation between the plates should be in the near-field region. We refer to the Supplemental Material for more details about possible setups~\cite{SM}.

In summary, by linking thermal radiation with nonequilibrium Floquet electron dynamics, we have shown that the global time modulation can induce significant radiative heat transfer between two metal objects at identical temperatures, arising from a unique photon exponential-staircase distribution caused by nonequilibrium electron occupation.
Due to its sensitivity to electronic properties, this effect enables the control of thermal emission and heat transfer in a manner analogous to tuning electronic transport, and can be extended to other systems with dielectric materials.
These findings offer new insights into active cooling, thermophotovoltaics, thermal imaging and  manipulation, and carrier dynamics probing.

\bigskip

\begin{acknowledgments}
\textit{Acknowledgments.--}
H.P. and J.-S.W. are supported by the Ministry of Education, Singapore, under the Academic Research Fund Tier 1 (FY2022), Grant No. A-8000990-00-00.
G.T. is supported by National Natural Science Foundation of China (Grants No. 12374048, and No. 12088101) and NSAF (Grant No. U2330401).
The computational work for this article was fully performed on resources of the National Supercomputing Centre, Singapore (\url{https://www.nscc.sg}).
\end{acknowledgments}

\bibliography{asymhr_floquet}

\begin{thebibliography}{85}%
\makeatletter
\providecommand \@ifxundefined [1]{%
 \@ifx{#1\undefined}
}%
\providecommand \@ifnum [1]{%
 \ifnum #1\expandafter \@firstoftwo
 \else \expandafter \@secondoftwo
 \fi
}%
\providecommand \@ifx [1]{%
 \ifx #1\expandafter \@firstoftwo
 \else \expandafter \@secondoftwo
 \fi
}%
\providecommand \natexlab [1]{#1}%
\providecommand \enquote  [1]{``#1''}%
\providecommand \bibnamefont  [1]{#1}%
\providecommand \bibfnamefont [1]{#1}%
\providecommand \citenamefont [1]{#1}%
\providecommand \href@noop [0]{\@secondoftwo}%
\providecommand \href [0]{\begingroup \@sanitize@url \@href}%
\providecommand \@href[1]{\@@startlink{#1}\@@href}%
\providecommand \@@href[1]{\endgroup#1\@@endlink}%
\providecommand \@sanitize@url [0]{\catcode `\\12\catcode `\$12\catcode `\&12\catcode `\#12\catcode `\^12\catcode `\_12\catcode `\%12\relax}%
\providecommand \@@startlink[1]{}%
\providecommand \@@endlink[0]{}%
\providecommand \url  [0]{\begingroup\@sanitize@url \@url }%
\providecommand \@url [1]{\endgroup\@href {#1}{\urlprefix }}%
\providecommand \urlprefix  [0]{URL }%
\providecommand \Eprint [0]{\href }%
\providecommand \doibase [0]{https://doi.org/}%
\providecommand \selectlanguage [0]{\@gobble}%
\providecommand \bibinfo  [0]{\@secondoftwo}%
\providecommand \bibfield  [0]{\@secondoftwo}%
\providecommand \translation [1]{[#1]}%
\providecommand \BibitemOpen [0]{}%
\providecommand \bibitemStop [0]{}%
\providecommand \bibitemNoStop [0]{.\EOS\space}%
\providecommand \EOS [0]{\spacefactor3000\relax}%
\providecommand \BibitemShut  [1]{\csname bibitem#1\endcsname}%
\let\auto@bib@innerbib\@empty
\bibitem [{\citenamefont {Harder}\ and\ \citenamefont {Green}(2003)}]{Harder2003}%
  \BibitemOpen
  \bibfield  {author} {\bibinfo {author} {\bibfnamefont {N.-P.}\ \bibnamefont {Harder}}\ and\ \bibinfo {author} {\bibfnamefont {M.~A.}\ \bibnamefont {Green}},\ }\bibfield  {title} {\bibinfo {title} {{Thermophotonics}},\ }\href {https://doi.org/10.1088/0268-1242/18/5/319} {\bibfield  {journal} {\bibinfo  {journal} {Semicond. Sci. Technol.}\ }\textbf {\bibinfo {volume} {18}},\ \bibinfo {pages} {S270} (\bibinfo {year} {2003})}\BibitemShut {NoStop}%
\bibitem [{\citenamefont {Fan}(2017)}]{Fan17}%
  \BibitemOpen
  \bibfield  {author} {\bibinfo {author} {\bibfnamefont {S.}~\bibnamefont {Fan}},\ }\bibfield  {title} {\bibinfo {title} {Thermal photonics and energy applications},\ }\href {https://doi.org/10.1016/j.joule.2017.07.012} {\bibfield  {journal} {\bibinfo  {journal} {Joule}\ }\textbf {\bibinfo {volume} {1}},\ \bibinfo {pages} {264} (\bibinfo {year} {2017})}\BibitemShut {NoStop}%
\bibitem [{\citenamefont {Klimchitskaya}\ \emph {et~al.}(2009)\citenamefont {Klimchitskaya}, \citenamefont {Mohideen},\ and\ \citenamefont {Mostepanenko}}]{Klimchitskaya2009}%
  \BibitemOpen
  \bibfield  {author} {\bibinfo {author} {\bibfnamefont {G.~L.}\ \bibnamefont {Klimchitskaya}}, \bibinfo {author} {\bibfnamefont {U.}~\bibnamefont {Mohideen}},\ and\ \bibinfo {author} {\bibfnamefont {V.~M.}\ \bibnamefont {Mostepanenko}},\ }\bibfield  {title} {\bibinfo {title} {{The Casimir force between real materials: Experiment and theory}},\ }\href {https://doi.org/10.1103/RevModPhys.81.1827} {\bibfield  {journal} {\bibinfo  {journal} {Rev. Mod. Phys.}\ }\textbf {\bibinfo {volume} {81}},\ \bibinfo {pages} {1827} (\bibinfo {year} {2009})}\BibitemShut {NoStop}%
\bibitem [{\citenamefont {Li}\ and\ \citenamefont {Fan}(2018)}]{Fan18}%
  \BibitemOpen
  \bibfield  {author} {\bibinfo {author} {\bibfnamefont {W.}~\bibnamefont {Li}}\ and\ \bibinfo {author} {\bibfnamefont {S.}~\bibnamefont {Fan}},\ }\bibfield  {title} {\bibinfo {title} {{Nanophotonic control of thermal radiation for energy applications}},\ }\href {https://doi.org/10.1364/oe.26.015995} {\bibfield  {journal} {\bibinfo  {journal} {Opt. Express}\ }\textbf {\bibinfo {volume} {26}},\ \bibinfo {pages} {15995} (\bibinfo {year} {2018})}\BibitemShut {NoStop}%
\bibitem [{\citenamefont {Baranov}\ \emph {et~al.}(2019)\citenamefont {Baranov}, \citenamefont {Xiao}, \citenamefont {Nechepurenko}, \citenamefont {Krasnok}, \citenamefont {Al{\`{u}}},\ and\ \citenamefont {Kats}}]{Baranov2019}%
  \BibitemOpen
  \bibfield  {author} {\bibinfo {author} {\bibfnamefont {D.~G.}\ \bibnamefont {Baranov}}, \bibinfo {author} {\bibfnamefont {Y.}~\bibnamefont {Xiao}}, \bibinfo {author} {\bibfnamefont {I.~A.}\ \bibnamefont {Nechepurenko}}, \bibinfo {author} {\bibfnamefont {A.}~\bibnamefont {Krasnok}}, \bibinfo {author} {\bibfnamefont {A.}~\bibnamefont {Al{\`{u}}}},\ and\ \bibinfo {author} {\bibfnamefont {M.~A.}\ \bibnamefont {Kats}},\ }\bibfield  {title} {\bibinfo {title} {{Nanophotonic engineering of far-field thermal emitters}},\ }\href {https://doi.org/10.1038/s41563-019-0363-y} {\bibfield  {journal} {\bibinfo  {journal} {Nat. Mater.}\ }\textbf {\bibinfo {volume} {18}},\ \bibinfo {pages} {920} (\bibinfo {year} {2019})}\BibitemShut {NoStop}%
\bibitem [{\citenamefont {Raman}\ \emph {et~al.}(2014)\citenamefont {Raman}, \citenamefont {Anoma}, \citenamefont {Zhu}, \citenamefont {Rephaeli},\ and\ \citenamefont {Fan}}]{Fan14}%
  \BibitemOpen
  \bibfield  {author} {\bibinfo {author} {\bibfnamefont {A.~P.}\ \bibnamefont {Raman}}, \bibinfo {author} {\bibfnamefont {M.~A.}\ \bibnamefont {Anoma}}, \bibinfo {author} {\bibfnamefont {L.}~\bibnamefont {Zhu}}, \bibinfo {author} {\bibfnamefont {E.}~\bibnamefont {Rephaeli}},\ and\ \bibinfo {author} {\bibfnamefont {S.}~\bibnamefont {Fan}},\ }\bibfield  {title} {\bibinfo {title} {{Passive radiative cooling below ambient air temperature under direct sunlight}},\ }\href {https://doi.org/10.1038/nature13883} {\bibfield  {journal} {\bibinfo  {journal} {Nature}\ }\textbf {\bibinfo {volume} {515}},\ \bibinfo {pages} {540} (\bibinfo {year} {2014})}\BibitemShut {NoStop}%
\bibitem [{\citenamefont {Lenert}\ \emph {et~al.}(2014)\citenamefont {Lenert}, \citenamefont {Bierman}, \citenamefont {Nam}, \citenamefont {Chan}, \citenamefont {Celanovi{\'{c}}}, \citenamefont {Solja{\v{c}}i{\'{c}}},\ and\ \citenamefont {Wang}}]{Lenert2014}%
  \BibitemOpen
  \bibfield  {author} {\bibinfo {author} {\bibfnamefont {A.}~\bibnamefont {Lenert}}, \bibinfo {author} {\bibfnamefont {D.~M.}\ \bibnamefont {Bierman}}, \bibinfo {author} {\bibfnamefont {Y.}~\bibnamefont {Nam}}, \bibinfo {author} {\bibfnamefont {W.~R.}\ \bibnamefont {Chan}}, \bibinfo {author} {\bibfnamefont {I.}~\bibnamefont {Celanovi{\'{c}}}}, \bibinfo {author} {\bibfnamefont {M.}~\bibnamefont {Solja{\v{c}}i{\'{c}}}},\ and\ \bibinfo {author} {\bibfnamefont {E.~N.}\ \bibnamefont {Wang}},\ }\bibfield  {title} {\bibinfo {title} {{A nanophotonic solar thermophotovoltaic device}},\ }\href {https://doi.org/10.1038/nnano.2013.286} {\bibfield  {journal} {\bibinfo  {journal} {Nat. Nanotechnol.}\ }\textbf {\bibinfo {volume} {9}},\ \bibinfo {pages} {126} (\bibinfo {year} {2014})}\BibitemShut {NoStop}%
\bibitem [{\citenamefont {Pan}\ \emph {et~al.}(2023)\citenamefont {Pan}, \citenamefont {Lu}, \citenamefont {Li}, \citenamefont {McBride}, \citenamefont {Juneja}, \citenamefont {Long}, \citenamefont {Lindsay}, \citenamefont {Caldwell},\ and\ \citenamefont {Li}}]{Pan23}%
  \BibitemOpen
  \bibfield  {author} {\bibinfo {author} {\bibfnamefont {Z.}~\bibnamefont {Pan}}, \bibinfo {author} {\bibfnamefont {G.}~\bibnamefont {Lu}}, \bibinfo {author} {\bibfnamefont {X.}~\bibnamefont {Li}}, \bibinfo {author} {\bibfnamefont {J.~R.}\ \bibnamefont {McBride}}, \bibinfo {author} {\bibfnamefont {R.}~\bibnamefont {Juneja}}, \bibinfo {author} {\bibfnamefont {M.}~\bibnamefont {Long}}, \bibinfo {author} {\bibfnamefont {L.}~\bibnamefont {Lindsay}}, \bibinfo {author} {\bibfnamefont {J.~D.}\ \bibnamefont {Caldwell}},\ and\ \bibinfo {author} {\bibfnamefont {D.}~\bibnamefont {Li}},\ }\bibfield  {title} {\bibinfo {title} {{Remarkable heat conduction mediated by non-equilibrium phonon polaritons}},\ }\href {https://doi.org/10.1038/s41586-023-06598-0} {\bibfield  {journal} {\bibinfo  {journal} {Nature}\ }\textbf {\bibinfo {volume} {623}},\ \bibinfo {pages} {307} (\bibinfo {year} {2023})}\BibitemShut {NoStop}%
\bibitem [{\citenamefont {Chen}\ \emph {et~al.}(2024)\citenamefont {Chen}, \citenamefont {Pacheco}, \citenamefont {Salihoglu},\ and\ \citenamefont {Xu}}]{Xu2024}%
  \BibitemOpen
  \bibfield  {author} {\bibinfo {author} {\bibfnamefont {Y.}~\bibnamefont {Chen}}, \bibinfo {author} {\bibfnamefont {M.~A.}\ \bibnamefont {Pacheco}}, \bibinfo {author} {\bibfnamefont {H.}~\bibnamefont {Salihoglu}},\ and\ \bibinfo {author} {\bibfnamefont {X.}~\bibnamefont {Xu}},\ }\bibfield  {title} {\bibinfo {title} {Greatly enhanced radiative transfer enabled by hyperbolic phonon polaritons in $\alpha$-moo3},\ }\href {https://doi.org/10.1002/adfm.202403719} {\bibfield  {journal} {\bibinfo  {journal} {Adv. Funct. Mater.}\ }\textbf {\bibinfo {volume} {34}},\ \bibinfo {pages} {2403719} (\bibinfo {year} {2024})}\BibitemShut {NoStop}%
\bibitem [{\citenamefont {Vincent}(2021)}]{Vincent2021}%
  \BibitemOpen
  \bibfield  {author} {\bibinfo {author} {\bibfnamefont {T.}~\bibnamefont {Vincent}},\ }\bibfield  {title} {\bibinfo {title} {{Scanning near-field infrared microscopy}},\ }\href {https://doi.org/10.1038/s42254-021-00337-y} {\bibfield  {journal} {\bibinfo  {journal} {Nat. Rev. Phys.}\ }\textbf {\bibinfo {volume} {3}},\ \bibinfo {pages} {537} (\bibinfo {year} {2021})}\BibitemShut {NoStop}%
\bibitem [{\citenamefont {Habibi}\ \emph {et~al.}(2025)\citenamefont {Habibi}, \citenamefont {Beardo},\ and\ \citenamefont {Cui}}]{Habibi2025}%
  \BibitemOpen
  \bibfield  {author} {\bibinfo {author} {\bibfnamefont {M.}~\bibnamefont {Habibi}}, \bibinfo {author} {\bibfnamefont {A.}~\bibnamefont {Beardo}},\ and\ \bibinfo {author} {\bibfnamefont {L.}~\bibnamefont {Cui}},\ }\bibfield  {title} {\bibinfo {title} {{Near-field thermal radiation as a probe of nanoscale hot electron and phonon transport}},\ }\href {https://doi.org/10.1021/acsnano.4c11893} {\bibfield  {journal} {\bibinfo  {journal} {ACS Nano}\ }\textbf {\bibinfo {volume} {19}},\ \bibinfo {pages} {6033} (\bibinfo {year} {2025})}\BibitemShut {NoStop}%
\bibitem [{\citenamefont {Rytov}(1953)}]{Rytov}%
  \BibitemOpen
  \bibfield  {author} {\bibinfo {author} {\bibfnamefont {S.~M.}\ \bibnamefont {Rytov}},\ }\href@noop {} {\emph {\bibinfo {title} {Theory of Electrical Fluctuation and Thermal Radiation}}}\ (\bibinfo  {publisher} {Academy of Science of USSR, Moscow},\ \bibinfo {year} {1953})\BibitemShut {NoStop}%
\bibitem [{\citenamefont {Polder}\ and\ \citenamefont {Van~Hove}(1971)}]{PvH}%
  \BibitemOpen
  \bibfield  {author} {\bibinfo {author} {\bibfnamefont {D.}~\bibnamefont {Polder}}\ and\ \bibinfo {author} {\bibfnamefont {M.}~\bibnamefont {Van~Hove}},\ }\bibfield  {title} {\bibinfo {title} {Theory of radiative heat transfer between closely spaced bodies},\ }\href {https://doi.org/10.1103/PhysRevB.4.3303} {\bibfield  {journal} {\bibinfo  {journal} {Phys. Rev. B}\ }\textbf {\bibinfo {volume} {4}},\ \bibinfo {pages} {3303} (\bibinfo {year} {1971})}\BibitemShut {NoStop}%
\bibitem [{\citenamefont {Callen}\ and\ \citenamefont {Welton}(1951)}]{Callen1951_FDT}%
  \BibitemOpen
  \bibfield  {author} {\bibinfo {author} {\bibfnamefont {H.~B.}\ \bibnamefont {Callen}}\ and\ \bibinfo {author} {\bibfnamefont {T.~A.}\ \bibnamefont {Welton}},\ }\bibfield  {title} {\bibinfo {title} {Irreversibility and generalized noise},\ }\href {https://doi.org/10.1103/PhysRev.83.34} {\bibfield  {journal} {\bibinfo  {journal} {Phys. Rev.}\ }\textbf {\bibinfo {volume} {83}},\ \bibinfo {pages} {34} (\bibinfo {year} {1951})}\BibitemShut {NoStop}%
\bibitem [{\citenamefont {Shaltout}\ \emph {et~al.}(2019)\citenamefont {Shaltout}, \citenamefont {Shalaev},\ and\ \citenamefont {Brongersma}}]{Shaltout19}%
  \BibitemOpen
  \bibfield  {author} {\bibinfo {author} {\bibfnamefont {A.~M.}\ \bibnamefont {Shaltout}}, \bibinfo {author} {\bibfnamefont {V.~M.}\ \bibnamefont {Shalaev}},\ and\ \bibinfo {author} {\bibfnamefont {M.~L.}\ \bibnamefont {Brongersma}},\ }\bibfield  {title} {\bibinfo {title} {{Spatiotemporal light control with active metasurfaces}},\ }\href {https://doi.org/10.1126/science.aat3100} {\bibfield  {journal} {\bibinfo  {journal} {Science}\ }\textbf {\bibinfo {volume} {364}},\ \bibinfo {pages} {eaat3100} (\bibinfo {year} {2019})}\BibitemShut {NoStop}%
\bibitem [{\citenamefont {Yin}\ \emph {et~al.}(2022)\citenamefont {Yin}, \citenamefont {Galiffi},\ and\ \citenamefont {Al{\`u}}}]{Yin22}%
  \BibitemOpen
  \bibfield  {author} {\bibinfo {author} {\bibfnamefont {S.}~\bibnamefont {Yin}}, \bibinfo {author} {\bibfnamefont {E.}~\bibnamefont {Galiffi}},\ and\ \bibinfo {author} {\bibfnamefont {A.}~\bibnamefont {Al{\`u}}},\ }\bibfield  {title} {\bibinfo {title} {Floquet metamaterials},\ }\href {https://doi.org/10.1186/s43593-022-00015-1} {\bibfield  {journal} {\bibinfo  {journal} {eLight}\ }\textbf {\bibinfo {volume} {2}},\ \bibinfo {pages} {8} (\bibinfo {year} {2022})}\BibitemShut {NoStop}%
\bibitem [{\citenamefont {Galiffi}\ \emph {et~al.}(2022)\citenamefont {Galiffi}, \citenamefont {Tirole}, \citenamefont {Yin}, \citenamefont {Li}, \citenamefont {Vezzoli}, \citenamefont {Huidobro}, \citenamefont {Silveirinha}, \citenamefont {Sapienza}, \citenamefont {Al{\`{u}}},\ and\ \citenamefont {Pendry}}]{Galiffi22}%
  \BibitemOpen
  \bibfield  {author} {\bibinfo {author} {\bibfnamefont {E.}~\bibnamefont {Galiffi}}, \bibinfo {author} {\bibfnamefont {R.}~\bibnamefont {Tirole}}, \bibinfo {author} {\bibfnamefont {S.}~\bibnamefont {Yin}}, \bibinfo {author} {\bibfnamefont {H.}~\bibnamefont {Li}}, \bibinfo {author} {\bibfnamefont {S.}~\bibnamefont {Vezzoli}}, \bibinfo {author} {\bibfnamefont {P.~A.}\ \bibnamefont {Huidobro}}, \bibinfo {author} {\bibfnamefont {M.~G.}\ \bibnamefont {Silveirinha}}, \bibinfo {author} {\bibfnamefont {R.}~\bibnamefont {Sapienza}}, \bibinfo {author} {\bibfnamefont {A.}~\bibnamefont {Al{\`{u}}}},\ and\ \bibinfo {author} {\bibfnamefont {J.~B.}\ \bibnamefont {Pendry}},\ }\bibfield  {title} {\bibinfo {title} {{Photonics of time-varying media}},\ }\href {https://doi.org/10.1117/1.AP.4.1.014002} {\bibfield  {journal} {\bibinfo  {journal} {Adv. Photonics}\ }\textbf {\bibinfo {volume} {4}},\ \bibinfo {pages} {014002} (\bibinfo {year} {2022})}\BibitemShut {NoStop}%
\bibitem [{\citenamefont {Horsley}\ and\ \citenamefont {Pendry}(2024)}]{Horsley2024}%
  \BibitemOpen
  \bibfield  {author} {\bibinfo {author} {\bibfnamefont {S.~A.~R.}\ \bibnamefont {Horsley}}\ and\ \bibinfo {author} {\bibfnamefont {J.~B.}\ \bibnamefont {Pendry}},\ }\bibfield  {title} {\bibinfo {title} {Traveling wave amplification in stationary gratings},\ }\href {https://doi.org/10.1103/PhysRevLett.133.156903} {\bibfield  {journal} {\bibinfo  {journal} {Phys. Rev. Lett.}\ }\textbf {\bibinfo {volume} {133}},\ \bibinfo {pages} {156903} (\bibinfo {year} {2024})}\BibitemShut {NoStop}%
\bibitem [{\citenamefont {Tsuji}\ \emph {et~al.}(2008)\citenamefont {Tsuji}, \citenamefont {Oka},\ and\ \citenamefont {Aoki}}]{Tsuji08}%
  \BibitemOpen
  \bibfield  {author} {\bibinfo {author} {\bibfnamefont {N.}~\bibnamefont {Tsuji}}, \bibinfo {author} {\bibfnamefont {T.}~\bibnamefont {Oka}},\ and\ \bibinfo {author} {\bibfnamefont {H.}~\bibnamefont {Aoki}},\ }\bibfield  {title} {\bibinfo {title} {Correlated electron systems periodically driven out of equilibrium: $\text{Floquet}+\text{DMFT}$ formalism},\ }\href {https://doi.org/10.1103/PhysRevB.78.235124} {\bibfield  {journal} {\bibinfo  {journal} {Phys. Rev. B}\ }\textbf {\bibinfo {volume} {78}},\ \bibinfo {pages} {235124} (\bibinfo {year} {2008})}\BibitemShut {NoStop}%
\bibitem [{\citenamefont {Kitagawa}\ \emph {et~al.}(2011)\citenamefont {Kitagawa}, \citenamefont {Oka}, \citenamefont {Brataas}, \citenamefont {Fu},\ and\ \citenamefont {Demler}}]{Kitagawa11}%
  \BibitemOpen
  \bibfield  {author} {\bibinfo {author} {\bibfnamefont {T.}~\bibnamefont {Kitagawa}}, \bibinfo {author} {\bibfnamefont {T.}~\bibnamefont {Oka}}, \bibinfo {author} {\bibfnamefont {A.}~\bibnamefont {Brataas}}, \bibinfo {author} {\bibfnamefont {L.}~\bibnamefont {Fu}},\ and\ \bibinfo {author} {\bibfnamefont {E.}~\bibnamefont {Demler}},\ }\bibfield  {title} {\bibinfo {title} {Transport properties of nonequilibrium systems under the application of light: Photoinduced quantum {H}all insulators without {L}andau levels},\ }\href {https://doi.org/10.1103/PhysRevB.84.235108} {\bibfield  {journal} {\bibinfo  {journal} {Phys. Rev. B}\ }\textbf {\bibinfo {volume} {84}},\ \bibinfo {pages} {235108} (\bibinfo {year} {2011})}\BibitemShut {NoStop}%
\bibitem [{\citenamefont {Seetharam}\ \emph {et~al.}(2015)\citenamefont {Seetharam}, \citenamefont {Bardyn}, \citenamefont {Lindner}, \citenamefont {Rudner},\ and\ \citenamefont {Refael}}]{Refael15}%
  \BibitemOpen
  \bibfield  {author} {\bibinfo {author} {\bibfnamefont {K.~I.}\ \bibnamefont {Seetharam}}, \bibinfo {author} {\bibfnamefont {C.-E.}\ \bibnamefont {Bardyn}}, \bibinfo {author} {\bibfnamefont {N.~H.}\ \bibnamefont {Lindner}}, \bibinfo {author} {\bibfnamefont {M.~S.}\ \bibnamefont {Rudner}},\ and\ \bibinfo {author} {\bibfnamefont {G.}~\bibnamefont {Refael}},\ }\bibfield  {title} {\bibinfo {title} {Controlled population of {F}loquet-{B}loch states via coupling to {B}ose and {F}ermi baths},\ }\href {https://doi.org/10.1103/PhysRevX.5.041050} {\bibfield  {journal} {\bibinfo  {journal} {Phys. Rev. X}\ }\textbf {\bibinfo {volume} {5}},\ \bibinfo {pages} {041050} (\bibinfo {year} {2015})}\BibitemShut {NoStop}%
\bibitem [{\citenamefont {Rudner}\ and\ \citenamefont {Lindner}(2020)}]{Rudner20}%
  \BibitemOpen
  \bibfield  {author} {\bibinfo {author} {\bibfnamefont {M.~S.}\ \bibnamefont {Rudner}}\ and\ \bibinfo {author} {\bibfnamefont {N.~H.}\ \bibnamefont {Lindner}},\ }\bibfield  {title} {\bibinfo {title} {Band structure engineering and non-equilibrium dynamics in {F}loquet topological insulators},\ }\href {https://doi.org/10.1038/s42254-020-0170-z} {\bibfield  {journal} {\bibinfo  {journal} {Nat. Rev. Phys.}\ }\textbf {\bibinfo {volume} {2}},\ \bibinfo {pages} {229} (\bibinfo {year} {2020})}\BibitemShut {NoStop}%
\bibitem [{\citenamefont {Merboldt}\ \emph {et~al.}(2024)\citenamefont {Merboldt}, \citenamefont {Schüler}, \citenamefont {Schmitt}, \citenamefont {Bange}, \citenamefont {Bennecke}, \citenamefont {Gadge}, \citenamefont {Pierz}, \citenamefont {Schumacher}, \citenamefont {Momeni}, \citenamefont {Steil}, \citenamefont {Manmana}, \citenamefont {Sentef}, \citenamefont {Reutzel},\ and\ \citenamefont {Mathias}}]{Floquet-grahene-24-1}%
  \BibitemOpen
  \bibfield  {author} {\bibinfo {author} {\bibfnamefont {M.}~\bibnamefont {Merboldt}}, \bibinfo {author} {\bibfnamefont {M.}~\bibnamefont {Schüler}}, \bibinfo {author} {\bibfnamefont {D.}~\bibnamefont {Schmitt}}, \bibinfo {author} {\bibfnamefont {J.~P.}\ \bibnamefont {Bange}}, \bibinfo {author} {\bibfnamefont {W.}~\bibnamefont {Bennecke}}, \bibinfo {author} {\bibfnamefont {K.}~\bibnamefont {Gadge}}, \bibinfo {author} {\bibfnamefont {K.}~\bibnamefont {Pierz}}, \bibinfo {author} {\bibfnamefont {H.~W.}\ \bibnamefont {Schumacher}}, \bibinfo {author} {\bibfnamefont {D.}~\bibnamefont {Momeni}}, \bibinfo {author} {\bibfnamefont {D.}~\bibnamefont {Steil}}, \bibinfo {author} {\bibfnamefont {S.~R.}\ \bibnamefont {Manmana}}, \bibinfo {author} {\bibfnamefont {M.}~\bibnamefont {Sentef}}, \bibinfo {author} {\bibfnamefont {M.}~\bibnamefont {Reutzel}},\ and\ \bibinfo {author} {\bibfnamefont {S.}~\bibnamefont {Mathias}},\ }\href@noop {} {\bibinfo {title} {Observation of {F}loquet states in graphene}} (\bibinfo {year}
  {2024}),\ \Eprint {https://arxiv.org/abs/arXiv:2404.12791} {arXiv:2404.12791} \BibitemShut {NoStop}%
\bibitem [{\citenamefont {Choi}\ \emph {et~al.}(2024)\citenamefont {Choi}, \citenamefont {Mogi}, \citenamefont {Giovannini}, \citenamefont {Azoury}, \citenamefont {Lv}, \citenamefont {Su}, \citenamefont {Hübener}, \citenamefont {Rubio},\ and\ \citenamefont {Gedik}}]{Floquet-graphene-24-2}%
  \BibitemOpen
  \bibfield  {author} {\bibinfo {author} {\bibfnamefont {D.}~\bibnamefont {Choi}}, \bibinfo {author} {\bibfnamefont {M.}~\bibnamefont {Mogi}}, \bibinfo {author} {\bibfnamefont {U.~D.}\ \bibnamefont {Giovannini}}, \bibinfo {author} {\bibfnamefont {D.}~\bibnamefont {Azoury}}, \bibinfo {author} {\bibfnamefont {B.}~\bibnamefont {Lv}}, \bibinfo {author} {\bibfnamefont {Y.}~\bibnamefont {Su}}, \bibinfo {author} {\bibfnamefont {H.}~\bibnamefont {Hübener}}, \bibinfo {author} {\bibfnamefont {A.}~\bibnamefont {Rubio}},\ and\ \bibinfo {author} {\bibfnamefont {N.}~\bibnamefont {Gedik}},\ }\href@noop {} {\bibinfo {title} {Direct observation of {F}loquet-{B}loch states in monolayer graphene}} (\bibinfo {year} {2024}),\ \Eprint {https://arxiv.org/abs/arXiv:2404.14392} {arXiv:2404.14392} \BibitemShut {NoStop}%
\bibitem [{\citenamefont {Sharabi}\ \emph {et~al.}(2021)\citenamefont {Sharabi}, \citenamefont {Lustig},\ and\ \citenamefont {Segev}}]{Sharabi21}%
  \BibitemOpen
  \bibfield  {author} {\bibinfo {author} {\bibfnamefont {Y.}~\bibnamefont {Sharabi}}, \bibinfo {author} {\bibfnamefont {E.}~\bibnamefont {Lustig}},\ and\ \bibinfo {author} {\bibfnamefont {M.}~\bibnamefont {Segev}},\ }\bibfield  {title} {\bibinfo {title} {Disordered photonic time crystals},\ }\href {https://doi.org/10.1103/PhysRevLett.126.163902} {\bibfield  {journal} {\bibinfo  {journal} {Phys. Rev. Lett.}\ }\textbf {\bibinfo {volume} {126}},\ \bibinfo {pages} {163902} (\bibinfo {year} {2021})}\BibitemShut {NoStop}%
\bibitem [{\citenamefont {Saha}\ \emph {et~al.}(2023)\citenamefont {Saha}, \citenamefont {Segal}, \citenamefont {Fruhling}, \citenamefont {Lustig}, \citenamefont {Segev}, \citenamefont {Boltasseva},\ and\ \citenamefont {Shalaev}}]{Saha23}%
  \BibitemOpen
  \bibfield  {author} {\bibinfo {author} {\bibfnamefont {S.}~\bibnamefont {Saha}}, \bibinfo {author} {\bibfnamefont {O.}~\bibnamefont {Segal}}, \bibinfo {author} {\bibfnamefont {C.}~\bibnamefont {Fruhling}}, \bibinfo {author} {\bibfnamefont {E.}~\bibnamefont {Lustig}}, \bibinfo {author} {\bibfnamefont {M.}~\bibnamefont {Segev}}, \bibinfo {author} {\bibfnamefont {A.}~\bibnamefont {Boltasseva}},\ and\ \bibinfo {author} {\bibfnamefont {V.~M.}\ \bibnamefont {Shalaev}},\ }\bibfield  {title} {\bibinfo {title} {Photonic time crystals: a materials perspective},\ }\href {https://doi.org/10.1364/OE.479257} {\bibfield  {journal} {\bibinfo  {journal} {Opt. Express}\ }\textbf {\bibinfo {volume} {31}},\ \bibinfo {pages} {8267} (\bibinfo {year} {2023})}\BibitemShut {NoStop}%
\bibitem [{\citenamefont {Lustig}\ \emph {et~al.}(2023)\citenamefont {Lustig}, \citenamefont {Segal}, \citenamefont {Saha}, \citenamefont {Fruhling}, \citenamefont {Shalaev}, \citenamefont {Boltasseva},\ and\ \citenamefont {Segev}}]{Lustig23}%
  \BibitemOpen
  \bibfield  {author} {\bibinfo {author} {\bibfnamefont {E.}~\bibnamefont {Lustig}}, \bibinfo {author} {\bibfnamefont {O.}~\bibnamefont {Segal}}, \bibinfo {author} {\bibfnamefont {S.}~\bibnamefont {Saha}}, \bibinfo {author} {\bibfnamefont {C.}~\bibnamefont {Fruhling}}, \bibinfo {author} {\bibfnamefont {V.~M.}\ \bibnamefont {Shalaev}}, \bibinfo {author} {\bibfnamefont {A.}~\bibnamefont {Boltasseva}},\ and\ \bibinfo {author} {\bibfnamefont {M.}~\bibnamefont {Segev}},\ }\bibfield  {title} {\bibinfo {title} {Photonic time-crystals - fundamental concepts},\ }\href {https://doi.org/10.1364/OE.479367} {\bibfield  {journal} {\bibinfo  {journal} {Opt. Express}\ }\textbf {\bibinfo {volume} {31}},\ \bibinfo {pages} {9165} (\bibinfo {year} {2023})}\BibitemShut {NoStop}%
\bibitem [{\citenamefont {He}\ \emph {et~al.}(2023)\citenamefont {He}, \citenamefont {Zhang}, \citenamefont {Qi}, \citenamefont {Bo},\ and\ \citenamefont {Li}}]{He23}%
  \BibitemOpen
  \bibfield  {author} {\bibinfo {author} {\bibfnamefont {H.}~\bibnamefont {He}}, \bibinfo {author} {\bibfnamefont {S.}~\bibnamefont {Zhang}}, \bibinfo {author} {\bibfnamefont {J.}~\bibnamefont {Qi}}, \bibinfo {author} {\bibfnamefont {F.}~\bibnamefont {Bo}},\ and\ \bibinfo {author} {\bibfnamefont {H.}~\bibnamefont {Li}},\ }\bibfield  {title} {\bibinfo {title} {Faraday rotation in nonreciprocal photonic time-crystals},\ }\href {https://doi.org/10.1063/5.0131818} {\bibfield  {journal} {\bibinfo  {journal} {Appl. Phys. Lett.}\ }\textbf {\bibinfo {volume} {122}} (\bibinfo {year} {2023})}\BibitemShut {NoStop}%
\bibitem [{\citenamefont {Yu}\ and\ \citenamefont {Fan}(2009)}]{Yu09}%
  \BibitemOpen
  \bibfield  {author} {\bibinfo {author} {\bibfnamefont {Z.}~\bibnamefont {Yu}}\ and\ \bibinfo {author} {\bibfnamefont {S.}~\bibnamefont {Fan}},\ }\bibfield  {title} {\bibinfo {title} {Complete optical isolation created by indirect interband photonic transitions},\ }\href {https://doi.org/10.1038/nphoton.2008.273} {\bibfield  {journal} {\bibinfo  {journal} {Nat. Photonics}\ }\textbf {\bibinfo {volume} {3}},\ \bibinfo {pages} {91} (\bibinfo {year} {2009})}\BibitemShut {NoStop}%
\bibitem [{\citenamefont {Hadad}\ \emph {et~al.}(2016)\citenamefont {Hadad}, \citenamefont {Soric},\ and\ \citenamefont {Alu}}]{Hadad16}%
  \BibitemOpen
  \bibfield  {author} {\bibinfo {author} {\bibfnamefont {Y.}~\bibnamefont {Hadad}}, \bibinfo {author} {\bibfnamefont {J.~C.}\ \bibnamefont {Soric}},\ and\ \bibinfo {author} {\bibfnamefont {A.}~\bibnamefont {Alu}},\ }\bibfield  {title} {\bibinfo {title} {Breaking temporal symmetries for emission and absorption},\ }\href {https://doi.org/10.1073/pnas.1517363113} {\bibfield  {journal} {\bibinfo  {journal} {Proc. Natl. Acad. Sci.}\ }\textbf {\bibinfo {volume} {113}},\ \bibinfo {pages} {3471} (\bibinfo {year} {2016})}\BibitemShut {NoStop}%
\bibitem [{\citenamefont {Sounas}\ and\ \citenamefont {Al{\`u}}(2017)}]{Sounas17}%
  \BibitemOpen
  \bibfield  {author} {\bibinfo {author} {\bibfnamefont {D.~L.}\ \bibnamefont {Sounas}}\ and\ \bibinfo {author} {\bibfnamefont {A.}~\bibnamefont {Al{\`u}}},\ }\bibfield  {title} {\bibinfo {title} {Non-reciprocal photonics based on time modulation},\ }\href {https://doi.org/10.1038/s41566-017-0051-x} {\bibfield  {journal} {\bibinfo  {journal} {Nat. Photonics}\ }\textbf {\bibinfo {volume} {11}},\ \bibinfo {pages} {774} (\bibinfo {year} {2017})}\BibitemShut {NoStop}%
\bibitem [{\citenamefont {Huidobro}\ \emph {et~al.}(2019)\citenamefont {Huidobro}, \citenamefont {Galiffi}, \citenamefont {Guenneau}, \citenamefont {Craster},\ and\ \citenamefont {Pendry}}]{Huidobro19}%
  \BibitemOpen
  \bibfield  {author} {\bibinfo {author} {\bibfnamefont {P.~A.}\ \bibnamefont {Huidobro}}, \bibinfo {author} {\bibfnamefont {E.}~\bibnamefont {Galiffi}}, \bibinfo {author} {\bibfnamefont {S.}~\bibnamefont {Guenneau}}, \bibinfo {author} {\bibfnamefont {R.~V.}\ \bibnamefont {Craster}},\ and\ \bibinfo {author} {\bibfnamefont {J.~B.}\ \bibnamefont {Pendry}},\ }\bibfield  {title} {\bibinfo {title} {Fresnel drag in space--time-modulated metamaterials},\ }\href {https://doi.org/10.1073/pnas.1915027116} {\bibfield  {journal} {\bibinfo  {journal} {Proc. Natl. Acad. Sci.}\ }\textbf {\bibinfo {volume} {116}},\ \bibinfo {pages} {24943} (\bibinfo {year} {2019})}\BibitemShut {NoStop}%
\bibitem [{\citenamefont {Galiffi}\ \emph {et~al.}(2019)\citenamefont {Galiffi}, \citenamefont {Huidobro},\ and\ \citenamefont {Pendry}}]{Galiffi19}%
  \BibitemOpen
  \bibfield  {author} {\bibinfo {author} {\bibfnamefont {E.}~\bibnamefont {Galiffi}}, \bibinfo {author} {\bibfnamefont {P.~A.}\ \bibnamefont {Huidobro}},\ and\ \bibinfo {author} {\bibfnamefont {J.~B.}\ \bibnamefont {Pendry}},\ }\bibfield  {title} {\bibinfo {title} {Broadband nonreciprocal amplification in luminal metamaterials},\ }\href {https://doi.org/10.1103/PhysRevLett.123.206101} {\bibfield  {journal} {\bibinfo  {journal} {Phys. Rev. Lett.}\ }\textbf {\bibinfo {volume} {123}},\ \bibinfo {pages} {206101} (\bibinfo {year} {2019})}\BibitemShut {NoStop}%
\bibitem [{\citenamefont {Wang}\ \emph {et~al.}(2020)\citenamefont {Wang}, \citenamefont {Ptitcyn}, \citenamefont {Asadchy}, \citenamefont {D\'{\i}az-Rubio}, \citenamefont {Mirmoosa}, \citenamefont {Fan},\ and\ \citenamefont {Tretyakov}}]{WangX20}%
  \BibitemOpen
  \bibfield  {author} {\bibinfo {author} {\bibfnamefont {X.}~\bibnamefont {Wang}}, \bibinfo {author} {\bibfnamefont {G.}~\bibnamefont {Ptitcyn}}, \bibinfo {author} {\bibfnamefont {V.~S.}\ \bibnamefont {Asadchy}}, \bibinfo {author} {\bibfnamefont {A.}~\bibnamefont {D\'{\i}az-Rubio}}, \bibinfo {author} {\bibfnamefont {M.~S.}\ \bibnamefont {Mirmoosa}}, \bibinfo {author} {\bibfnamefont {S.}~\bibnamefont {Fan}},\ and\ \bibinfo {author} {\bibfnamefont {S.~A.}\ \bibnamefont {Tretyakov}},\ }\bibfield  {title} {\bibinfo {title} {Nonreciprocity in bianisotropic systems with uniform time modulation},\ }\href {https://doi.org/10.1103/PhysRevLett.125.266102} {\bibfield  {journal} {\bibinfo  {journal} {Phys. Rev. Lett.}\ }\textbf {\bibinfo {volume} {125}},\ \bibinfo {pages} {266102} (\bibinfo {year} {2020})}\BibitemShut {NoStop}%
\bibitem [{\citenamefont {Coppens}\ and\ \citenamefont {Valentine}(2017)}]{coppens17}%
  \BibitemOpen
  \bibfield  {author} {\bibinfo {author} {\bibfnamefont {Z.~J.}\ \bibnamefont {Coppens}}\ and\ \bibinfo {author} {\bibfnamefont {J.~G.}\ \bibnamefont {Valentine}},\ }\bibfield  {title} {\bibinfo {title} {Spatial and temporal modulation of thermal emission},\ }\href {https://doi.org/10.1002/adma.201701275} {\bibfield  {journal} {\bibinfo  {journal} {Adv. Mater.}\ }\textbf {\bibinfo {volume} {29}},\ \bibinfo {pages} {1701275} (\bibinfo {year} {2017})}\BibitemShut {NoStop}%
\bibitem [{\citenamefont {Kou}\ and\ \citenamefont {Minnich}(2018)}]{Kou18}%
  \BibitemOpen
  \bibfield  {author} {\bibinfo {author} {\bibfnamefont {J.}~\bibnamefont {Kou}}\ and\ \bibinfo {author} {\bibfnamefont {A.~J.}\ \bibnamefont {Minnich}},\ }\bibfield  {title} {\bibinfo {title} {Dynamic optical control of near-field radiative transfer},\ }\href {https://doi.org/10.1364/OE.26.00A729} {\bibfield  {journal} {\bibinfo  {journal} {Opt. Express}\ }\textbf {\bibinfo {volume} {26}},\ \bibinfo {pages} {A729} (\bibinfo {year} {2018})}\BibitemShut {NoStop}%
\bibitem [{\citenamefont {Li}\ \emph {et~al.}(2019)\citenamefont {Li}, \citenamefont {Fern\'andez-Alc\'azar}, \citenamefont {Ellis}, \citenamefont {Shapiro},\ and\ \citenamefont {Kottos}}]{Li19}%
  \BibitemOpen
  \bibfield  {author} {\bibinfo {author} {\bibfnamefont {H.}~\bibnamefont {Li}}, \bibinfo {author} {\bibfnamefont {L.~J.}\ \bibnamefont {Fern\'andez-Alc\'azar}}, \bibinfo {author} {\bibfnamefont {F.}~\bibnamefont {Ellis}}, \bibinfo {author} {\bibfnamefont {B.}~\bibnamefont {Shapiro}},\ and\ \bibinfo {author} {\bibfnamefont {T.}~\bibnamefont {Kottos}},\ }\bibfield  {title} {\bibinfo {title} {Adiabatic thermal radiation pumps for thermal photonics},\ }\href {https://doi.org/10.1103/PhysRevLett.123.165901} {\bibfield  {journal} {\bibinfo  {journal} {Phys. Rev. Lett.}\ }\textbf {\bibinfo {volume} {123}},\ \bibinfo {pages} {165901} (\bibinfo {year} {2019})}\BibitemShut {NoStop}%
\bibitem [{\citenamefont {Fern\'andez-Alc\'azar}\ \emph {et~al.}(2021{\natexlab{a}})\citenamefont {Fern\'andez-Alc\'azar}, \citenamefont {Li}, \citenamefont {Nafari},\ and\ \citenamefont {Kottos}}]{Li21_ACS}%
  \BibitemOpen
  \bibfield  {author} {\bibinfo {author} {\bibfnamefont {L.~J.}\ \bibnamefont {Fern\'andez-Alc\'azar}}, \bibinfo {author} {\bibfnamefont {H.}~\bibnamefont {Li}}, \bibinfo {author} {\bibfnamefont {M.}~\bibnamefont {Nafari}},\ and\ \bibinfo {author} {\bibfnamefont {T.}~\bibnamefont {Kottos}},\ }\bibfield  {title} {\bibinfo {title} {Implementation of optimal thermal radiation pumps using adiabatically modulated photonic cavities},\ }\href {https://doi.org/10.1021/acsphotonics.1c00896} {\bibfield  {journal} {\bibinfo  {journal} {ACS Photonics}\ }\textbf {\bibinfo {volume} {8}},\ \bibinfo {pages} {2973} (\bibinfo {year} {2021}{\natexlab{a}})}\BibitemShut {NoStop}%
\bibitem [{\citenamefont {Picardi}\ \emph {et~al.}(2023)\citenamefont {Picardi}, \citenamefont {Nimje},\ and\ \citenamefont {Papadakis}}]{Picardi23}%
  \BibitemOpen
  \bibfield  {author} {\bibinfo {author} {\bibfnamefont {M.~F.}\ \bibnamefont {Picardi}}, \bibinfo {author} {\bibfnamefont {K.~N.}\ \bibnamefont {Nimje}},\ and\ \bibinfo {author} {\bibfnamefont {G.~T.}\ \bibnamefont {Papadakis}},\ }\bibfield  {title} {\bibinfo {title} {Dynamic modulation of thermal emission—{A} tutorial},\ }\href {https://doi.org/10.1063/5.0134951} {\bibfield  {journal} {\bibinfo  {journal} {J. Appl. Phys.}\ }\textbf {\bibinfo {volume} {133}},\ \bibinfo {pages} {111101} (\bibinfo {year} {2023})}\BibitemShut {NoStop}%
\bibitem [{\citenamefont {V{\'a}zquez-Lozano}\ and\ \citenamefont {Liberal}(2023)}]{Lozano23}%
  \BibitemOpen
  \bibfield  {author} {\bibinfo {author} {\bibfnamefont {J.~E.}\ \bibnamefont {V{\'a}zquez-Lozano}}\ and\ \bibinfo {author} {\bibfnamefont {I.}~\bibnamefont {Liberal}},\ }\bibfield  {title} {\bibinfo {title} {Incandescent temporal metamaterials},\ }\href {https://doi.org/10.1038/s41467-023-40281-2} {\bibfield  {journal} {\bibinfo  {journal} {Nat. Commun.}\ }\textbf {\bibinfo {volume} {14}},\ \bibinfo {pages} {4606} (\bibinfo {year} {2023})}\BibitemShut {NoStop}%
\bibitem [{\citenamefont {Latella}\ \emph {et~al.}(2018)\citenamefont {Latella}, \citenamefont {Messina}, \citenamefont {Rubi},\ and\ \citenamefont {Ben-Abdallah}}]{Shuttling18}%
  \BibitemOpen
  \bibfield  {author} {\bibinfo {author} {\bibfnamefont {I.}~\bibnamefont {Latella}}, \bibinfo {author} {\bibfnamefont {R.}~\bibnamefont {Messina}}, \bibinfo {author} {\bibfnamefont {J.~M.}\ \bibnamefont {Rubi}},\ and\ \bibinfo {author} {\bibfnamefont {P.}~\bibnamefont {Ben-Abdallah}},\ }\bibfield  {title} {\bibinfo {title} {Radiative heat shuttling},\ }\href {https://doi.org/10.1103/PhysRevLett.121.023903} {\bibfield  {journal} {\bibinfo  {journal} {Phys. Rev. Lett.}\ }\textbf {\bibinfo {volume} {121}},\ \bibinfo {pages} {023903} (\bibinfo {year} {2018})}\BibitemShut {NoStop}%
\bibitem [{\citenamefont {Buddhiraju}\ \emph {et~al.}(2020)\citenamefont {Buddhiraju}, \citenamefont {Li},\ and\ \citenamefont {Fan}}]{Fan20}%
  \BibitemOpen
  \bibfield  {author} {\bibinfo {author} {\bibfnamefont {S.}~\bibnamefont {Buddhiraju}}, \bibinfo {author} {\bibfnamefont {W.}~\bibnamefont {Li}},\ and\ \bibinfo {author} {\bibfnamefont {S.}~\bibnamefont {Fan}},\ }\bibfield  {title} {\bibinfo {title} {Photonic refrigeration from time-modulated thermal emission},\ }\href {https://doi.org/10.1103/PhysRevLett.124.077402} {\bibfield  {journal} {\bibinfo  {journal} {Phys. Rev. Lett.}\ }\textbf {\bibinfo {volume} {124}},\ \bibinfo {pages} {077402} (\bibinfo {year} {2020})}\BibitemShut {NoStop}%
\bibitem [{\citenamefont {Yu}\ and\ \citenamefont {Fan}(2023)}]{Fan23}%
  \BibitemOpen
  \bibfield  {author} {\bibinfo {author} {\bibfnamefont {R.}~\bibnamefont {Yu}}\ and\ \bibinfo {author} {\bibfnamefont {S.}~\bibnamefont {Fan}},\ }\bibfield  {title} {\bibinfo {title} {Manipulating coherence of near-field thermal radiation in time-modulated systems},\ }\href {https://doi.org/10.1103/PhysRevLett.130.096902} {\bibfield  {journal} {\bibinfo  {journal} {Phys. Rev. Lett.}\ }\textbf {\bibinfo {volume} {130}},\ \bibinfo {pages} {096902} (\bibinfo {year} {2023})}\BibitemShut {NoStop}%
\bibitem [{\citenamefont {Yu}\ and\ \citenamefont {Fan}(2024)}]{Fan24}%
  \BibitemOpen
  \bibfield  {author} {\bibinfo {author} {\bibfnamefont {R.}~\bibnamefont {Yu}}\ and\ \bibinfo {author} {\bibfnamefont {S.}~\bibnamefont {Fan}},\ }\bibfield  {title} {\bibinfo {title} {{Time-modulated near-field radiative heat transfer}},\ }\href {https://pnas.org/doi/10.1073/pnas.2401514121} {\bibfield  {journal} {\bibinfo  {journal} {Proc. Natl. Acad. Sci.}\ }\textbf {\bibinfo {volume} {121}} (\bibinfo {year} {2024})}\BibitemShut {NoStop}%
\bibitem [{\citenamefont {Fern\'andez-Alc\'azar}\ \emph {et~al.}(2021{\natexlab{b}})\citenamefont {Fern\'andez-Alc\'azar}, \citenamefont {Kononchuk}, \citenamefont {Li},\ and\ \citenamefont {Kottos}}]{Li21}%
  \BibitemOpen
  \bibfield  {author} {\bibinfo {author} {\bibfnamefont {L.~J.}\ \bibnamefont {Fern\'andez-Alc\'azar}}, \bibinfo {author} {\bibfnamefont {R.}~\bibnamefont {Kononchuk}}, \bibinfo {author} {\bibfnamefont {H.}~\bibnamefont {Li}},\ and\ \bibinfo {author} {\bibfnamefont {T.}~\bibnamefont {Kottos}},\ }\bibfield  {title} {\bibinfo {title} {Extreme nonreciprocal near-field thermal radiation via {F}loquet photonics},\ }\href {https://doi.org/10.1103/PhysRevLett.126.204101} {\bibfield  {journal} {\bibinfo  {journal} {Phys. Rev. Lett.}\ }\textbf {\bibinfo {volume} {126}},\ \bibinfo {pages} {204101} (\bibinfo {year} {2021}{\natexlab{b}})}\BibitemShut {NoStop}%
\bibitem [{\citenamefont {Biehs}\ and\ \citenamefont {Agarwal}(2023)}]{Biehs22}%
  \BibitemOpen
  \bibfield  {author} {\bibinfo {author} {\bibfnamefont {S.-A.}\ \bibnamefont {Biehs}}\ and\ \bibinfo {author} {\bibfnamefont {G.~S.}\ \bibnamefont {Agarwal}},\ }\bibfield  {title} {\bibinfo {title} {Breakdown of detailed balance for thermal radiation by synthetic fields},\ }\href {https://doi.org/10.1103/PhysRevLett.130.110401} {\bibfield  {journal} {\bibinfo  {journal} {Phys. Rev. Lett.}\ }\textbf {\bibinfo {volume} {130}},\ \bibinfo {pages} {110401} (\bibinfo {year} {2023})}\BibitemShut {NoStop}%
\bibitem [{\citenamefont {Biehs}\ \emph {et~al.}(2023)\citenamefont {Biehs}, \citenamefont {Rodriguez-Lopez}, \citenamefont {Antezza},\ and\ \citenamefont {Agarwal}}]{Biehs23_2}%
  \BibitemOpen
  \bibfield  {author} {\bibinfo {author} {\bibfnamefont {S.-A.}\ \bibnamefont {Biehs}}, \bibinfo {author} {\bibfnamefont {P.}~\bibnamefont {Rodriguez-Lopez}}, \bibinfo {author} {\bibfnamefont {M.}~\bibnamefont {Antezza}},\ and\ \bibinfo {author} {\bibfnamefont {G.~S.}\ \bibnamefont {Agarwal}},\ }\bibfield  {title} {\bibinfo {title} {Nonreciprocal heat flux via synthetic fields in linear quantum systems},\ }\href {https://doi.org/10.1103/PhysRevA.108.042201} {\bibfield  {journal} {\bibinfo  {journal} {Phys. Rev. A}\ }\textbf {\bibinfo {volume} {108}},\ \bibinfo {pages} {042201} (\bibinfo {year} {2023})}\BibitemShut {NoStop}%
\bibitem [{\citenamefont {Aoki}\ \emph {et~al.}(2014)\citenamefont {Aoki}, \citenamefont {Tsuji}, \citenamefont {Eckstein}, \citenamefont {Kollar}, \citenamefont {Oka},\ and\ \citenamefont {Werner}}]{DMFT-RMP14}%
  \BibitemOpen
  \bibfield  {author} {\bibinfo {author} {\bibfnamefont {H.}~\bibnamefont {Aoki}}, \bibinfo {author} {\bibfnamefont {N.}~\bibnamefont {Tsuji}}, \bibinfo {author} {\bibfnamefont {M.}~\bibnamefont {Eckstein}}, \bibinfo {author} {\bibfnamefont {M.}~\bibnamefont {Kollar}}, \bibinfo {author} {\bibfnamefont {T.}~\bibnamefont {Oka}},\ and\ \bibinfo {author} {\bibfnamefont {P.}~\bibnamefont {Werner}},\ }\bibfield  {title} {\bibinfo {title} {Nonequilibrium dynamical mean-field theory and its applications},\ }\href {https://doi.org/10.1103/RevModPhys.86.779} {\bibfield  {journal} {\bibinfo  {journal} {Rev. Mod. Phys.}\ }\textbf {\bibinfo {volume} {86}},\ \bibinfo {pages} {779} (\bibinfo {year} {2014})}\BibitemShut {NoStop}%
\bibitem [{\citenamefont {Tang}\ and\ \citenamefont {Wang}(2024)}]{GT24}%
  \BibitemOpen
  \bibfield  {author} {\bibinfo {author} {\bibfnamefont {G.}~\bibnamefont {Tang}}\ and\ \bibinfo {author} {\bibfnamefont {J.-S.}\ \bibnamefont {Wang}},\ }\bibfield  {title} {\bibinfo {title} {Modulating near-field thermal transfer through temporal drivings: A quantum many-body theory},\ }\href {https://doi.org/10.1103/PhysRevB.109.085428} {\bibfield  {journal} {\bibinfo  {journal} {Phys. Rev. B}\ }\textbf {\bibinfo {volume} {109}},\ \bibinfo {pages} {085428} (\bibinfo {year} {2024})}\BibitemShut {NoStop}%
\bibitem [{\citenamefont {Chen}\ \emph {et~al.}(2015)\citenamefont {Chen}, \citenamefont {Santhanam}, \citenamefont {Sandhu}, \citenamefont {Zhu},\ and\ \citenamefont {Fan}}]{Chen2015}%
  \BibitemOpen
  \bibfield  {author} {\bibinfo {author} {\bibfnamefont {K.}~\bibnamefont {Chen}}, \bibinfo {author} {\bibfnamefont {P.}~\bibnamefont {Santhanam}}, \bibinfo {author} {\bibfnamefont {S.}~\bibnamefont {Sandhu}}, \bibinfo {author} {\bibfnamefont {L.}~\bibnamefont {Zhu}},\ and\ \bibinfo {author} {\bibfnamefont {S.}~\bibnamefont {Fan}},\ }\bibfield  {title} {\bibinfo {title} {{Heat-flux control and solid-state cooling by regulating chemical potential of photons in near-field electromagnetic heat transfer}},\ }\href {https://doi.org/10.1103/PhysRevB.91.134301} {\bibfield  {journal} {\bibinfo  {journal} {Phys. Rev. B}\ }\textbf {\bibinfo {volume} {91}},\ \bibinfo {pages} {134301} (\bibinfo {year} {2015})}\BibitemShut {NoStop}%
\bibitem [{\citenamefont {Zhu}\ \emph {et~al.}(2019)\citenamefont {Zhu}, \citenamefont {Fiorino}, \citenamefont {Thompson}, \citenamefont {Mittapally}, \citenamefont {Meyhofer},\ and\ \citenamefont {Reddy}}]{Zhu2019}%
  \BibitemOpen
  \bibfield  {author} {\bibinfo {author} {\bibfnamefont {L.}~\bibnamefont {Zhu}}, \bibinfo {author} {\bibfnamefont {A.}~\bibnamefont {Fiorino}}, \bibinfo {author} {\bibfnamefont {D.}~\bibnamefont {Thompson}}, \bibinfo {author} {\bibfnamefont {R.}~\bibnamefont {Mittapally}}, \bibinfo {author} {\bibfnamefont {E.}~\bibnamefont {Meyhofer}},\ and\ \bibinfo {author} {\bibfnamefont {P.}~\bibnamefont {Reddy}},\ }\bibfield  {title} {\bibinfo {title} {{Near-field photonic cooling through control of the chemical potential of photons}},\ }\href {https://doi.org/10.1038/s41586-019-0918-8} {\bibfield  {journal} {\bibinfo  {journal} {Nature}\ }\textbf {\bibinfo {volume} {566}},\ \bibinfo {pages} {239} (\bibinfo {year} {2019})}\BibitemShut {NoStop}%
\bibitem [{\citenamefont {Matsyshyn}\ \emph {et~al.}(2023)\citenamefont {Matsyshyn}, \citenamefont {Song}, \citenamefont {Villadiego},\ and\ \citenamefont {Shi}}]{Matsyshyn23}%
  \BibitemOpen
  \bibfield  {author} {\bibinfo {author} {\bibfnamefont {O.}~\bibnamefont {Matsyshyn}}, \bibinfo {author} {\bibfnamefont {J.~C.~W.}\ \bibnamefont {Song}}, \bibinfo {author} {\bibfnamefont {I.~S.}\ \bibnamefont {Villadiego}},\ and\ \bibinfo {author} {\bibfnamefont {L.-k.}\ \bibnamefont {Shi}},\ }\bibfield  {title} {\bibinfo {title} {{F}ermi-{D}irac staircase occupation of {F}loquet bands and current rectification inside the optical gap of metals: {A}n exact approach},\ }\href {https://doi.org/10.1103/PhysRevB.107.195135} {\bibfield  {journal} {\bibinfo  {journal} {Phys. Rev. B}\ }\textbf {\bibinfo {volume} {107}},\ \bibinfo {pages} {195135} (\bibinfo {year} {2023})}\BibitemShut {NoStop}%
\bibitem [{\citenamefont {Shi}\ \emph {et~al.}(2024)\citenamefont {Shi}, \citenamefont {Matsyshyn}, \citenamefont {Song},\ and\ \citenamefont {Villadiego}}]{Floquet-Fermi}%
  \BibitemOpen
  \bibfield  {author} {\bibinfo {author} {\bibfnamefont {L.-k.}\ \bibnamefont {Shi}}, \bibinfo {author} {\bibfnamefont {O.}~\bibnamefont {Matsyshyn}}, \bibinfo {author} {\bibfnamefont {J.~C.~W.}\ \bibnamefont {Song}},\ and\ \bibinfo {author} {\bibfnamefont {I.~S.}\ \bibnamefont {Villadiego}},\ }\bibfield  {title} {\bibinfo {title} {{F}loquet {F}ermi liquid},\ }\href {https://doi.org/10.1103/PhysRevLett.132.146402} {\bibfield  {journal} {\bibinfo  {journal} {Phys. Rev. Lett.}\ }\textbf {\bibinfo {volume} {132}},\ \bibinfo {pages} {146402} (\bibinfo {year} {2024})}\BibitemShut {NoStop}%
\bibitem [{\citenamefont {Kumari}\ \emph {et~al.}(2024)\citenamefont {Kumari}, \citenamefont {Seradjeh},\ and\ \citenamefont {Kundu}}]{kumari_josephson-current_2023}%
  \BibitemOpen
  \bibfield  {author} {\bibinfo {author} {\bibfnamefont {R.}~\bibnamefont {Kumari}}, \bibinfo {author} {\bibfnamefont {B.}~\bibnamefont {Seradjeh}},\ and\ \bibinfo {author} {\bibfnamefont {A.}~\bibnamefont {Kundu}},\ }\bibfield  {title} {\bibinfo {title} {Josephson-current signatures of unpaired {F}loquet {M}ajorana fermions},\ }\href {https://doi.org/10.1103/PhysRevLett.133.196601} {\bibfield  {journal} {\bibinfo  {journal} {Phys. Rev. Lett.}\ }\textbf {\bibinfo {volume} {133}},\ \bibinfo {pages} {196601} (\bibinfo {year} {2024})}\BibitemShut {NoStop}%
\bibitem [{\citenamefont {Mercad\'e}\ \emph {et~al.}(2021)\citenamefont {Mercad\'e}, \citenamefont {Pelka}, \citenamefont {Burgwal}, \citenamefont {Xuereb}, \citenamefont {Mart\'{\i}nez},\ and\ \citenamefont {Verhagen}}]{Laura2021}%
  \BibitemOpen
  \bibfield  {author} {\bibinfo {author} {\bibfnamefont {L.}~\bibnamefont {Mercad\'e}}, \bibinfo {author} {\bibfnamefont {K.}~\bibnamefont {Pelka}}, \bibinfo {author} {\bibfnamefont {R.}~\bibnamefont {Burgwal}}, \bibinfo {author} {\bibfnamefont {A.}~\bibnamefont {Xuereb}}, \bibinfo {author} {\bibfnamefont {A.}~\bibnamefont {Mart\'{\i}nez}},\ and\ \bibinfo {author} {\bibfnamefont {E.}~\bibnamefont {Verhagen}},\ }\bibfield  {title} {\bibinfo {title} {Floquet phonon lasing in multimode optomechanical systems},\ }\href {https://doi.org/10.1103/PhysRevLett.127.073601} {\bibfield  {journal} {\bibinfo  {journal} {Phys. Rev. Lett.}\ }\textbf {\bibinfo {volume} {127}},\ \bibinfo {pages} {073601} (\bibinfo {year} {2021})}\BibitemShut {NoStop}%
\bibitem [{\citenamefont {Wang}\ and\ \citenamefont {Peng}(2017)}]{JSW0}%
  \BibitemOpen
  \bibfield  {author} {\bibinfo {author} {\bibfnamefont {J.-S.}\ \bibnamefont {Wang}}\ and\ \bibinfo {author} {\bibfnamefont {J.}~\bibnamefont {Peng}},\ }\bibfield  {title} {\bibinfo {title} {Capacitor physics in ultra-near-field heat transfer},\ }\href {https://doi.org/10.1209/0295-5075/118/24001} {\bibfield  {journal} {\bibinfo  {journal} {Europhys. Lett.}\ }\textbf {\bibinfo {volume} {118}},\ \bibinfo {pages} {24001} (\bibinfo {year} {2017})}\BibitemShut {NoStop}%
\bibitem [{\citenamefont {Jiang}\ and\ \citenamefont {Wang}(2017)}]{JSW1}%
  \BibitemOpen
  \bibfield  {author} {\bibinfo {author} {\bibfnamefont {J.-H.}\ \bibnamefont {Jiang}}\ and\ \bibinfo {author} {\bibfnamefont {J.-S.}\ \bibnamefont {Wang}},\ }\bibfield  {title} {\bibinfo {title} {Caroli formalism in near-field heat transfer between parallel graphene sheets},\ }\href {https://doi.org/10.1103/PhysRevB.96.155437} {\bibfield  {journal} {\bibinfo  {journal} {Phys. Rev. B}\ }\textbf {\bibinfo {volume} {96}},\ \bibinfo {pages} {155437} (\bibinfo {year} {2017})}\BibitemShut {NoStop}%
\bibitem [{\citenamefont {Zhu}\ \emph {et~al.}(2020)\citenamefont {Zhu}, \citenamefont {Zhang}, \citenamefont {Gao},\ and\ \citenamefont {Wang}}]{JSW2}%
  \BibitemOpen
  \bibfield  {author} {\bibinfo {author} {\bibfnamefont {T.}~\bibnamefont {Zhu}}, \bibinfo {author} {\bibfnamefont {Z.-Q.}\ \bibnamefont {Zhang}}, \bibinfo {author} {\bibfnamefont {Z.}~\bibnamefont {Gao}},\ and\ \bibinfo {author} {\bibfnamefont {J.-S.}\ \bibnamefont {Wang}},\ }\bibfield  {title} {\bibinfo {title} {First-principles method to study near-field radiative heat transfer},\ }\href {https://doi.org/10.1103/PhysRevApplied.14.024080} {\bibfield  {journal} {\bibinfo  {journal} {Phys. Rev. Appl.}\ }\textbf {\bibinfo {volume} {14}},\ \bibinfo {pages} {024080} (\bibinfo {year} {2020})}\BibitemShut {NoStop}%
\bibitem [{\citenamefont {Wise}\ \emph {et~al.}(2022)\citenamefont {Wise}, \citenamefont {Roubinowitz}, \citenamefont {Belzig},\ and\ \citenamefont {Basko}}]{Wise22}%
  \BibitemOpen
  \bibfield  {author} {\bibinfo {author} {\bibfnamefont {J.~L.}\ \bibnamefont {Wise}}, \bibinfo {author} {\bibfnamefont {N.}~\bibnamefont {Roubinowitz}}, \bibinfo {author} {\bibfnamefont {W.}~\bibnamefont {Belzig}},\ and\ \bibinfo {author} {\bibfnamefont {D.~M.}\ \bibnamefont {Basko}},\ }\bibfield  {title} {\bibinfo {title} {Signature of resonant modes in radiative heat current noise spectrum},\ }\href {https://doi.org/10.1103/PhysRevB.106.165407} {\bibfield  {journal} {\bibinfo  {journal} {Phys. Rev. B}\ }\textbf {\bibinfo {volume} {106}},\ \bibinfo {pages} {165407} (\bibinfo {year} {2022})}\BibitemShut {NoStop}%
\bibitem [{\citenamefont {Chudnovskiy}\ \emph {et~al.}(2023)\citenamefont {Chudnovskiy}, \citenamefont {Levchenko},\ and\ \citenamefont {Kamenev}}]{Kamenev23}%
  \BibitemOpen
  \bibfield  {author} {\bibinfo {author} {\bibfnamefont {A.~L.}\ \bibnamefont {Chudnovskiy}}, \bibinfo {author} {\bibfnamefont {A.}~\bibnamefont {Levchenko}},\ and\ \bibinfo {author} {\bibfnamefont {A.}~\bibnamefont {Kamenev}},\ }\bibfield  {title} {\bibinfo {title} {Coulomb drag and heat transfer in strange metals},\ }\href {https://doi.org/10.1103/PhysRevLett.131.096501} {\bibfield  {journal} {\bibinfo  {journal} {Phys. Rev. Lett.}\ }\textbf {\bibinfo {volume} {131}},\ \bibinfo {pages} {096501} (\bibinfo {year} {2023})}\BibitemShut {NoStop}%
\bibitem [{\citenamefont {Wang}\ \emph {et~al.}(2023)\citenamefont {Wang}, \citenamefont {Peng}, \citenamefont {Zhang}, \citenamefont {Zhang},\ and\ \citenamefont {Zhu}}]{JSW23}%
  \BibitemOpen
  \bibfield  {author} {\bibinfo {author} {\bibfnamefont {J.-S.}\ \bibnamefont {Wang}}, \bibinfo {author} {\bibfnamefont {J.}~\bibnamefont {Peng}}, \bibinfo {author} {\bibfnamefont {Z.-Q.}\ \bibnamefont {Zhang}}, \bibinfo {author} {\bibfnamefont {Y.-M.}\ \bibnamefont {Zhang}},\ and\ \bibinfo {author} {\bibfnamefont {T.}~\bibnamefont {Zhu}},\ }\bibfield  {title} {\bibinfo {title} {{Transport in electron-photon systems}},\ }\href {https://doi.org/10.1007/s11467-023-1260-z} {\bibfield  {journal} {\bibinfo  {journal} {Front. Phys.}\ }\textbf {\bibinfo {volume} {18}},\ \bibinfo {pages} {43602} (\bibinfo {year} {2023})}\BibitemShut {NoStop}%
\bibitem [{\citenamefont {Wang}\ and\ \citenamefont {Antezza}(2024)}]{JSW24}%
  \BibitemOpen
  \bibfield  {author} {\bibinfo {author} {\bibfnamefont {J.-S.}\ \bibnamefont {Wang}}\ and\ \bibinfo {author} {\bibfnamefont {M.}~\bibnamefont {Antezza}},\ }\bibfield  {title} {\bibinfo {title} {Photon mediated transport of energy, linear momentum, and angular momentum in fullerene and graphene systems beyond local equilibrium},\ }\href {https://doi.org/10.1103/PhysRevB.109.125105} {\bibfield  {journal} {\bibinfo  {journal} {Phys. Rev. B}\ }\textbf {\bibinfo {volume} {109}},\ \bibinfo {pages} {125105} (\bibinfo {year} {2024})}\BibitemShut {NoStop}%
\bibitem [{SM()}]{SM}%
  \BibitemOpen
  \href@noop {} {\bibinfo {title} {{See Supplemental Material for details of derivations (photon self-energies, energy currents, and effective electron/photon distributions), proofs, and possible experimental setups, which includes Refs.~\cite{ peierls_zur_1933-3, rodriguez-vega_low-frequency_2021, Kohn2001, sipe_new_1987, joulain_surface_2005, yap_radiative_2017, Langreth, rostami_gauge_2021, shirley_solution_1965, breuer_quantum_1989}.}}}\BibitemShut {Stop}%
\bibitem [{\citenamefont {Haug}\ and\ \citenamefont {Jauho}(2008)}]{Haug_Jauho}%
  \BibitemOpen
  \bibfield  {author} {\bibinfo {author} {\bibfnamefont {H.}~\bibnamefont {Haug}}\ and\ \bibinfo {author} {\bibfnamefont {A.-P.}\ \bibnamefont {Jauho}},\ }\href@noop {} {\emph {\bibinfo {title} {Quantum Kinetics in Transport and Optics of Semiconductors}}},\ Vol.~\bibinfo {volume} {2}\ (\bibinfo  {publisher} {Springer},\ \bibinfo {year} {2008})\BibitemShut {NoStop}%
\bibitem [{\citenamefont {Lifshitz}\ and\ \citenamefont {Pitaevskii}(2013)}]{Lifshitz_book}%
  \BibitemOpen
  \bibfield  {author} {\bibinfo {author} {\bibfnamefont {E.~M.}\ \bibnamefont {Lifshitz}}\ and\ \bibinfo {author} {\bibfnamefont {L.~P.}\ \bibnamefont {Pitaevskii}},\ }\href@noop {} {\emph {\bibinfo {title} {{Statistical Physics: Theory of the Condensed State}}}},\ Vol.~\bibinfo {volume} {9}\ (\bibinfo  {publisher} {Elsevier},\ \bibinfo {year} {2013})\BibitemShut {NoStop}%
\bibitem [{\citenamefont {Kay}(2022)}]{kay_quantum_2021}%
  \BibitemOpen
  \bibfield  {author} {\bibinfo {author} {\bibfnamefont {B.~S.}\ \bibnamefont {Kay}},\ }\bibfield  {title} {\bibinfo {title} {{Quantum Electrostatics, Gauss's Law, and a Product Picture for Quantum Electrodynamics; or, the Temporal Gauge Revised}},\ }\href {https://doi.org/10.1007/s10701-021-00512-2} {\bibfield  {journal} {\bibinfo  {journal} {Found. Phys.}\ }\textbf {\bibinfo {volume} {52}},\ \bibinfo {pages} {6} (\bibinfo {year} {2022})}\BibitemShut {NoStop}%
\bibitem [{\citenamefont {Babuty}\ \emph {et~al.}(2013)\citenamefont {Babuty}, \citenamefont {Joulain}, \citenamefont {Chapuis}, \citenamefont {Greffet},\ and\ \citenamefont {{De Wilde}}}]{Babuty2013}%
  \BibitemOpen
  \bibfield  {author} {\bibinfo {author} {\bibfnamefont {A.}~\bibnamefont {Babuty}}, \bibinfo {author} {\bibfnamefont {K.}~\bibnamefont {Joulain}}, \bibinfo {author} {\bibfnamefont {P.-O.}\ \bibnamefont {Chapuis}}, \bibinfo {author} {\bibfnamefont {J.-J.}\ \bibnamefont {Greffet}},\ and\ \bibinfo {author} {\bibfnamefont {Y.}~\bibnamefont {{De Wilde}}},\ }\bibfield  {title} {\bibinfo {title} {Blackbody spectrum revisited in the near field},\ }\href {https://doi.org/10.1103/PhysRevLett.110.146103} {\bibfield  {journal} {\bibinfo  {journal} {Phys. Rev. Lett.}\ }\textbf {\bibinfo {volume} {110}},\ \bibinfo {pages} {146103} (\bibinfo {year} {2013})}\BibitemShut {NoStop}%
\bibitem [{\citenamefont {Zare}\ \emph {et~al.}(2019)\citenamefont {Zare}, \citenamefont {Tripp},\ and\ \citenamefont {Edalatpour}}]{Saman2019}%
  \BibitemOpen
  \bibfield  {author} {\bibinfo {author} {\bibfnamefont {S.}~\bibnamefont {Zare}}, \bibinfo {author} {\bibfnamefont {C.~P.}\ \bibnamefont {Tripp}},\ and\ \bibinfo {author} {\bibfnamefont {S.}~\bibnamefont {Edalatpour}},\ }\bibfield  {title} {\bibinfo {title} {Measurement of near-field thermal emission spectra using an internal reflection element},\ }\href {https://doi.org/10.1103/PhysRevB.100.235450} {\bibfield  {journal} {\bibinfo  {journal} {Phys. Rev. B}\ }\textbf {\bibinfo {volume} {100}},\ \bibinfo {pages} {235450} (\bibinfo {year} {2019})}\BibitemShut {NoStop}%
\bibitem [{\citenamefont {Hwang}\ and\ \citenamefont {Das~Sarma}(2007)}]{Hwang2007}%
  \BibitemOpen
  \bibfield  {author} {\bibinfo {author} {\bibfnamefont {E.~H.}\ \bibnamefont {Hwang}}\ and\ \bibinfo {author} {\bibfnamefont {S.}~\bibnamefont {Das~Sarma}},\ }\bibfield  {title} {\bibinfo {title} {Dielectric function, screening, and plasmons in two-dimensional graphene},\ }\href {https://doi.org/10.1103/PhysRevB.75.205418} {\bibfield  {journal} {\bibinfo  {journal} {Phys. Rev. B}\ }\textbf {\bibinfo {volume} {75}},\ \bibinfo {pages} {205418} (\bibinfo {year} {2007})}\BibitemShut {NoStop}%
\bibitem [{\citenamefont {Hu}\ \emph {et~al.}(2008)\citenamefont {Hu}, \citenamefont {Narayanaswamy}, \citenamefont {Chen},\ and\ \citenamefont {Chen}}]{Hu2008}%
  \BibitemOpen
  \bibfield  {author} {\bibinfo {author} {\bibfnamefont {L.}~\bibnamefont {Hu}}, \bibinfo {author} {\bibfnamefont {A.}~\bibnamefont {Narayanaswamy}}, \bibinfo {author} {\bibfnamefont {X.}~\bibnamefont {Chen}},\ and\ \bibinfo {author} {\bibfnamefont {G.}~\bibnamefont {Chen}},\ }\bibfield  {title} {\bibinfo {title} {{Near-field thermal radiation between two closely spaced glass plates exceeding Planck's blackbody radiation law}},\ }\href {https://doi.org/10.1063/1.2905286} {\bibfield  {journal} {\bibinfo  {journal} {Appl. Phys. Lett.}\ }\textbf {\bibinfo {volume} {92}},\ \bibinfo {pages} {133106} (\bibinfo {year} {2008})}\BibitemShut {NoStop}%
\bibitem [{\citenamefont {Song}\ \emph {et~al.}(2016)\citenamefont {Song}, \citenamefont {Thompson}, \citenamefont {Fiorino}, \citenamefont {Ganjeh}, \citenamefont {Reddy},\ and\ \citenamefont {Meyhofer}}]{Song16}%
  \BibitemOpen
  \bibfield  {author} {\bibinfo {author} {\bibfnamefont {B.}~\bibnamefont {Song}}, \bibinfo {author} {\bibfnamefont {D.}~\bibnamefont {Thompson}}, \bibinfo {author} {\bibfnamefont {A.}~\bibnamefont {Fiorino}}, \bibinfo {author} {\bibfnamefont {Y.}~\bibnamefont {Ganjeh}}, \bibinfo {author} {\bibfnamefont {P.}~\bibnamefont {Reddy}},\ and\ \bibinfo {author} {\bibfnamefont {E.}~\bibnamefont {Meyhofer}},\ }\bibfield  {title} {\bibinfo {title} {Radiative heat conductances between dielectric and metallic parallel plates with nanoscale gaps},\ }\href {https://doi.org/10.1038/nnano.2016.17} {\bibfield  {journal} {\bibinfo  {journal} {Nat. Nanotechnol.}\ }\textbf {\bibinfo {volume} {11}},\ \bibinfo {pages} {509} (\bibinfo {year} {2016})}\BibitemShut {NoStop}%
\bibitem [{\citenamefont {Zhang}\ \emph {et~al.}(2024)\citenamefont {Zhang}, \citenamefont {Dang}, \citenamefont {Li}, \citenamefont {Iqbal}, \citenamefont {Jin}, \citenamefont {Choudhury}, \citenamefont {Antezza}, \citenamefont {Xu},\ and\ \citenamefont {Ma}}]{Zhang2024}%
  \BibitemOpen
  \bibfield  {author} {\bibinfo {author} {\bibfnamefont {S.}~\bibnamefont {Zhang}}, \bibinfo {author} {\bibfnamefont {Y.}~\bibnamefont {Dang}}, \bibinfo {author} {\bibfnamefont {X.}~\bibnamefont {Li}}, \bibinfo {author} {\bibfnamefont {N.}~\bibnamefont {Iqbal}}, \bibinfo {author} {\bibfnamefont {Y.}~\bibnamefont {Jin}}, \bibinfo {author} {\bibfnamefont {P.~K.}\ \bibnamefont {Choudhury}}, \bibinfo {author} {\bibfnamefont {M.}~\bibnamefont {Antezza}}, \bibinfo {author} {\bibfnamefont {J.}~\bibnamefont {Xu}},\ and\ \bibinfo {author} {\bibfnamefont {Y.}~\bibnamefont {Ma}},\ }\bibfield  {title} {\bibinfo {title} {{Measurement of near-field thermal radiation between multilayered metamaterials}},\ }\href {https://doi.org/10.1103/PhysRevApplied.21.024054} {\bibfield  {journal} {\bibinfo  {journal} {Phys. Rev. Appl.}\ }\textbf {\bibinfo {volume} {21}},\ \bibinfo {pages} {24054} (\bibinfo {year} {2024})}\BibitemShut {NoStop}%
\bibitem [{\citenamefont {Li}\ \emph {et~al.}(2024)\citenamefont {Li}, \citenamefont {Dang}, \citenamefont {Zhang}, \citenamefont {Li}, \citenamefont {Chen}, \citenamefont {Choudhury}, \citenamefont {Jin}, \citenamefont {Xu}, \citenamefont {Ben-Abdallah}, \citenamefont {Ju},\ and\ \citenamefont {Ma}}]{Li2024}%
  \BibitemOpen
  \bibfield  {author} {\bibinfo {author} {\bibfnamefont {Y.}~\bibnamefont {Li}}, \bibinfo {author} {\bibfnamefont {Y.}~\bibnamefont {Dang}}, \bibinfo {author} {\bibfnamefont {S.}~\bibnamefont {Zhang}}, \bibinfo {author} {\bibfnamefont {X.}~\bibnamefont {Li}}, \bibinfo {author} {\bibfnamefont {T.}~\bibnamefont {Chen}}, \bibinfo {author} {\bibfnamefont {P.~K.}\ \bibnamefont {Choudhury}}, \bibinfo {author} {\bibfnamefont {Y.}~\bibnamefont {Jin}}, \bibinfo {author} {\bibfnamefont {J.}~\bibnamefont {Xu}}, \bibinfo {author} {\bibfnamefont {P.}~\bibnamefont {Ben-Abdallah}}, \bibinfo {author} {\bibfnamefont {B.-F.}\ \bibnamefont {Ju}},\ and\ \bibinfo {author} {\bibfnamefont {Y.}~\bibnamefont {Ma}},\ }\bibfield  {title} {\bibinfo {title} {{Observation of heat pumping effect by radiative shuttling}},\ }\href {https://doi.org/10.1038/s41467-024-49802-z} {\bibfield  {journal} {\bibinfo  {journal} {Nat. Commun.}\ }\textbf {\bibinfo {volume} {15}},\ \bibinfo {pages} {5465} (\bibinfo {year} {2024})}\BibitemShut {NoStop}%
\bibitem [{\citenamefont {Thompson}\ \emph {et~al.}(2020)\citenamefont {Thompson}, \citenamefont {Zhu}, \citenamefont {Meyhofer},\ and\ \citenamefont {Reddy}}]{Thompson2020}%
  \BibitemOpen
  \bibfield  {author} {\bibinfo {author} {\bibfnamefont {D.}~\bibnamefont {Thompson}}, \bibinfo {author} {\bibfnamefont {L.}~\bibnamefont {Zhu}}, \bibinfo {author} {\bibfnamefont {E.}~\bibnamefont {Meyhofer}},\ and\ \bibinfo {author} {\bibfnamefont {P.}~\bibnamefont {Reddy}},\ }\bibfield  {title} {\bibinfo {title} {{Nanoscale radiative thermal switching via multi-body effects}},\ }\href {https://doi.org/10.1038/s41565-019-0595-7} {\bibfield  {journal} {\bibinfo  {journal} {Nat. Nanotechnol.}\ }\textbf {\bibinfo {volume} {15}},\ \bibinfo {pages} {99} (\bibinfo {year} {2020})}\BibitemShut {NoStop}%
\bibitem [{\citenamefont {Lim}\ \emph {et~al.}(2024)\citenamefont {Lim}, \citenamefont {Majumder}, \citenamefont {Mittapally}, \citenamefont {Gutierrez}, \citenamefont {Luan}, \citenamefont {Meyhofer},\ and\ \citenamefont {Reddy}}]{Lim2024}%
  \BibitemOpen
  \bibfield  {author} {\bibinfo {author} {\bibfnamefont {J.~W.}\ \bibnamefont {Lim}}, \bibinfo {author} {\bibfnamefont {A.}~\bibnamefont {Majumder}}, \bibinfo {author} {\bibfnamefont {R.}~\bibnamefont {Mittapally}}, \bibinfo {author} {\bibfnamefont {A.-R.}\ \bibnamefont {Gutierrez}}, \bibinfo {author} {\bibfnamefont {Y.}~\bibnamefont {Luan}}, \bibinfo {author} {\bibfnamefont {E.}~\bibnamefont {Meyhofer}},\ and\ \bibinfo {author} {\bibfnamefont {P.}~\bibnamefont {Reddy}},\ }\bibfield  {title} {\bibinfo {title} {{A nanoscale photonic thermal transistor for sub-second heat flow switching}},\ }\href {https://doi.org/10.1038/s41467-024-49936-0} {\bibfield  {journal} {\bibinfo  {journal} {Nat. Commun.}\ }\textbf {\bibinfo {volume} {15}},\ \bibinfo {pages} {5584} (\bibinfo {year} {2024})}\BibitemShut {NoStop}%
\bibitem [{\citenamefont {Peierls}(1933)}]{peierls_zur_1933-3}%
  \BibitemOpen
  \bibfield  {author} {\bibinfo {author} {\bibfnamefont {R.}~\bibnamefont {Peierls}},\ }\bibfield  {title} {\bibinfo {title} {{Zur Theorie des Diamagnetismus von Leitungselektronen}},\ }\href {https://doi.org/10.1007/BF01342591} {\bibfield  {journal} {\bibinfo  {journal} {Z. Phys.}\ }\textbf {\bibinfo {volume} {80}},\ \bibinfo {pages} {763} (\bibinfo {year} {1933})}\BibitemShut {NoStop}%
\bibitem [{\citenamefont {{Rodriguez-Vega}}\ \emph {et~al.}(2021)\citenamefont {{Rodriguez-Vega}}, \citenamefont {Vogl},\ and\ \citenamefont {Fiete}}]{rodriguez-vega_low-frequency_2021}%
  \BibitemOpen
  \bibfield  {author} {\bibinfo {author} {\bibfnamefont {M.}~\bibnamefont {{Rodriguez-Vega}}}, \bibinfo {author} {\bibfnamefont {M.}~\bibnamefont {Vogl}},\ and\ \bibinfo {author} {\bibfnamefont {G.~A.}\ \bibnamefont {Fiete}},\ }\bibfield  {title} {\bibinfo {title} {Low-frequency and {{Moir{\'e}}}--{{Floquet}} engineering: {{A}} review},\ }\href {https://doi.org/10.1016/j.aop.2021.168434} {\bibfield  {journal} {\bibinfo  {journal} {Ann. Phys.}\ }\textbf {\bibinfo {volume} {435}},\ \bibinfo {pages} {168434} (\bibinfo {year} {2021})}\BibitemShut {NoStop}%
\bibitem [{\citenamefont {Kohn}(2001)}]{Kohn2001}%
  \BibitemOpen
  \bibfield  {author} {\bibinfo {author} {\bibfnamefont {W.}~\bibnamefont {Kohn}},\ }\bibfield  {title} {\bibinfo {title} {Periodic thermodynamics},\ }\href {https://doi.org/10.1023/A:1010327828445} {\bibfield  {journal} {\bibinfo  {journal} {J. Stat. Phys.}\ }\textbf {\bibinfo {volume} {103}},\ \bibinfo {pages} {417} (\bibinfo {year} {2001})}\BibitemShut {NoStop}%
\bibitem [{\citenamefont {Sipe}(1987)}]{sipe_new_1987}%
  \BibitemOpen
  \bibfield  {author} {\bibinfo {author} {\bibfnamefont {J.~E.}\ \bibnamefont {Sipe}},\ }\bibfield  {title} {\bibinfo {title} {New {G}reen-function formalism for surface optics},\ }\href {https://doi.org/10.1364/JOSAB.4.000481} {\bibfield  {journal} {\bibinfo  {journal} {J. Opt. Soc. Am. B}\ }\textbf {\bibinfo {volume} {4}},\ \bibinfo {pages} {481} (\bibinfo {year} {1987})}\BibitemShut {NoStop}%
\bibitem [{\citenamefont {Joulain}\ \emph {et~al.}(2005)\citenamefont {Joulain}, \citenamefont {Mulet}, \citenamefont {Marquier}, \citenamefont {Carminati},\ and\ \citenamefont {Greffet}}]{joulain_surface_2005}%
  \BibitemOpen
  \bibfield  {author} {\bibinfo {author} {\bibfnamefont {K.}~\bibnamefont {Joulain}}, \bibinfo {author} {\bibfnamefont {J.-P.}\ \bibnamefont {Mulet}}, \bibinfo {author} {\bibfnamefont {F.}~\bibnamefont {Marquier}}, \bibinfo {author} {\bibfnamefont {R.}~\bibnamefont {Carminati}},\ and\ \bibinfo {author} {\bibfnamefont {J.-J.}\ \bibnamefont {Greffet}},\ }\bibfield  {title} {\bibinfo {title} {Surface electromagnetic waves thermally excited: {R}adiative heat transfer, coherence properties and {C}asimir forces revisited in the near field},\ }\href {https://doi.org/10.1016/j.surfrep.2004.12.002} {\bibfield  {journal} {\bibinfo  {journal} {Surf. Sci. Rep.}\ }\textbf {\bibinfo {volume} {57}},\ \bibinfo {pages} {59} (\bibinfo {year} {2005})}\BibitemShut {NoStop}%
\bibitem [{\citenamefont {Yap}\ and\ \citenamefont {Wang}(2017)}]{yap_radiative_2017}%
  \BibitemOpen
  \bibfield  {author} {\bibinfo {author} {\bibfnamefont {H.~H.}\ \bibnamefont {Yap}}\ and\ \bibinfo {author} {\bibfnamefont {J.-S.}\ \bibnamefont {Wang}},\ }\bibfield  {title} {\bibinfo {title} {Radiative heat transfer as a {L}andauer-{B}\"uttiker problem},\ }\href {https://doi.org/10.1103/PhysRevE.95.012126} {\bibfield  {journal} {\bibinfo  {journal} {Phys. Rev. E}\ }\textbf {\bibinfo {volume} {95}},\ \bibinfo {pages} {012126} (\bibinfo {year} {2017})}\BibitemShut {NoStop}%
\bibitem [{\citenamefont {Langreth}\ and\ \citenamefont {Nordlander}(1991)}]{Langreth}%
  \BibitemOpen
  \bibfield  {author} {\bibinfo {author} {\bibfnamefont {D.~C.}\ \bibnamefont {Langreth}}\ and\ \bibinfo {author} {\bibfnamefont {P.}~\bibnamefont {Nordlander}},\ }\bibfield  {title} {\bibinfo {title} {Derivation of a master equation for charge-transfer processes in atom-surface collisions},\ }\href {https://doi.org/10.1103/PhysRevB.43.2541} {\bibfield  {journal} {\bibinfo  {journal} {Phys. Rev. B}\ }\textbf {\bibinfo {volume} {43}},\ \bibinfo {pages} {2541} (\bibinfo {year} {1991})}\BibitemShut {NoStop}%
\bibitem [{\citenamefont {Rostami}\ \emph {et~al.}(2021)\citenamefont {Rostami}, \citenamefont {Katsnelson}, \citenamefont {Vignale},\ and\ \citenamefont {Polini}}]{rostami_gauge_2021}%
  \BibitemOpen
  \bibfield  {author} {\bibinfo {author} {\bibfnamefont {H.}~\bibnamefont {Rostami}}, \bibinfo {author} {\bibfnamefont {M.~I.}\ \bibnamefont {Katsnelson}}, \bibinfo {author} {\bibfnamefont {G.}~\bibnamefont {Vignale}},\ and\ \bibinfo {author} {\bibfnamefont {M.}~\bibnamefont {Polini}},\ }\bibfield  {title} {\bibinfo {title} {Gauge invariance and {{Ward}} identities in nonlinear response theory},\ }\href {https://doi.org/10.1016/j.aop.2021.168523} {\bibfield  {journal} {\bibinfo  {journal} {Ann. Phys.}\ }\textbf {\bibinfo {volume} {431}},\ \bibinfo {pages} {168523} (\bibinfo {year} {2021})}\BibitemShut {NoStop}%
\bibitem [{\citenamefont {Shirley}(1965)}]{shirley_solution_1965}%
  \BibitemOpen
  \bibfield  {author} {\bibinfo {author} {\bibfnamefont {J.~H.}\ \bibnamefont {Shirley}},\ }\bibfield  {title} {\bibinfo {title} {Solution of the {S}chr\"odinger equation with a {H}amiltonian periodic in time},\ }\href {https://doi.org/10.1103/PhysRev.138.B979} {\bibfield  {journal} {\bibinfo  {journal} {Phys. Rev.}\ }\textbf {\bibinfo {volume} {138}},\ \bibinfo {pages} {B979} (\bibinfo {year} {1965})}\BibitemShut {NoStop}%
\bibitem [{\citenamefont {Breuer}\ and\ \citenamefont {Holthaus}(1989)}]{breuer_quantum_1989}%
  \BibitemOpen
  \bibfield  {author} {\bibinfo {author} {\bibfnamefont {H.~P.}\ \bibnamefont {Breuer}}\ and\ \bibinfo {author} {\bibfnamefont {M.}~\bibnamefont {Holthaus}},\ }\bibfield  {title} {\bibinfo {title} {Quantum phases and {{Landau-Zener}} transitions in oscillating fields},\ }\href {https://doi.org/10.1016/0375-9601(89)90132-1} {\bibfield  {journal} {\bibinfo  {journal} {Phys. Lett. A}\ }\textbf {\bibinfo {volume} {140}},\ \bibinfo {pages} {507} (\bibinfo {year} {1989})}\BibitemShut {NoStop}%
\end{thebibliography}%


\begin{thebibliography}{27}%
\makeatletter
\providecommand \@ifxundefined [1]{%
 \@ifx{#1\undefined}
}%
\providecommand \@ifnum [1]{%
 \ifnum #1\expandafter \@firstoftwo
 \else \expandafter \@secondoftwo
 \fi
}%
\providecommand \@ifx [1]{%
 \ifx #1\expandafter \@firstoftwo
 \else \expandafter \@secondoftwo
 \fi
}%
\providecommand \natexlab [1]{#1}%
\providecommand \enquote  [1]{``#1''}%
\providecommand \bibnamefont  [1]{#1}%
\providecommand \bibfnamefont [1]{#1}%
\providecommand \citenamefont [1]{#1}%
\providecommand \href@noop [0]{\@secondoftwo}%
\providecommand \href [0]{\begingroup \@sanitize@url \@href}%
\providecommand \@href[1]{\@@startlink{#1}\@@href}%
\providecommand \@@href[1]{\endgroup#1\@@endlink}%
\providecommand \@sanitize@url [0]{\catcode `\\12\catcode `\$12\catcode `\&12\catcode `\#12\catcode `\^12\catcode `\_12\catcode `\%12\relax}%
\providecommand \@@startlink[1]{}%
\providecommand \@@endlink[0]{}%
\providecommand \url  [0]{\begingroup\@sanitize@url \@url }%
\providecommand \@url [1]{\endgroup\@href {#1}{\urlprefix }}%
\providecommand \urlprefix  [0]{URL }%
\providecommand \Eprint [0]{\href }%
\providecommand \doibase [0]{https://doi.org/}%
\providecommand \selectlanguage [0]{\@gobble}%
\providecommand \bibinfo  [0]{\@secondoftwo}%
\providecommand \bibfield  [0]{\@secondoftwo}%
\providecommand \translation [1]{[#1]}%
\providecommand \BibitemOpen [0]{}%
\providecommand \bibitemStop [0]{}%
\providecommand \bibitemNoStop [0]{.\EOS\space}%
\providecommand \EOS [0]{\spacefactor3000\relax}%
\providecommand \BibitemShut  [1]{\csname bibitem#1\endcsname}%
\let\auto@bib@innerbib\@empty
\bibitem [{\citenamefont {Wang}\ \emph {et~al.}(2023)\citenamefont {Wang}, \citenamefont {Peng}, \citenamefont {Zhang}, \citenamefont {Zhang},\ and\ \citenamefont {Zhu}}]{JSW23}%
  \BibitemOpen
  \bibfield  {author} {\bibinfo {author} {\bibfnamefont {J.-S.}\ \bibnamefont {Wang}}, \bibinfo {author} {\bibfnamefont {J.}~\bibnamefont {Peng}}, \bibinfo {author} {\bibfnamefont {Z.-Q.}\ \bibnamefont {Zhang}}, \bibinfo {author} {\bibfnamefont {Y.-M.}\ \bibnamefont {Zhang}},\ and\ \bibinfo {author} {\bibfnamefont {T.}~\bibnamefont {Zhu}},\ }\bibfield  {title} {\bibinfo {title} {{Transport in electron-photon systems}},\ }\href {https://doi.org/10.1007/s11467-023-1260-z} {\bibfield  {journal} {\bibinfo  {journal} {Front. Phys.}\ }\textbf {\bibinfo {volume} {18}},\ \bibinfo {pages} {43602} (\bibinfo {year} {2023})}\BibitemShut {NoStop}%
\bibitem [{\citenamefont {Matsyshyn}\ \emph {et~al.}(2023)\citenamefont {Matsyshyn}, \citenamefont {Song}, \citenamefont {Villadiego},\ and\ \citenamefont {Shi}}]{Matsyshyn23}%
  \BibitemOpen
  \bibfield  {author} {\bibinfo {author} {\bibfnamefont {O.}~\bibnamefont {Matsyshyn}}, \bibinfo {author} {\bibfnamefont {J.~C.~W.}\ \bibnamefont {Song}}, \bibinfo {author} {\bibfnamefont {I.~S.}\ \bibnamefont {Villadiego}},\ and\ \bibinfo {author} {\bibfnamefont {L.-k.}\ \bibnamefont {Shi}},\ }\bibfield  {title} {\bibinfo {title} {{F}ermi-{D}irac staircase occupation of {F}loquet bands and current rectification inside the optical gap of metals: {A}n exact approach},\ }\href {https://doi.org/10.1103/PhysRevB.107.195135} {\bibfield  {journal} {\bibinfo  {journal} {Phys. Rev. B}\ }\textbf {\bibinfo {volume} {107}},\ \bibinfo {pages} {195135} (\bibinfo {year} {2023})}\BibitemShut {NoStop}%
\bibitem [{\citenamefont {Lifshitz}\ and\ \citenamefont {Pitaevskii}(2013)}]{Lifshitz_book}%
  \BibitemOpen
  \bibfield  {author} {\bibinfo {author} {\bibfnamefont {E.~M.}\ \bibnamefont {Lifshitz}}\ and\ \bibinfo {author} {\bibfnamefont {L.~P.}\ \bibnamefont {Pitaevskii}},\ }\href@noop {} {\emph {\bibinfo {title} {{Statistical Physics: Theory of the Condensed State}}}},\ Vol.~\bibinfo {volume} {9}\ (\bibinfo  {publisher} {Elsevier},\ \bibinfo {year} {2013})\BibitemShut {NoStop}%
\bibitem [{\citenamefont {Kay}(2022)}]{kay_quantum_2021}%
  \BibitemOpen
  \bibfield  {author} {\bibinfo {author} {\bibfnamefont {B.~S.}\ \bibnamefont {Kay}},\ }\bibfield  {title} {\bibinfo {title} {{Quantum Electrostatics, Gauss's Law, and a Product Picture for Quantum Electrodynamics; or, the Temporal Gauge Revised}},\ }\href {https://doi.org/10.1007/s10701-021-00512-2} {\bibfield  {journal} {\bibinfo  {journal} {Found. Phys.}\ }\textbf {\bibinfo {volume} {52}},\ \bibinfo {pages} {6} (\bibinfo {year} {2022})}\BibitemShut {NoStop}%
\bibitem [{\citenamefont {Peierls}(1933)}]{peierls_zur_1933-3}%
  \BibitemOpen
  \bibfield  {author} {\bibinfo {author} {\bibfnamefont {R.}~\bibnamefont {Peierls}},\ }\bibfield  {title} {\bibinfo {title} {{Zur Theorie des Diamagnetismus von Leitungselektronen}},\ }\href {https://doi.org/10.1007/BF01342591} {\bibfield  {journal} {\bibinfo  {journal} {Z. Phys.}\ }\textbf {\bibinfo {volume} {80}},\ \bibinfo {pages} {763} (\bibinfo {year} {1933})}\BibitemShut {NoStop}%
\bibitem [{\citenamefont {Tsuji}\ \emph {et~al.}(2008)\citenamefont {Tsuji}, \citenamefont {Oka},\ and\ \citenamefont {Aoki}}]{Tsuji08}%
  \BibitemOpen
  \bibfield  {author} {\bibinfo {author} {\bibfnamefont {N.}~\bibnamefont {Tsuji}}, \bibinfo {author} {\bibfnamefont {T.}~\bibnamefont {Oka}},\ and\ \bibinfo {author} {\bibfnamefont {H.}~\bibnamefont {Aoki}},\ }\bibfield  {title} {\bibinfo {title} {Correlated electron systems periodically driven out of equilibrium: $\text{Floquet}+\text{DMFT}$ formalism},\ }\href {https://doi.org/10.1103/PhysRevB.78.235124} {\bibfield  {journal} {\bibinfo  {journal} {Phys. Rev. B}\ }\textbf {\bibinfo {volume} {78}},\ \bibinfo {pages} {235124} (\bibinfo {year} {2008})}\BibitemShut {NoStop}%
\bibitem [{\citenamefont {Aoki}\ \emph {et~al.}(2014)\citenamefont {Aoki}, \citenamefont {Tsuji}, \citenamefont {Eckstein}, \citenamefont {Kollar}, \citenamefont {Oka},\ and\ \citenamefont {Werner}}]{DMFT-RMP14}%
  \BibitemOpen
  \bibfield  {author} {\bibinfo {author} {\bibfnamefont {H.}~\bibnamefont {Aoki}}, \bibinfo {author} {\bibfnamefont {N.}~\bibnamefont {Tsuji}}, \bibinfo {author} {\bibfnamefont {M.}~\bibnamefont {Eckstein}}, \bibinfo {author} {\bibfnamefont {M.}~\bibnamefont {Kollar}}, \bibinfo {author} {\bibfnamefont {T.}~\bibnamefont {Oka}},\ and\ \bibinfo {author} {\bibfnamefont {P.}~\bibnamefont {Werner}},\ }\bibfield  {title} {\bibinfo {title} {Nonequilibrium dynamical mean-field theory and its applications},\ }\href {https://doi.org/10.1103/RevModPhys.86.779} {\bibfield  {journal} {\bibinfo  {journal} {Rev. Mod. Phys.}\ }\textbf {\bibinfo {volume} {86}},\ \bibinfo {pages} {779} (\bibinfo {year} {2014})}\BibitemShut {NoStop}%
\bibitem [{\citenamefont {{Rodriguez-Vega}}\ \emph {et~al.}(2021)\citenamefont {{Rodriguez-Vega}}, \citenamefont {Vogl},\ and\ \citenamefont {Fiete}}]{rodriguez-vega_low-frequency_2021}%
  \BibitemOpen
  \bibfield  {author} {\bibinfo {author} {\bibfnamefont {M.}~\bibnamefont {{Rodriguez-Vega}}}, \bibinfo {author} {\bibfnamefont {M.}~\bibnamefont {Vogl}},\ and\ \bibinfo {author} {\bibfnamefont {G.~A.}\ \bibnamefont {Fiete}},\ }\bibfield  {title} {\bibinfo {title} {Low-frequency and {{Moir{\'e}}}--{{Floquet}} engineering: {{A}} review},\ }\href {https://doi.org/10.1016/j.aop.2021.168434} {\bibfield  {journal} {\bibinfo  {journal} {Ann. Phys.}\ }\textbf {\bibinfo {volume} {435}},\ \bibinfo {pages} {168434} (\bibinfo {year} {2021})}\BibitemShut {NoStop}%
\bibitem [{\citenamefont {Tang}\ and\ \citenamefont {Wang}(2024)}]{GT24}%
  \BibitemOpen
  \bibfield  {author} {\bibinfo {author} {\bibfnamefont {G.}~\bibnamefont {Tang}}\ and\ \bibinfo {author} {\bibfnamefont {J.-S.}\ \bibnamefont {Wang}},\ }\bibfield  {title} {\bibinfo {title} {Modulating near-field thermal transfer through temporal drivings: A quantum many-body theory},\ }\href {https://doi.org/10.1103/PhysRevB.109.085428} {\bibfield  {journal} {\bibinfo  {journal} {Phys. Rev. B}\ }\textbf {\bibinfo {volume} {109}},\ \bibinfo {pages} {085428} (\bibinfo {year} {2024})}\BibitemShut {NoStop}%
\bibitem [{\citenamefont {Haug}\ and\ \citenamefont {Jauho}(2008)}]{Haug_Jauho}%
  \BibitemOpen
  \bibfield  {author} {\bibinfo {author} {\bibfnamefont {H.}~\bibnamefont {Haug}}\ and\ \bibinfo {author} {\bibfnamefont {A.-P.}\ \bibnamefont {Jauho}},\ }\href@noop {} {\emph {\bibinfo {title} {Quantum Kinetics in Transport and Optics of Semiconductors}}},\ Vol.~\bibinfo {volume} {2}\ (\bibinfo  {publisher} {Springer},\ \bibinfo {year} {2008})\BibitemShut {NoStop}%
\bibitem [{\citenamefont {Kohn}(2001)}]{Kohn2001}%
  \BibitemOpen
  \bibfield  {author} {\bibinfo {author} {\bibfnamefont {W.}~\bibnamefont {Kohn}},\ }\bibfield  {title} {\bibinfo {title} {Periodic thermodynamics},\ }\href {https://doi.org/10.1023/A:1010327828445} {\bibfield  {journal} {\bibinfo  {journal} {J. Stat. Phys.}\ }\textbf {\bibinfo {volume} {103}},\ \bibinfo {pages} {417} (\bibinfo {year} {2001})}\BibitemShut {NoStop}%
\bibitem [{\citenamefont {Sipe}(1987)}]{sipe_new_1987}%
  \BibitemOpen
  \bibfield  {author} {\bibinfo {author} {\bibfnamefont {J.~E.}\ \bibnamefont {Sipe}},\ }\bibfield  {title} {\bibinfo {title} {New {G}reen-function formalism for surface optics},\ }\href {https://doi.org/10.1364/JOSAB.4.000481} {\bibfield  {journal} {\bibinfo  {journal} {J. Opt. Soc. Am. B}\ }\textbf {\bibinfo {volume} {4}},\ \bibinfo {pages} {481} (\bibinfo {year} {1987})}\BibitemShut {NoStop}%
\bibitem [{\citenamefont {Joulain}\ \emph {et~al.}(2005)\citenamefont {Joulain}, \citenamefont {Mulet}, \citenamefont {Marquier}, \citenamefont {Carminati},\ and\ \citenamefont {Greffet}}]{joulain_surface_2005}%
  \BibitemOpen
  \bibfield  {author} {\bibinfo {author} {\bibfnamefont {K.}~\bibnamefont {Joulain}}, \bibinfo {author} {\bibfnamefont {J.-P.}\ \bibnamefont {Mulet}}, \bibinfo {author} {\bibfnamefont {F.}~\bibnamefont {Marquier}}, \bibinfo {author} {\bibfnamefont {R.}~\bibnamefont {Carminati}},\ and\ \bibinfo {author} {\bibfnamefont {J.-J.}\ \bibnamefont {Greffet}},\ }\bibfield  {title} {\bibinfo {title} {Surface electromagnetic waves thermally excited: {R}adiative heat transfer, coherence properties and {C}asimir forces revisited in the near field},\ }\href {https://doi.org/10.1016/j.surfrep.2004.12.002} {\bibfield  {journal} {\bibinfo  {journal} {Surf. Sci. Rep.}\ }\textbf {\bibinfo {volume} {57}},\ \bibinfo {pages} {59} (\bibinfo {year} {2005})}\BibitemShut {NoStop}%
\bibitem [{\citenamefont {Yap}\ and\ \citenamefont {Wang}(2017)}]{yap_radiative_2017}%
  \BibitemOpen
  \bibfield  {author} {\bibinfo {author} {\bibfnamefont {H.~H.}\ \bibnamefont {Yap}}\ and\ \bibinfo {author} {\bibfnamefont {J.-S.}\ \bibnamefont {Wang}},\ }\bibfield  {title} {\bibinfo {title} {Radiative heat transfer as a {L}andauer-{B}\"uttiker problem},\ }\href {https://doi.org/10.1103/PhysRevE.95.012126} {\bibfield  {journal} {\bibinfo  {journal} {Phys. Rev. E}\ }\textbf {\bibinfo {volume} {95}},\ \bibinfo {pages} {012126} (\bibinfo {year} {2017})}\BibitemShut {NoStop}%
\bibitem [{\citenamefont {Langreth}\ and\ \citenamefont {Nordlander}(1991)}]{Langreth}%
  \BibitemOpen
  \bibfield  {author} {\bibinfo {author} {\bibfnamefont {D.~C.}\ \bibnamefont {Langreth}}\ and\ \bibinfo {author} {\bibfnamefont {P.}~\bibnamefont {Nordlander}},\ }\bibfield  {title} {\bibinfo {title} {Derivation of a master equation for charge-transfer processes in atom-surface collisions},\ }\href {https://doi.org/10.1103/PhysRevB.43.2541} {\bibfield  {journal} {\bibinfo  {journal} {Phys. Rev. B}\ }\textbf {\bibinfo {volume} {43}},\ \bibinfo {pages} {2541} (\bibinfo {year} {1991})}\BibitemShut {NoStop}%
\bibitem [{\citenamefont {Wang}\ and\ \citenamefont {Antezza}(2024)}]{JSW24}%
  \BibitemOpen
  \bibfield  {author} {\bibinfo {author} {\bibfnamefont {J.-S.}\ \bibnamefont {Wang}}\ and\ \bibinfo {author} {\bibfnamefont {M.}~\bibnamefont {Antezza}},\ }\bibfield  {title} {\bibinfo {title} {Photon mediated transport of energy, linear momentum, and angular momentum in fullerene and graphene systems beyond local equilibrium},\ }\href {https://doi.org/10.1103/PhysRevB.109.125105} {\bibfield  {journal} {\bibinfo  {journal} {Phys. Rev. B}\ }\textbf {\bibinfo {volume} {109}},\ \bibinfo {pages} {125105} (\bibinfo {year} {2024})}\BibitemShut {NoStop}%
\bibitem [{\citenamefont {Rostami}\ \emph {et~al.}(2021)\citenamefont {Rostami}, \citenamefont {Katsnelson}, \citenamefont {Vignale},\ and\ \citenamefont {Polini}}]{rostami_gauge_2021}%
  \BibitemOpen
  \bibfield  {author} {\bibinfo {author} {\bibfnamefont {H.}~\bibnamefont {Rostami}}, \bibinfo {author} {\bibfnamefont {M.~I.}\ \bibnamefont {Katsnelson}}, \bibinfo {author} {\bibfnamefont {G.}~\bibnamefont {Vignale}},\ and\ \bibinfo {author} {\bibfnamefont {M.}~\bibnamefont {Polini}},\ }\bibfield  {title} {\bibinfo {title} {Gauge invariance and {{Ward}} identities in nonlinear response theory},\ }\href {https://doi.org/10.1016/j.aop.2021.168523} {\bibfield  {journal} {\bibinfo  {journal} {Ann. Phys.}\ }\textbf {\bibinfo {volume} {431}},\ \bibinfo {pages} {168523} (\bibinfo {year} {2021})}\BibitemShut {NoStop}%
\bibitem [{\citenamefont {Shirley}(1965)}]{shirley_solution_1965}%
  \BibitemOpen
  \bibfield  {author} {\bibinfo {author} {\bibfnamefont {J.~H.}\ \bibnamefont {Shirley}},\ }\bibfield  {title} {\bibinfo {title} {Solution of the {S}chr\"odinger equation with a {H}amiltonian periodic in time},\ }\href {https://doi.org/10.1103/PhysRev.138.B979} {\bibfield  {journal} {\bibinfo  {journal} {Phys. Rev.}\ }\textbf {\bibinfo {volume} {138}},\ \bibinfo {pages} {B979} (\bibinfo {year} {1965})}\BibitemShut {NoStop}%
\bibitem [{\citenamefont {Breuer}\ and\ \citenamefont {Holthaus}(1989)}]{breuer_quantum_1989}%
  \BibitemOpen
  \bibfield  {author} {\bibinfo {author} {\bibfnamefont {H.~P.}\ \bibnamefont {Breuer}}\ and\ \bibinfo {author} {\bibfnamefont {M.}~\bibnamefont {Holthaus}},\ }\bibfield  {title} {\bibinfo {title} {Quantum phases and {{Landau-Zener}} transitions in oscillating fields},\ }\href {https://doi.org/10.1016/0375-9601(89)90132-1} {\bibfield  {journal} {\bibinfo  {journal} {Phys. Lett. A}\ }\textbf {\bibinfo {volume} {140}},\ \bibinfo {pages} {507} (\bibinfo {year} {1989})}\BibitemShut {NoStop}%
\bibitem [{\citenamefont {Seetharam}\ \emph {et~al.}(2015)\citenamefont {Seetharam}, \citenamefont {Bardyn}, \citenamefont {Lindner}, \citenamefont {Rudner},\ and\ \citenamefont {Refael}}]{Refael15}%
  \BibitemOpen
  \bibfield  {author} {\bibinfo {author} {\bibfnamefont {K.~I.}\ \bibnamefont {Seetharam}}, \bibinfo {author} {\bibfnamefont {C.-E.}\ \bibnamefont {Bardyn}}, \bibinfo {author} {\bibfnamefont {N.~H.}\ \bibnamefont {Lindner}}, \bibinfo {author} {\bibfnamefont {M.~S.}\ \bibnamefont {Rudner}},\ and\ \bibinfo {author} {\bibfnamefont {G.}~\bibnamefont {Refael}},\ }\bibfield  {title} {\bibinfo {title} {Controlled population of {F}loquet-{B}loch states via coupling to {B}ose and {F}ermi baths},\ }\href {https://doi.org/10.1103/PhysRevX.5.041050} {\bibfield  {journal} {\bibinfo  {journal} {Phys. Rev. X}\ }\textbf {\bibinfo {volume} {5}},\ \bibinfo {pages} {041050} (\bibinfo {year} {2015})}\BibitemShut {NoStop}%
\bibitem [{\citenamefont {Kumari}\ \emph {et~al.}(2024)\citenamefont {Kumari}, \citenamefont {Seradjeh},\ and\ \citenamefont {Kundu}}]{kumari_josephson-current_2023}%
  \BibitemOpen
  \bibfield  {author} {\bibinfo {author} {\bibfnamefont {R.}~\bibnamefont {Kumari}}, \bibinfo {author} {\bibfnamefont {B.}~\bibnamefont {Seradjeh}},\ and\ \bibinfo {author} {\bibfnamefont {A.}~\bibnamefont {Kundu}},\ }\bibfield  {title} {\bibinfo {title} {Josephson-current signatures of unpaired {F}loquet {M}ajorana fermions},\ }\href {https://doi.org/10.1103/PhysRevLett.133.196601} {\bibfield  {journal} {\bibinfo  {journal} {Phys. Rev. Lett.}\ }\textbf {\bibinfo {volume} {133}},\ \bibinfo {pages} {196601} (\bibinfo {year} {2024})}\BibitemShut {NoStop}%
\bibitem [{\citenamefont {Hu}\ \emph {et~al.}(2008)\citenamefont {Hu}, \citenamefont {Narayanaswamy}, \citenamefont {Chen},\ and\ \citenamefont {Chen}}]{Hu2008}%
  \BibitemOpen
  \bibfield  {author} {\bibinfo {author} {\bibfnamefont {L.}~\bibnamefont {Hu}}, \bibinfo {author} {\bibfnamefont {A.}~\bibnamefont {Narayanaswamy}}, \bibinfo {author} {\bibfnamefont {X.}~\bibnamefont {Chen}},\ and\ \bibinfo {author} {\bibfnamefont {G.}~\bibnamefont {Chen}},\ }\bibfield  {title} {\bibinfo {title} {{Near-field thermal radiation between two closely spaced glass plates exceeding Planck's blackbody radiation law}},\ }\href {https://doi.org/10.1063/1.2905286} {\bibfield  {journal} {\bibinfo  {journal} {Appl. Phys. Lett.}\ }\textbf {\bibinfo {volume} {92}},\ \bibinfo {pages} {133106} (\bibinfo {year} {2008})}\BibitemShut {NoStop}%
\bibitem [{\citenamefont {Song}\ \emph {et~al.}(2016)\citenamefont {Song}, \citenamefont {Thompson}, \citenamefont {Fiorino}, \citenamefont {Ganjeh}, \citenamefont {Reddy},\ and\ \citenamefont {Meyhofer}}]{Song16}%
  \BibitemOpen
  \bibfield  {author} {\bibinfo {author} {\bibfnamefont {B.}~\bibnamefont {Song}}, \bibinfo {author} {\bibfnamefont {D.}~\bibnamefont {Thompson}}, \bibinfo {author} {\bibfnamefont {A.}~\bibnamefont {Fiorino}}, \bibinfo {author} {\bibfnamefont {Y.}~\bibnamefont {Ganjeh}}, \bibinfo {author} {\bibfnamefont {P.}~\bibnamefont {Reddy}},\ and\ \bibinfo {author} {\bibfnamefont {E.}~\bibnamefont {Meyhofer}},\ }\bibfield  {title} {\bibinfo {title} {Radiative heat conductances between dielectric and metallic parallel plates with nanoscale gaps},\ }\href {https://doi.org/10.1038/nnano.2016.17} {\bibfield  {journal} {\bibinfo  {journal} {Nat. Nanotechnol.}\ }\textbf {\bibinfo {volume} {11}},\ \bibinfo {pages} {509} (\bibinfo {year} {2016})}\BibitemShut {NoStop}%
\bibitem [{\citenamefont {Zhang}\ \emph {et~al.}(2024)\citenamefont {Zhang}, \citenamefont {Dang}, \citenamefont {Li}, \citenamefont {Iqbal}, \citenamefont {Jin}, \citenamefont {Choudhury}, \citenamefont {Antezza}, \citenamefont {Xu},\ and\ \citenamefont {Ma}}]{Zhang2024}%
  \BibitemOpen
  \bibfield  {author} {\bibinfo {author} {\bibfnamefont {S.}~\bibnamefont {Zhang}}, \bibinfo {author} {\bibfnamefont {Y.}~\bibnamefont {Dang}}, \bibinfo {author} {\bibfnamefont {X.}~\bibnamefont {Li}}, \bibinfo {author} {\bibfnamefont {N.}~\bibnamefont {Iqbal}}, \bibinfo {author} {\bibfnamefont {Y.}~\bibnamefont {Jin}}, \bibinfo {author} {\bibfnamefont {P.~K.}\ \bibnamefont {Choudhury}}, \bibinfo {author} {\bibfnamefont {M.}~\bibnamefont {Antezza}}, \bibinfo {author} {\bibfnamefont {J.}~\bibnamefont {Xu}},\ and\ \bibinfo {author} {\bibfnamefont {Y.}~\bibnamefont {Ma}},\ }\bibfield  {title} {\bibinfo {title} {{Measurement of near-field thermal radiation between multilayered metamaterials}},\ }\href {https://doi.org/10.1103/PhysRevApplied.21.024054} {\bibfield  {journal} {\bibinfo  {journal} {Phys. Rev. Appl.}\ }\textbf {\bibinfo {volume} {21}},\ \bibinfo {pages} {24054} (\bibinfo {year} {2024})}\BibitemShut {NoStop}%
\bibitem [{\citenamefont {Li}\ \emph {et~al.}(2024)\citenamefont {Li}, \citenamefont {Dang}, \citenamefont {Zhang}, \citenamefont {Li}, \citenamefont {Chen}, \citenamefont {Choudhury}, \citenamefont {Jin}, \citenamefont {Xu}, \citenamefont {Ben-Abdallah}, \citenamefont {Ju},\ and\ \citenamefont {Ma}}]{Li2024}%
  \BibitemOpen
  \bibfield  {author} {\bibinfo {author} {\bibfnamefont {Y.}~\bibnamefont {Li}}, \bibinfo {author} {\bibfnamefont {Y.}~\bibnamefont {Dang}}, \bibinfo {author} {\bibfnamefont {S.}~\bibnamefont {Zhang}}, \bibinfo {author} {\bibfnamefont {X.}~\bibnamefont {Li}}, \bibinfo {author} {\bibfnamefont {T.}~\bibnamefont {Chen}}, \bibinfo {author} {\bibfnamefont {P.~K.}\ \bibnamefont {Choudhury}}, \bibinfo {author} {\bibfnamefont {Y.}~\bibnamefont {Jin}}, \bibinfo {author} {\bibfnamefont {J.}~\bibnamefont {Xu}}, \bibinfo {author} {\bibfnamefont {P.}~\bibnamefont {Ben-Abdallah}}, \bibinfo {author} {\bibfnamefont {B.-F.}\ \bibnamefont {Ju}},\ and\ \bibinfo {author} {\bibfnamefont {Y.}~\bibnamefont {Ma}},\ }\bibfield  {title} {\bibinfo {title} {{Observation of heat pumping effect by radiative shuttling}},\ }\href {https://doi.org/10.1038/s41467-024-49802-z} {\bibfield  {journal} {\bibinfo  {journal} {Nat. Commun.}\ }\textbf {\bibinfo {volume} {15}},\ \bibinfo {pages} {5465} (\bibinfo {year} {2024})}\BibitemShut {NoStop}%
\bibitem [{\citenamefont {Thompson}\ \emph {et~al.}(2020)\citenamefont {Thompson}, \citenamefont {Zhu}, \citenamefont {Meyhofer},\ and\ \citenamefont {Reddy}}]{Thompson2020}%
  \BibitemOpen
  \bibfield  {author} {\bibinfo {author} {\bibfnamefont {D.}~\bibnamefont {Thompson}}, \bibinfo {author} {\bibfnamefont {L.}~\bibnamefont {Zhu}}, \bibinfo {author} {\bibfnamefont {E.}~\bibnamefont {Meyhofer}},\ and\ \bibinfo {author} {\bibfnamefont {P.}~\bibnamefont {Reddy}},\ }\bibfield  {title} {\bibinfo {title} {{Nanoscale radiative thermal switching via multi-body effects}},\ }\href {https://doi.org/10.1038/s41565-019-0595-7} {\bibfield  {journal} {\bibinfo  {journal} {Nat. Nanotechnol.}\ }\textbf {\bibinfo {volume} {15}},\ \bibinfo {pages} {99} (\bibinfo {year} {2020})}\BibitemShut {NoStop}%
\bibitem [{\citenamefont {Lim}\ \emph {et~al.}(2024)\citenamefont {Lim}, \citenamefont {Majumder}, \citenamefont {Mittapally}, \citenamefont {Gutierrez}, \citenamefont {Luan}, \citenamefont {Meyhofer},\ and\ \citenamefont {Reddy}}]{Lim2024}%
  \BibitemOpen
  \bibfield  {author} {\bibinfo {author} {\bibfnamefont {J.~W.}\ \bibnamefont {Lim}}, \bibinfo {author} {\bibfnamefont {A.}~\bibnamefont {Majumder}}, \bibinfo {author} {\bibfnamefont {R.}~\bibnamefont {Mittapally}}, \bibinfo {author} {\bibfnamefont {A.-R.}\ \bibnamefont {Gutierrez}}, \bibinfo {author} {\bibfnamefont {Y.}~\bibnamefont {Luan}}, \bibinfo {author} {\bibfnamefont {E.}~\bibnamefont {Meyhofer}},\ and\ \bibinfo {author} {\bibfnamefont {P.}~\bibnamefont {Reddy}},\ }\bibfield  {title} {\bibinfo {title} {{A nanoscale photonic thermal transistor for sub-second heat flow switching}},\ }\href {https://doi.org/10.1038/s41467-024-49936-0} {\bibfield  {journal} {\bibinfo  {journal} {Nat. Commun.}\ }\textbf {\bibinfo {volume} {15}},\ \bibinfo {pages} {5584} (\bibinfo {year} {2024})}\BibitemShut {NoStop}%
\end{thebibliography}%

\end{document}


\title{
Supplemental Material for 
``Asymmetry-induced radiative heat transfer in Floquet systems''
}

\author{Hui Pan}
\email{panhui@nus.edu.sg}
\affiliation{Department of Physics, National University of Singapore, Singapore 117551, Republic of Singapore}

\author{Yuhua Ren}
\affiliation{Department of Physics, National University of Singapore, Singapore 117551, Republic of Singapore}

\author{Gaomin Tang}
\email{gmtang@gscaep.ac.cn}
\affiliation{Graduate School of China Academy of Engineering Physics, Beijing 100193, China}

\author{Jian-Sheng Wang}
\email{phywjs@nus.edu.sg}
\affiliation{Department of Physics, National University of Singapore, Singapore 117551, Republic of Singapore}

\bigskip

\maketitle

\tableofcontents

\subsection{Model and Hamiltonian} 
We investigate the radiative heat transfer between two objects due to current fluctuations. The energy transport between the two subsystems separated by a vacuum gap is mediated by an electromagnetic (EM) field. The total Hamiltonian of the open quantum system can be partitioned as~\cite{JSW23}
\begin{equation}
  \hat{H}_{\rm tot} = \hat{H}_e + \hat{H}_{\rm ph} + \hat{H}_{\rm int} .
\end{equation}
The electronic Hamiltonian is given by
\begin{equation}
    \hat{H}_e = \sum_\alpha \big( \hat{H}_\alpha + \hat{H}_{\alpha B} + \hat{V}_{\alpha B} + \hat{V}^\dag_{\alpha B} \big) ,
\end{equation}
where $\alpha=L,R$ denotes the left and right subsystems. 
In the tight-binding model, the Hamiltonian of object $\alpha$ in second quantization is
\begin{equation} \label{H_e}
  \hat{H}_\alpha = \sum_{i, j\in \alpha} c_i^\dag H_{ij} c_j ,
\end{equation}
where $i$ and $j$ run over all the electron sites with coordinates $\bm{r}_i$ and $\bm{r}_j$ of object $\alpha$.

The electronic reservoir, which provides the dissipation channel for the energy current~\cite{Matsyshyn23}, is modeled with the Hamiltonian 
\begin{equation}
  \hat{H}_{\alpha B} = \sum_{k} \varepsilon_{k\alpha} d_{k\alpha}^\dag d_{k\alpha} .
\end{equation}
The coupling between object $\alpha$ and its corresponding reservoir is given by
\begin{equation}
  \hat{V}_{\alpha B} = \sum_{k;i\in\alpha} t_{k\alpha,i} d_{k\alpha}^\dag c_i .
\end{equation}
The degrees of freedom of the reservoirs can be analytically integrated out and act as self-energies in the electron Green's functions. 

The Hamiltonian of the EM field depends on the gauge chosen. In the temporal gauge~\cite{Lifshitz_book, kay_quantum_2021}, the Hamiltonian is given in terms of the vector potential $\bm{A}$ by
\begin{equation} \label{H_ph}
  \hat{H}_{\rm ph} = \frac{1}{2} \int dV \left[ \epsilon_0
  \big( \partial_t \bm{A} \big)^2
    + \mu_0^{-1} \big( \nabla \times \bm{A} \big)^2 \right] ,
\end{equation}
where $\epsilon_0$ and $\mu_0$ are the vacuum
permittivity and permeability.
The practical advantage of the temporal gauge is the reduced number of components in the physical quantities that are needed to be calculated. 

The interaction between the electrons and the EM field is described by the Peierls substitution~\cite{peierls_zur_1933-3}. Equation~\eqref{H_e} is modified as
\begin{equation} \label{Peierls}
  \hat{H}_\alpha + \hat{H}_{\rm int} = \sum_{i, j\in \alpha} c_i^\dag  H_{ij} c_j \exp\left[ \frac{e_0}{i\hbar} \int_{\bm{r}_j}^{{\bm{r}_i}} d\bm{r} \cdot \bm{A}(\bm{r}) \right] ,
\end{equation}
where $e_0 > 0$ denotes the elementary charge. 
To isolate $\hat{H}_{\rm int}$, we approximate the line integral with the trapezoidal rule
\begin{equation}
  \int_{\bm{r}_j}^{{\bm{r}_i}} d\bm{r} \cdot \bm{A}(\bm{r}) 
  \approx \frac{1}{2} (\bm{A}_i + \bm{A}_j) \cdot (\bm{r}_i - \bm{r}_j) ,
\end{equation}
where we have used the notation $\bm{A}_i \equiv \bm{A}(\bm{r}_i)$. 
Performing a Taylor expansion to the exponential in Eq.~\eqref{Peierls} up to second order in the lattice constant, we can write the interaction Hamiltonian as
\begin{align}
  \hat{H}_{\rm int} &= \sum_{i,j,l,\mu} c_i^\dag M_{ij}^{l\mu} c_j A_{l \mu}
  \label{H_int} + \frac{1}{2} \sum_{i,j,l,l',\mu,\nu} c_i^\dag N_{ij}^{l\mu, l'\nu} c_j A_{l \mu} A_{l' \nu} \notag \\
  &\equiv -\sum_l \bm{I}_l \cdot \bm{A}_l + \hat{H}_{\rm dia},
\end{align}
where the tensor $M$ and $N$ are, respectively, expressed as
\begin{align}
  & M_{ij}^{l\mu} = \frac{e_0}{2} (\delta_{il} + \delta_{jl}) V_{ij}^\mu 
  \label{M_tensor} , \\
  & N_{ij}^{l\mu, l'\nu} = \frac{e_0}{2i\hbar} M_{ij}^{l\mu} (r_i^\nu - r_j^\nu) (\delta_{il'} + \delta_{jl'})
  \label{N_tensor} ,
\end{align}
and the velocity matrix $\bm{V}$ has the elements
\begin{equation} \label{V_site}
  V^\mu_{ij} = \frac{1}{i\hbar} H_{ij} \big( r^\mu_i - r^\mu_j \big) ,
\end{equation}
which represents the velocity of the electron hopping from site $j$ to $i$. 
Here, the indices $i$, $j$, and $l$ denote the electron site, and $\mu =x,y,z$ labels the Cartesian directions.

\subsection{Floquet representation}\label{Floquet}
For a two-time function with the symmetry $G(t_1 + 2\pi/\Omega, t_2 + 2\pi/\Omega) = G(t_1, t_2)$, we define its Floquet representation~\cite{Tsuji08,DMFT-RMP14,rodriguez-vega_low-frequency_2021} by
\begin{align}
     & G(t_1, t_2) \to \bm{G}(\omega) \notag \\
     & G_{mn}(\omega) \equiv \frac{\Omega}{2\pi} \int_{0}^{2\pi/\Omega} d t_1 \, \int_{-\infty}^\infty d t_2 \,
     G(t_1, t_2)
     e^{i (\omega_m t_1 - \omega_n t_2)}  \label{floquet_rep} ,
\end{align}
where $m, n \in \mathbb{Z}$. $G_{mn}(\omega)$ can be interpreted as the $(m,n)$ entries of an infinite block matrix $\bm{G}(\omega)$, and such extended Floquet matrices will be denoted in bold.
Here, we have defined $\omega_m = m\Omega+\omega$.
A periodic function of one argument, $H(t_1)$, can be upgraded into a two-time function by considering $H(t_1) \delta(t_1-t_2)$, and its Floquet representation $\bm{H}$ is a block Toeplitz matrix. 
The inverse transformation to return to the time domain is given by
\begin{equation}
    G(t_1,t_2) = \sum_{mn} \int_{\rm BZ} \frac{d\omega}{2\pi} G_{mn}(\omega) e^{-i (\omega_m t_1 - \omega_n t_2)} \label{floquet_rep_inv} .
\end{equation}
The integration region in Eq.~\eqref{floquet_rep_inv}, the Floquet-Brillouin zone (BZ), is chosen to be $-\Omega/2 < \omega \leq \Omega/2$. 
Eq.~\eqref{floquet_rep} allows for $\omega$ to be outside of the range given by BZ, but no new information is conveyed because of the modular property 
\begin{equation}
    G_{mn}(\omega + k \Omega) = G_{m+k,n+k}(\omega) \label{modular_property} .
\end{equation}
Nevertheless, Eq.~\eqref{modular_property} is useful in numerical computations as the middle element (``00'' block) of the Floquet matrix has the smallest truncation errors.
 
Below, we list several important properties of the Floquet representation that will be used later. Differentiation in the time domain becomes multiplication in Floquet representation. 
\begin{align}
    i \hbar \frac{\partial}{\partial t_1} G(t_1, t_2) &\to \bm{E}(\omega) \bm{G}(\omega) , \\
    i \hbar \frac{\partial}{\partial t_2} G(t_1, t_2) &\to - \bm{G}(\omega) \bm{E}(\omega) ,
\end{align}
where the Floquet matrix $\bm{E}(\omega)$ associated with differentiation is defined as the block diagonal matrix with elements 
\begin{equation}
    E_{mn}(\omega) = \hbar (\omega + m \Omega) \delta_{mn} I \label{floquet_E} .
\end{equation}
We also have the analog of the convolution theorems~\cite{Tsuji08,GT24} in the Floquet representation. If $A$ and $B$ both have the same discrete time translational symmetry, then
\begin{equation}
  \int dt' A(t_1,t') B(t',t_2) \to \bm{A}(\omega) \bm{B}(\omega) \label{CAB} .
\end{equation}
and
\begin{align}
  &C(t_1, t_2) = A(t_1,t_2) B(t_2,t_1) \notag \\
  &\to C_{mn}(\omega) = \sum_{l p}\int_{\rm BZ} \frac{d \omega'}{2\pi} A_{lp}(\omega') B_{(p-n)(l-m)}(\omega'-\omega) \label{CAB_1} .
\end{align}
Eq.~\eqref{CAB_1} is useful for obtaining the polarization functions, $\Pi$, in the Floquet representation.

\subsection{Floquet electron Green's function} 
The nonequilibrium electron Green's function on the Keldysh contour is defined as~\cite{Haug_Jauho}
\begin{equation} \label{electron_GF_keldysh}
    G_{ij}(\tau, \tau') = \frac{1}{i\hbar} \left< {\cal T} c_{i}(\tau) c_{j}^\dag(\tau') \right> ,
\end{equation}
where ${\cal T}$ is the contour time-ordering operator and $c_i, (c_i^\dag)$ are the annihilation (creation) operators in the Heisenberg picture. 
In this Supplemental Material, the time variables in Greek letters sit on the Keldysh contour while those in Latin letters are the normal ones. 
In the absence of coupling to the EM field, the retarded component of $G$ satisfies the Dyson equation
\begin{align}
  \big[i\hbar \partial_t - H(t) \big] G^r(t,t') = \delta(t-t') I  + \int dt_1 \Sigma^r(t,t_1) G^r(t_1, t') \label{gr_time} ,
\end{align}
where $\delta(t)$ is the Dirac-delta function, and $I$ the identity matrix.
The self-energy term denoted by $\Sigma^r$ describes the presence of a bath.
We use the wide-band approximation, where each site is coupled to the bath independently of energy, so that the retarded self-energy in the energy domain takes the form of 
\begin{equation}
    \Sigma^r(\omega) = - i \eta
\end{equation}
with $\hbar/\eta$ having the interpretation of the electron relaxation time.

Due to Floquet's theorem, we have the discrete time-translational symmetry in $G^r$, which allows us to write Eq.~\eqref{gr_time} in the Floquet representation as
\begin{equation} \label{gr_floquet}
    \bm{G}^r(\omega) = \big[ \bm{E}(\omega) + i\eta \bm{I} - \bm{H} \big]^{-1}, 
\end{equation}
where $\bm{I}$ is the identity matrix and $\bm{E}(\omega)$ is previously defined in Eq.~\eqref{floquet_E}.
The advanced component is obtained with $G^a_{mn}(\omega) = \big[G^r_{nm}(\omega)\big]^\dag$.
The electron distribution is described by the lesser Green's function $G^<$ with
\begin{equation}
  G^<_{ij}(t,t') = -\frac{1}{i\hbar} \left< c_j^\dag(t') c_i(t) \right> .
\end{equation}
In a Floquet steady state~\cite{Kohn2001, Matsyshyn23}, the lesser Green's function can be given by the Keldysh equation as
\begin{equation} \label{keldysh_time}
  G^<(t,t') = \int d t_1 \int d t_2 \, G^r(t,t_1) \Sigma^<(t_1 -t_2) G^a(t_2,t') .
\end{equation}
In the Floquet representation, Eq.~\eqref{keldysh_time} is written as
\begin{equation} \label{keldysh_floquet}
  \bm{G}^<(\omega) = \bm{G}^r(\omega)  \bm{\Sigma}^<(\omega) \bm{G}^a(\omega) , 
\end{equation}
where $\bm{\Sigma}^<$ is the lesser component of the bath self-energy.
We assume that the bath remains in thermal equilibrium and thus obeys the fluctuation-dissipation theorem,
\begin{equation}
  \Sigma^<_{mn}(\omega) = -f(\omega_m) \big[\Sigma^r_{mn}(\omega) - \Sigma^a_{mn}(\omega) \big] ,
\end{equation}
with
\begin{equation}
  f(\omega) = \frac{1}{\exp\left[(\hbar\omega-\mu_s)/(k_B T_0)\right] + 1}
\end{equation}
being the Fermi-Dirac function at temperature $T_0$ and chemical potential $\mu_s$.

\subsection{Free photon Green's function}
In the temporal gauge, the Maxwell-Ampere law relates the current $\bm{j}$ to the vector potential $\bm{A}$ by a linear operator, $v^{-1}$, given by
\begin{equation} \label{A_EOM}
  -\bm{j} = - \epsilon_0 \frac{\partial^2}{\partial t^2} \bm{A} - \mu_0^{-1} \nabla\times\nabla \times \bm{A} \equiv v^{-1} \bm{A} .  
\end{equation}
For the present problem with planar geometry, the expression of $v^r(t, \bm{r})$, the retarded Green's function corresponding to the operator $v^{-1}$, can be found in the momentum and frequency domain to be~\cite{Lifshitz_book, sipe_new_1987, joulain_surface_2005, yap_radiative_2017}
\begin{equation}
    v^r(\omega, \bm{q}, z) 
    = \frac{\mu_0}{a^2 k_0^2} \left(
    \frac{e^{i\gamma |z|} }{2 i \gamma}
    \begin{bmatrix}
        k_0^2- q_x^2 & - q_x q_y & - s q_x \gamma \\
        - q_x q_y & k_0^2- q_y^2 & - s q_y \gamma \\
        - s q_x \gamma & - s q_y \gamma & k_0^2- \gamma^2 
    \end{bmatrix} 
    + \delta(z) \hat{\bm{z}}\hat{\bm{z}} 
    \right) \label{vr} ,
\end{equation}
where $s={\rm sgn}(z)$, $a$ is the lattice constant, $\bm{q}=(q_x,q_y)$ the 2D wave vector, and $k_0 = \omega/c$. The photon propagation constant $\gamma$ is defined as $\gamma = \sqrt{k_0^2 - q^2}$ for the propagating mode ($k_0>q$), and $\gamma = i\sqrt{q^2 - k_0^2}$ for the evanescent mode ($k_0<q$), where $q$ is the wave vector length.
In numerical calculations, we use the discrete Fourier transform instead of the continuous version.
The factor of $1/a^2$ in Eq.~\eqref{vr} is due to the normalization convention for space discrete Fourier transform.
Since the spatial integral is replaced by a sum over all sites, the value of the 1D Dirac-delta in Eq.~\eqref{vr} is represented by $1/a$.

\subsection{Nonequilibrium photon Green's function and self-energies}
The full photon Green's function is defined by the vector potential $\bm{A}$ on the Keldysh contour as~\cite{Lifshitz_book, JSW23} 
\begin{equation} \label{photon_GF_keldysh}
  D_{\mu\nu}(\bm{r}\tau, \bm{r}'\tau') = \frac{1}{i\hbar} \big< {\cal T}
  A_{\mu}(\bm{r}\tau) A_{\nu}(\bm{r}'\tau') \big> ,
\end{equation}
with $\mu ,\nu  = x,y,z$ denoting the spatial directions. 
The retarded and lesser components of the photon Green's function are, respectively, given by
\begin{align}
  & D_{\mu\nu}^r(\bm{r}t,\bm{r}'t') = \frac{1}{i\hbar} \theta(t-t') \big< \big[
  A_\mu(\bm{r}t), A_\nu(\bm{r}'t') \big] \big> , \\
  & D^<_{\mu\nu}(\bm{r}t, \bm{r}'t') = \frac{1}{i\hbar}
  \big< A_\nu(\bm{r}' t')A_\mu(\bm{r}t)\big> . 
\end{align}
The vector potential are quantum operators in the Heisenberg picture, and the square brackets represent the commutator.
The full Green's function $D$ is related to the polarization $\Pi$ and the free Green's function $v$ by the Dyson equation on the Keldysh contour as
\begin{equation}
  D(\bm{r}\tau,\bm{r}'\tau') = v(\bm{r}\tau, \bm{r}'\tau') + \sum_{ij} \int d\tau_1 \int d\tau_2 \, v(\bm{r}\tau,\bm{r}_i\tau_1) \Pi_{ij}(\tau_1,\tau_2)
  D(\bm{r}_j\tau_2,\bm{r}'\tau') \label{Dyson_D} .
\end{equation}
From Langreth's theorem~\cite{Langreth}, the retarded component of $D$ satisfies the Dyson equation, written in Floquet representation as 
\begin{equation}
    \bm{D}^r = \bm{v} + \bm{v} \bm{\Pi}^r \bm{D}^r .
\end{equation}
The lesser photon Green's function describes the photon distribution and is again related to the retarded one by the Keldysh equation
\begin{equation}
    \bm{D}^< = \bm{D}^r \big(\bm{\Pi}^< + \bm{\Pi}_{\infty}^< \big) \bm{D}^a \label{keldysh_D_floquet} . 
\end{equation}
The extra term $\Pi^<_\infty$ called the bath at infinity represents energy dissipation into the vacuum.
It can be formally written as
\begin{equation}
  \Pi^r_\infty = \frac{1}{2} \left[ - v^{-1} + (v^{-1})^\dag \right] .
\end{equation}
In frequency and momentum domain, $\Pi^a_\infty = (\Pi^r_\infty)^\dagger$ and $\Pi^<_\infty$ is determined by the fluctuation-dissipation theorem at zero temperature. 

Based on the interaction Hamiltonian [Eq.~\eqref{H_int}], 
the total polarization $\Pi$ is a sum of a paramagnetic term $\Pi^{\rm RPA}$ and a diamagnetic term $\Pi^{\rm dia}$, 
\begin{equation}
    \Pi_{l\mu, l'\nu}(\tau, \tau') = \Pi^{\rm RPA}_{l\mu, l'\nu}(\tau, \tau') + \Pi^{\rm dia}_{l\mu, l'\nu}(\tau, \tau') .
\end{equation}    
The paramagnetic term $\Pi^{\rm RPA}$ under the random phase approximation (RPA) is
\begin{equation} \label{Pi_RPA}
  \Pi _{l\mu ,l'\nu }^{{\text{RPA}}}\left( {\tau ,\tau '} \right) = \frac{1}{{i\hbar }}\big\langle {{\cal T}{I_{l\mu }}\left( \tau  \right){I_{l'\nu }}\left( {\tau '} \right)} \big\rangle , 
\end{equation}    
and $\Pi^{\rm dia}$ is given later [Eq.~\eqref{Pir_dia}] in momentum space. 
The retarded photon self-energy $\Pi^r$ can be interpreted as a linear response of the induced current in matter, $\bm{j}^{\rm ind} = - \Pi^r \bm{A}$.
The photon self-energy $\Pi$ is obtained through a diagrammatic expansion, with the interaction from the Peierls substitution Hamiltonian, Eq.~\eqref{H_int}, taking the contributions linear in $\bm{A}$ with the RPA and quadratic in $\bm{A}$ for the diamagnetic term. 
The expressions of the self-energies under the RPA can be obtained from two electron Green's functions with
\begin{equation} \label{GG}
  \Pi^{\rm{RPA}}_{l\mu ,l'\nu}(\tau, \tau') = -2i\hbar \sum_{ijkp} M^{l\mu}_{ij} G_{jk}(\tau,\tau') M^{l'\nu}_{kp} G_{pi}(\tau',\tau) .
\end{equation}
where $M^{i\mu}_{pq}$ is the local current operator between site $p$ and site $q$ as defined in Eq.~\eqref{M_tensor}. 
The additional factor of $2$ accounts for the spin degeneracy of the electrons.  
Its components are obtained using Langreth's theorem as
\begin{subequations}
    \label{GG_t}
    \begin{align}
        \Pi_{l\mu , l'\nu}^{\rm{RPA},<}(t,t') &= -2 i\hbar \, \mathrm{tr} \big[M^{l\mu} G^<(t,t') M^{l' \nu} G^>(t', t) \big] 
        \label{GG_<_t} , \\
        \Pi_{l\mu , l'\nu}^{\rm{RPA},>}(t,t') &= -2 i\hbar \, \mathrm{tr} \big[M^{l\mu} G^>(t,t') M^{l' \nu} G^<(t', t) \big] 
        \label{GG_>_t} , \\
        \Pi_{l\mu , l'\nu}^{{\rm RPA},r}(t,t') &= -2 i\hbar \, \mathrm{tr} \big[M^{l\mu} G^r(t,t') M^{l' \nu} G^<(t', t)  + M^{l\mu} G^<(t,t') M^{l' \nu} G^a(t', t) \big] 
        \label{GG_r_t} , \\
        \Pi_{l\mu , l'\nu}^{{\rm RPA},a}(t,t') &= -2 i\hbar \, \mathrm{tr} \big[M^{l\mu} G^a(t,t') M^{l' \nu} G^<(t', t) + M^{l\mu} G^<(t,t') M^{l' \nu} G^r(t', t) \big]
        \label{GG_a_t} .
    \end{align}
\end{subequations}
One benefit of the trace notation [Eqs.~\eqref{GG_<_t}-\eqref{GG_a_t}] instead of summing over indices [Eq.~\eqref{GG}] is the freedom in the choice of basis states.
For systems with translational invariance, working in the momentum basis is usually preferred.
In momentum space, Eq.~\eqref{M_tensor} is
\begin{align} \label{M_tensor_q}
    M^{\bm{q} \mu}_{\bm{k}\bm{k}'} &= \frac{1}{N^{3/2}} \sum_{mnl} e^{i (-\bm{k} \cdot \bm{r}_m + \bm{k}' \cdot \bm{r}_n - \bm{q} \cdot \bm{r}_l)} M^{l\mu}_{mn} \notag \\
    &= \frac{e_0}{\sqrt{N}} \frac{V^{\mu}(\bm{k}) + V^{\mu}(\bm{k}')}{2} \delta_{\bm{q}, \bm{k}'-\bm{k}} ,
\end{align}
with $N$ being the total number of lattice sites, and the expression for $V^{\mu}(\bm{k})$ is given later in Eq.~\eqref{velocity_squre}. 
The  polarization functions are also diagonal in the momentum basis, in the sense that they only require one $\bm{q}$ index instead of two. 
Taking the lesser component as an example, we have  
\begin{align}
    \quad \Pi_{\mu\nu}^{\rm{RPA},<}(\bm{q}, t, t') &= \frac{1}{N} \sum_{l, l'} e^{-i \bm{q} \cdot (\bm{r}_l - \bm{r}_{l'})}\Pi^<_{\mu\nu}(\bm{r}_l t, \bm{r}_{l'}  t')  \notag \\
    &= -2 i\hbar \sum_{\bm{k}} 
    M^{\bm{q} \mu}_{\bm{k}-\bm{q}, \bm{k}}
    G^<_{\bm{k}}(t,t')
    M^{-\bm{q} \nu}_{\bm{k}, \bm{k}-\bm{q}}
    G^>_{\bm{k}-\bm{q}}(t', t)  \label{Pi_q_space} .
\end{align}

The retarded component $\Pi^r$ has a contribution due to the diamagnetic part of the current~\cite{JSW24}. In the long wavelength limit, the diamagnetic part of $\Pi$ can be determined by gauge invariance~\cite{rostami_gauge_2021} to be
\begin{equation}\label{Pir_dia}
    \Pi_{\mu\nu}^{\rm{dia}}(\bm{q}, t, t') = 
    - \delta(t-t') \frac{\Omega}{2\pi} \int_0^{2\pi / \Omega} d t_1 \int_{-\infty}^\infty d t_2 \Pi^{\rm{RPA}}_{\mu\nu}\left(\bm{0}, t_1, t_2\right) .
\end{equation}

As a function of the electron Green's function $G$, $\Pi$ is diagonal in subsystem space with
\begin{equation} \label{Pi_L_R}
  \Pi = 
  \begin{bmatrix}
    \Pi_L & 0 \\ 0 & \Pi_R 
  \end{bmatrix} 
\end{equation}
Since $\Pi$ is only non-zero on the lattice sites, the relevant coordinates of $D$ in calculating energy current are those of the electrons, as shown in Eq.~\eqref{It}. 
Therefore, it is convenient to partition $v$ and $D$ in numerical calculation using subsystem space as shown below with
\begin{equation}
  v = 
  \begin{bmatrix}
    v_{LL} & v_{LR} \\ v_{RL} & v_{RR}
  \end{bmatrix} , \qquad 
  D = 
  \begin{bmatrix}
    D_{LL} & D_{LR} \\ D_{RL} & D_{RR}
  \end{bmatrix} .
\end{equation}

To obtain the Floquet representation of the polarization function, we utilize the Floquet convolution theorem [Eq.~\eqref{CAB_1}]. 
E.g., Eq.~\eqref{Pi_q_space} can be rewritten as
\begin{equation} 
    \big[ 
    \bm{\Pi}^{\textrm{RPA}, <}_{\mu\nu}(\bm{q}, \omega) \big]_{mn} 
    = -2i\hbar \sum_{\bm{k}} 
    \int_{\rm BZ} \frac{d\omega'}{2\pi} \sum_{l p} 
    M^{\bm{q} \mu}_{\bm{k}-\bm{q}, \bm{k}}
    \big[
    \bm{G}^<_{\bm{k}}(\omega')
    \big]_{lp} 
    M^{-\bm{q} \nu}_{\bm{k}, \bm{k}-\bm{q}}
    \big[
    \bm{G}^>_{\bm{k}-\bm{q}}(\omega'-\omega)
    \big]_{(p-n)(l-m)} 
    \label{Pi} .
\end{equation}

\subsection{Derivation of the expression of photon energy currents}
Here, we derive the energy current from the perspective of the EM field.
We start from Poynting's theorem
\begin{equation} \label{Poynting}
  - \partial_t u = \nabla \cdot \bm{S} + \bm{j} \cdot \bm{E} ,
\end{equation}
where $\bm{S}$ is the Poynting vector and $u=\left(\epsilon_0 E^2 + B^2/\mu_0\right)/2$ is the energy density.
By integrating over the volume around object $\alpha$ and using Gauss's theorem, Eq.~\eqref{Poynting} can be rewritten as
\begin{equation}
    I_\alpha(t) \equiv \oint_{\alpha} \left< \bm{S} \right> \cdot d\bm{\Sigma} 
    = - \int_{\alpha} d\bm{r} \left< \bm{j} \cdot \bm{E} \right>
    - \int_{\alpha} d\bm{r} \left< \partial_t u \right> ,
\end{equation}
The ensemble average is defined as $\left< \cdots \right> = \rm{tr}(\hat{\rho}\cdots)$, where $\hat{\rho}$ is the density-matrix operator. 
The right-hand side consists of two terms: the first term describes Joule heating by the electric current, while the second term describes the energy change rate of the field.

In steady state, the time average of the second term vanishes, and thus the average energy current flowing out of an object is solely due to Joule heating.
The energy current can be expressed in terms of the photon Green's function as
\begin{align}
  I_\alpha(t) 
  &= - \iint_{\alpha} d\bm{r}d\bm{r}' \bigg[ \frac{\partial\ }{\partial t'} \sum_{\mu,\nu} v^{-1}_{\mu\nu}(\bm{r} t, \bm{r}' t) \left< A_\nu(\bm{r}' t) A_\mu(\bm{r} t') \right> \bigg] \bigg|_{t'=t}  \notag \\
  &= - \frac{i\hbar}{2}\iint_{\alpha} d\bm{r}d\bm{r}' \bigg\{ \frac{\partial\ }{\partial t'} {\rm tr} \left[ v^{-1}(\bm{r} t, \bm{r}' t) D^K(\bm{r}' t, \bm{r} t')\right] \bigg\} \bigg|_{t'=t} \label{It} .
\end{align}
where the trace is taken over the directions $\mu$.
The reason for using the Keldysh Green's function $D^K$ is that the physical observable of the energy current must be real. 
The product of two arbitrary Hermitian operators need not be Hermitian, whereas the symmetrized version does ensure a Hermitian result. 
Using the Keldysh equation and the Dyson equation, the energy current can be further expressed in terms of the self-energies $\Pi$ and the Green's functions $D$ as
\begin{equation} \label{It1}
  I_\alpha(t) 
  = - \frac{i\hbar}{2} \iint_{\alpha} d\bm{r}d\bm{r}' \int dt_1 \frac{\partial\ }{\partial t'}{\rm tr} \Big[ \Pi_\alpha^r(\bm{r} t, \bm{r}' t_1)D^K(\bm{r}' t_1, \bm{r} t')  + \Pi_\alpha^K(\bm{r} t, \bm{r}' t_1) D^a(\bm{r}' t_1, \bm{r} t') \Big] \Big|_{t'=t} .
\end{equation}
The above expression is valid for arbitrary time dependence. 
For the special case of periodic drive, the energy current averaged over one period is a meaningful quantity.
We can calculate the average current using the Floquet representation (Sec.~\ref{Floquet}) with
\begin{equation} 
  \bar{I}_\alpha = \int_{\rm BZ} \frac{d\omega}{4\pi} 
  {\rm Tr}\left[\bm{E} \left( \bm{\Pi}^r_\alpha \bm{D}^K + \bm{D}^K_\alpha \bm{\Pi}^a \right) \right] \label{IE} .
\end{equation}
where the symbol ``${\rm Tr}$'' denotes tracing over the directions $\mu$, the electron sites and the Floquet indices. 
Eq.~\eqref{IE} is a generalization of the Meir-Wingreen formula~\cite{JSW24} for the energy transport in Floquet space under periodic modulation. 
Using the symmetries of the Floquet representation, one can alternatively express the energy current as
\begin{equation}
  \bar{I}_\alpha = - \int_{\rm BZ} \frac{d\omega}{4\pi} 
  {\rm Tr}\left[\bm{E} \left( \bm{D}^r \bm{\Pi}^K_\alpha + \bm{D}^K \bm{\Pi}^a_\alpha \right) \right] .
\end{equation}
An alternative approach from the perspective of electron transport is shown in Ref.~\cite{GT24}, in which a similar expression in form as Eq.~\eqref{IE} is also given by considering the Coulomb interaction.

\subsection{Effective distribution of the electrons in Floquet systems}
The solutions of the Schr\"{o}dinger equation with a time-periodic Hamiltonian are linear combinations of the Floquet states~\cite{shirley_solution_1965, breuer_quantum_1989, Kohn2001, Tsuji08, rodriguez-vega_low-frequency_2021}
\begin{equation}
  \ket{\Psi_a(t)} = e^{-i \varepsilon_a t/\hbar} \ket{u_a(t)},
\end{equation}
where $\varepsilon_a$ is the quasienergy defined up to integer multiples of $\hbar\Omega$. 
Floquet theorem states that $\ket{u_a(t)} = \ket{u_a(t+2\pi/\Omega)}$ has the same periodicity as $H(t)$, and thus admits a Fourier series expansion, 
\begin{equation}
    \ket{u_a(t)} = \sum_n e^{-in\Omega t} \ket{u_a^n}. 
\end{equation}
The Floquet states are normalized with $\sum_n\langle u^n_{a}|u^n_{b}\rangle=\delta_{ab}$. 
By analyzing the Fourier components of the Schr\"odinger equation, we get the following eigenvalue equation in Floquet representation, 
\begin{equation} \label{shirley_floquet}
    (\varepsilon_a - p \hbar \Omega) \ket{\bm{u}_a^p} = (\bm{H} - \hbar\bm{\Omega}) \ket{\bm{u}_a^p},
\end{equation}
where $\ket{\bm{u}_a^p}$ is a vertically-stacked column vector in Floquet space with the entry in the $n$th block-row being $(\ket{\bm{u}^p_a})_n = \ket{u_a^{n-p}}$. 
Notice that if $\ket{\bm{u}_a^0}$ is an eigenvector with eigenvalue $\varepsilon_a$, then $\ket{\bm{u}_a^p}$ is an eigenvector with eigenvalue $(\varepsilon_a - p \hbar \Omega)$.
The index $p$ keeps track of such harmonics arising from the same Floquet state, which we consider to have the same $a$.
Assuming that the eigenvectors are complete,  Eq.~\eqref{gr_floquet} admits the following spectral decomposition
\begin{equation} 
    \bm{G}^r(\omega) = \sum_{a, p} \frac{\ket{\bm{u}_a^p} \bra{\bm{u}_a^p}}{\hbar (\omega + p \Omega) + i\eta - \varepsilon_a} . 
\end{equation}
$\bm{G}^a(\omega)$ is expanded in the same way with the final expression differing by a minus sign in front of $\eta$. 
We can compute $\bm{G}^<(\omega)$ using Eq.~\eqref{keldysh_floquet}, and their elements are
\begin{align} \label{G<_mode}
    G^<_{mn}(\omega) 
    = 2i\eta \sum_{\substack{l, a, b, \\ p, q}} \frac{\ket{u^{m-p}_{a}} \bra{u^{l-p}_{a}}}{\hbar (\omega + p \Omega) + i\eta - \varepsilon_a} f(\omega_l) \frac{\ket{u^{l-q}_{b}} \bra{u^{n-q}_{b}}}{\hbar (\omega + q \Omega) - i\eta - \varepsilon_b}.
\end{align}

The lesser Green's function encodes information about the electron steady-state distribution under a periodic drive. 
In the case of a weak coupling to the bath $\eta \to 0^+$, Eq.~\eqref{G<_mode} can be greatly simplified by
\begin{align} \label{G<_mode_simplified}
    G^<_{mn}(\omega) 
    &= 2\pi i \sum_{a, p} \ket{u^{m-p}_{a}} \bra{u^{n-p}_{a}} \delta(\hbar\omega_p - \varepsilon_a) \sum_l \braket{u^{l-p}_{a} | u^{l-p}_{a}} f(\omega_l) \notag \\
    &= 2\pi i \sum_{a, p} \ket{u^{m-p}_{a}} \bra{u^{n-p}_{a}} \delta(\hbar\omega_p - \varepsilon_a) \bar{f}_a,
\end{align}
with the effective electron distribution $\bar{f}$ as~\cite{Refael15, kumari_josephson-current_2023, Matsyshyn23} 
\begin{equation} \label{eff_fermi}
  \bar{f}_a \equiv \sum_l \braket{u^{l}_{a} | u^{l}_{a}}
  f(\varepsilon_a/\hbar + l \Omega).
\end{equation}
The effective electron distribution accounts for all the energy levels differing by multiples of $\hbar\Omega$ from the $\varepsilon_a$ Floquet state. 
The greater electron Green's function $G^>$ can be treated in the same way with
\begin{equation}\label{G>_mode_simplified}
    G^>_{mn}(\omega) 
    = 2\pi i \sum_{a, p} \ket{u^{m-p}_{a}} \bra{u^{n-p}_{a}} \delta(\hbar\omega_p - \varepsilon_a) \big( \bar{f}_a -1 \big). 
\end{equation}

\subsection{Effective photon distribution and its asymptotic properties}
We now derive the analytical formula of the effective photon distribution for a metal plate under driving in the $\eta \to 0^+$ limit.
For the square-lattice model, the tight-binding electron Hamiltonian with a periodic potential modulation is diagonal in momentum space with the dispersion given by
\begin{equation}
    H\left(\bm{k},t\right) = \varepsilon_{\bm{k}} + \mu (t) = -2 t_0 \left[\cos\left(k_x a\right) + \cos\left(k_y a\right)\right] + 2\mu_d \cos \left(\Omega t\right)
\end{equation}
where $a$ is the lattice constant, $t_0$ means the nearest-neighbor hopping parameter, and the components of wave vector $k_x, k_y \in (-\pi/a, \pi/a]$. 
We consider the case where the system is periodically modulated by a time-varying potential $2\mu_d \cos (\Omega t)$.
The velocity operator $\bm{V}$ in Eq.~\eqref{GG} is also diagonal in momentum space with the elements given by
\begin{equation}
    \upsilon^\mu(\bm{k}) = \frac{2at}{\hbar} \sin (k_\mu a) \label{velocity_squre} .
\end{equation}
With the velocity matrix, the local current density operator needed to compute $\Pi$ [Eqs.~\eqref{GG_<_t}-\eqref{GG_a_t}] can be defined using Eq.~\eqref{M_tensor}.

Our driving potential is chosen in a way that the Hamiltonian commutes with itself at all times. 
In the Floquet representation, we mentioned that $G^r$ can be obtained by numerical diagonalization, but it can also be expanded exactly using the Jacobi-Anger expansion.
Under a uniform drive, the weights in Eq.~\eqref{eff_fermi} do not depend on the Floquet index and are given by
\begin{equation}
    \braket{u^{l}_{a} | u^{l}_{a}} = \left[J_l(\kappa) \right]^2, 
\end{equation}
where $\kappa = 2\mu_d/(\hbar\Omega)$ and $J_l$ is the $l$th order Bessel function of the first kind. 
As such, we can rewrite Eq.~\eqref{G<_mode_simplified} in time domain and momentum space as
\begin{equation}
    G^<_{\bm{k}}(t, t') = \frac{-\bar{f}_{\bm{k}}}{i\hbar} \Psi_{\bm{k}}(t) \Psi_{\bm{k}}(t')^*, 
\end{equation}
where
\begin{equation}
  \Psi_{\bm{k}}(t) = \exp\left(-i \varepsilon_{\bm{k}} t /\hbar \right)
  \exp\left[-i \kappa \cos(\Omega t) \right]  
\end{equation}
is a complex number that describes the phase of the quantum state labeled by $\bm{k}$.

Having computed the electron Green's functions, the polarization can be obtained using Eq.~\eqref{Pi_q_space}, and the results are
\begin{subequations}
\begin{align}
    &\Pi^{\rm{RPA},<}_{\mu\nu}(\bm{q}, t, t') = \frac{2}{i\hbar} \frac{e^2}{N} \sum_{\bm k} 
    \exp\left[\frac{(\varepsilon_{\bm{k}} - \varepsilon_{\bm{k}-\bm{q}}) (t-t')}{i\hbar}\right] 
    \bar{f}_{\bm{k}} (1-\bar{f}_{\bm{k}-\bm{q}})
    \frac{\upsilon^\mu_{\bm k} + \upsilon^\mu_{\bm{k}-\bm{q}}}{2} 
    \frac{\upsilon^\nu_{\bm k} + \upsilon^\nu_{\bm{k}-\bm{q}}}{2} , \label{Pi_<_simplified} \\
    &\Pi^{\rm{RPA},>}_{\mu\nu}(\bm{q}, t, t') = \frac{2}{i\hbar} \frac{e^2}{N} \sum_{\bm k} 
    \exp\left[\frac{(\varepsilon_{\bm{k}} - \varepsilon_{\bm{k}-\bm{q}}) (t-t')}{i\hbar}\right] 
    \bar{f}_{\bm{k}-\bm{q}} (1-\bar{f}_{\bm{k}})
    \frac{\upsilon^\mu_{\bm k} + \upsilon^\mu_{\bm{k}-\bm{q}}}{2} 
    \frac{\upsilon^\nu_{\bm k} + \upsilon^\nu_{\bm{k}-\bm{q}}}{2} . \label{Pi_>_simplified} 
\end{align}
\end{subequations}
The summation should be replaced with an integral in the limit where $N \to \infty$. When compared to the equilibrium version, the only difference is that the equilibrium distribution $f(\varepsilon_{\bm{k}}/\hbar)$ has been replaced with the effective distribution $\bar{f}_{\bm{k}}$. 
Since all the $\Pi$ above are functions of $t-t'$, the Floquet representation is unnecessary here and we can do the usual Fourier transform in $t-t'$.

It is also useful to define the spectral function as
\begin{equation}
    \mathcal{A}_{\mu\nu}(\bm{q}, \omega) 
    \equiv {\rm Im} \left[\Pi^{{\rm RPA},r}_{\mu\nu}(\bm{q}, \omega) - \Pi^{{\rm RPA},a}_{\mu\nu}(\bm{q}, \omega) \right] = 2\operatorname{Im} \left[ {\Pi _{\mu \nu }^{{\text{RPA}},r}(\bm{q},\omega )} \right].
\end{equation}
For the equilibrium case, we have the fluctuation-dissipation theorem
\begin{subequations}
\begin{align}
    \Pi^{\rm{RPA},<}_{\mu\nu}(\bm{q}, \omega) &= i\mathcal{A}_{\mu\nu}(\bm{q}, \omega) N(\omega, T), \\
    \Pi^{\rm{RPA},>}_{\mu\nu}(\bm{q}, \omega) &= i\mathcal{A}_{\mu\nu}(\bm{q}, \omega) [N(\omega, T)+1]
\end{align}
\end{subequations}
with $N(\omega ,T) = 1/\left[ {\exp (\hbar \omega /{k_B}T) - 1} \right]$ the Bose-Einstein distribution.
For the nonequilibrium case with drivings, the \emph{effective} distribution of the photon occupation can be defined as
\begin{equation}
  \bar N(\bm{q},\omega ) = \frac{{\operatorname{Tr} \left[\operatorname{Im} {\bm{\Pi} ^ < }(\bm{q},\omega )\right]}}{{\operatorname{Tr} \left[ \bm{\mathcal{A}}(\bm{q},\omega )\right]}},
\end{equation}
which we find can be further expressed analytically in terms of a weighted sum of the effective electron distribution $\bar{f}$ as
\begin{equation}\label{SEq:eff-ph-dist}
    \bar N(\bm{q},\omega ) = \frac{{\sum\nolimits_{\bm{k}} {{W_{{\bm{k}},{\bm{q}}}}\left( \omega  \right)\bar f\left( {{\varepsilon _{\bm{k}}}} \right)\left[ {1 - \bar f\left( {{\varepsilon _{\bm{k}}} - \hbar \omega } \right)} \right]} }}{{\sum\nolimits_{\bm{k}} {{W_{{\bm{k}},{\bm{q}}}}\left( \omega  \right)\left[ {\bar f\left( {{\varepsilon _{\bm{k}}} - \hbar \omega } \right) - \bar f\left( {{\varepsilon _{\bm{k}}}} \right)} \right]} }}
\end{equation}
with ${W_{{\bm{k}},{\bm{q}}}}\left( \omega  \right) \equiv {\left| {{\bm{\upsilon} _{\bm{k}}} + {\bm{\upsilon} _{{\bm{k}} - {\bm{q}}}}} \right|^2}\delta \left( {{\varepsilon _{\bm{k}}} - {\varepsilon _{{\bm{k}} - {\bm{q}}}} - \hbar\omega } \right)$ the weight function.

\begin{figure}[htbp]
    \centering
    \includegraphics[width=0.8\linewidth]{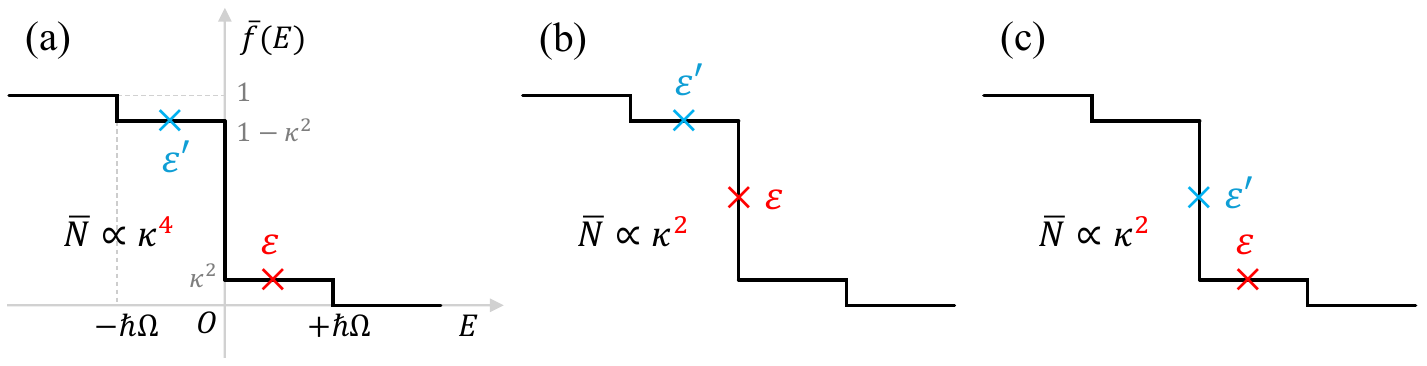}
    \caption{Illustrations of typical patterns of electron occupation in a two-level system and the corresponding asymptotics of $\bar{N}$ respect to $\kappa \equiv 2\mu_d/\hbar\Omega$ for $\kappa \ll 1$. (a) shows a typical pattern, which gives the asymptotics $\bar{N}\propto\kappa^4$; (b) and (c) contribute the lowest-order also most important asymptotics $\bar{N}\propto\kappa^2$, sharing same axes with (a).}
    \label{Fig:figs1}
\end{figure}

Now, we show the asymptotic behavior $\bar{N} \propto \mu_d^2$ for $\kappa  = 2{\mu _d}/\hbar \Omega  \ll 1$, using a two-level model.
For the simple model with the site distance of $a$, the hopping parameter of $t_0$, and the two energies of $\varepsilon$ and $\varepsilon'$ ($\varepsilon > \varepsilon'$), Eq.~\eqref{SEq:eff-ph-dist} can be simplified to $\bar N = \bar f(\varepsilon )\left[ {1 - \bar f\left( {\varepsilon '} \right)} \right]/\left[ {\bar f\left( {\varepsilon '} \right) - \bar f(\varepsilon )} \right]$ by canceling the weighted function $W$, as the $\bm{k}$-dependent electron velocity $\bm{\upsilon}_k$ is replaced by constant $a t_0/\hbar$.
We try to find the lowest-order dependence of $\bar{N}$ on $\kappa$, since we have assumed $\kappa \ll 1$.
Figure~\ref{Fig:figs1}(a) shows a typical pattern of the two-energy levels, where $\varepsilon$ and $\varepsilon '$ are located at the two sides of the Fermi energy, respectively. We note that $\bar f \sim {\kappa ^2}$ when $\kappa \ll 1$, which gives the jump of the first step. In this case, we obtain $\bar{N}\propto\kappa^4$, which does not have the lowest-order dependence on $\kappa$. Figure~\ref{Fig:figs1}(b) and~\ref{Fig:figs1}(c) show another two possible patterns where one of the energy levels is located at the Fermi surface because of the broadening of the Fermi level at finite temperatures. In these two cases, we obtain $\bar{N} \propto \kappa^2$ and consequently $\bar{N}\propto\mu_d^2$.

\subsection{Dependence of heat flux on hopping parameter difference}
\begin{figure}[htbp]
    \centering
    \includegraphics[width=0.9\linewidth]{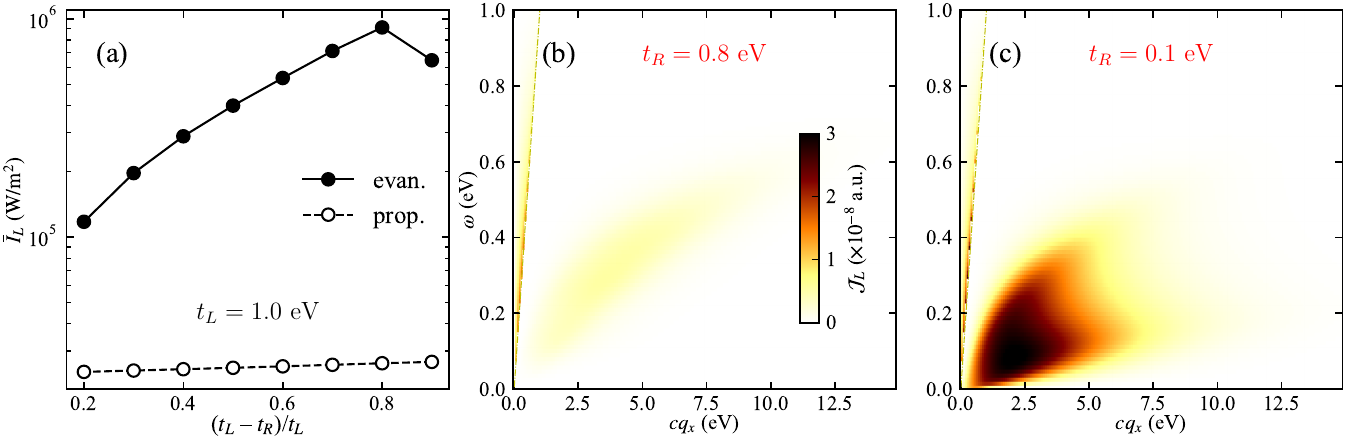}
    \caption{(a) The heat flux changes with the relative hopping difference $(t_L - t_R)/t_L$ with $t_R = 1.0$ eV. The heat flux maps of the left plate with (b) $t_R = 0.8$ eV and (c) $t_R = 0.1$ eV. (b) and (c) share the same color bar. }
    \label{Fig:figs2}
\end{figure}

As shown in Fig.~\ref{Fig:figs2}(a), the heat flux increases with the hopping parameter difference $t_L - t_R$ first. As the difference increases further, the heat flux begins to decrease when $t_R$ = 0.1 eV.
Figures ~\ref{Fig:figs2}(b) and~\ref{Fig:figs2}(c) show the heat-flux maps when $t_R = 0.8$ eV and 0.1 eV, respectively. In contrast to Fig.~\ref{Fig:figs2}(b), the transmission region of evanescent mode ($c|\bm{q}| > \omega$) in Fig.~\ref{Fig:figs2}(c) splits into two plasmon dispersion regions, which is due to the large mismatch of plasmon frequency between the two plates.

\subsection{Possible experimental setups}
\begin{figure}[htbp]
    \centering
    \includegraphics[width=0.7\linewidth]{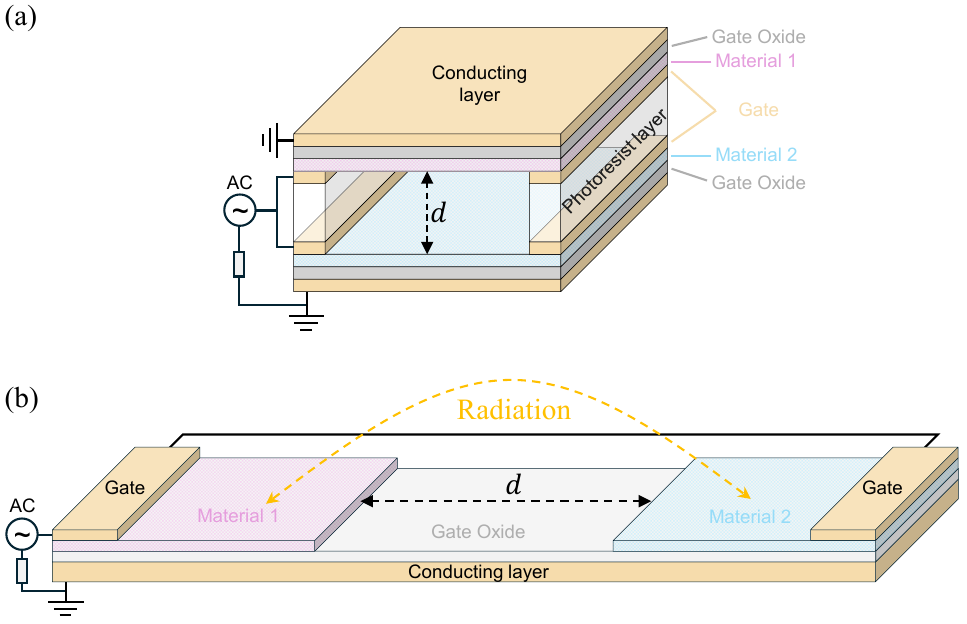}
    \caption{Illustrations of possible experimental setups with (a) face-to-face and (b) side-by-side geometries.}
    \label{Fig:figs3}
\end{figure}

Following the discussion about the possibility of experimental realizations in the main text, here we provide two possible experimental setups, as illustrated in Fig.~\ref{Fig:figs3}(a) and~\ref{Fig:figs3}(b), with face-to-face and side-by-side geometries.
The distance $d$ between two material layers in Fig.~\ref{Fig:figs3}(a) can be controlled by the thickness of the photoresist layers.
A periodic chemical potential driving is realized by applying an out-of-plane AC electric field via a gate voltage with the amplitude $V_g = 2\mu_d/e_0$.
As the barrier and the heat bath of electrons, the gate-oxide layer should be in contact with heaters and temperature sensors to maintain the identical bath temperatures both sides.
The observation of the asymmetry-induced heat transfer requires to exclude the influence of other factors such as the Joule heating from the equivalent series resistance of capacitors and the heat conduction across other contacts (e.g., the gate oxide and photoresist layers).
Given the advances in the experimental measurements of the near-field heat transfer in mesoscopic to macroscopic planar systems~\cite{Hu2008, Song16, Zhang2024, Li2024, Thompson2020, Lim2024}, we believe such setups can be realized in experiments.

\bibliography{asymhr_floquet}{}